\documentclass{aa}
\usepackage{appendix}
\usepackage{latexsym}
\usepackage{graphicx}
\usepackage{natbib}
\usepackage{lscape}
\usepackage{afterpage}  
\usepackage{booktabs}
\usepackage{xcolor}

\begin{document}

\title{The multi--phase ISM in the nearby composite AGN-SB galaxy NGC 4945\thanks{Observations based on {\it Herschel} and the {\it Atacama Pathfinder EXperiment (APEX)} data. {\it Herschel} is an ESA space observatory with science instruments provided by European--led Principal Investigator consortia and with important participation from NASA. {\it APEX} is a collaboration between the Max Planck Institut f$\ddot{u}$r Radioastronomie, the European Southern Observatory, and the Onsala Space Observatory.}: large (parsecs) scale mechanical heating }

\author{Enrica Bellocchi\inst{1}, Jes\'us Mart\'in-Pintado\inst{2}, Rolf G\"usten\inst{3}, Miguel Angel Requena-Torres\inst{4}, Andrew Harris\inst{4}, Paul van der Werf\inst{5}, Frank Israel\inst{5}, Axel Weiss\inst{3}, Carsten Kramer\inst{6}, Santiago Garc\'ia-Burillo\inst{7}, J\"urgen Stutzki\inst{8}} 
\institute{$^1$ Centro de Astrobiolog\'ia ({\tt CAB, CSIC--INTA}), {\tt ESAC} Campus, 28692 Villanueva de la Ca\~nada, Madrid, Spain\\
$^2$ Centro de Astrobiolog\'ia, ({\tt CAB, CSIC--INTA}), Departamento de Astrof\'isica, Cra. de Ajalvir Km.~4, 28850 - Torrej\'on de Ardoz, Madrid, Spain\\ 
$^3$ Max--Planck--Institut f\"ur Radioastronomie, Auf dem H\"ugel 69, 53121 Bonn, Germany \\
$^4$  Department of Astronomy, University of Maryland, College Park, MD 20742, USA\\
$^5$ Leiden Observatory, Leiden University, PO Box 9513, 2300 RA Leiden, The Netherlands\\
$^6$ Instituto Radioastronom\'ia Milim\'etrica, Av. Divina Pastora 7, N\'ucleo Central, E--18012 Granada, Spain\\
$^7$ Observatorio de Madrid, {\tt OAN--IGN}, Alfonso XII, 3, E--28014 Madrid, Spain\\
$^8$ I.Physikalisches Institut der Universit\"at zu K\"oln Z\"ulpicher Str. 77, D-50937 K\"oln, Germany \\
 \email{enrica.bellocchi@gmail.com}\\                }
 \date{Received 20 February 2020 / Accepted 29 July 2020}

\abstract 
{
Understanding the dominant heating mechanism in the nuclei of galaxies is crucial to understand star formation in starbursts ({SB}), active galactic nuclei ({AGN}) phenomena and the relationship between the star formation and {AGN} activity in galaxies. The analysis of the carbon monoxide ($^{12}$CO) rotational ladder {\it versus} the infrared continuum emission (hereafter, $^{12}$CO/IR) in galaxies with different type of activity have shown important differences between them. 
}
{
We aim at carrying out a comprehensive study of the nearby composite AGN-SB galaxy, NGC 4945, using spectroscopic and photometric data from the {\it Herschel} satellite. In particular, we want to characterize the thermal structure in this galaxy by a multi-transitions analysis of the spatial distribution of the $^{12}$CO emission at different spatial scales. We also want to establish the dominant heating mechanism at work in the inner region of this object at smaller spatial scales ($\lesssim$200 pc).
}
{
We present far-infrared (FIR) and sub-millimeter (sub-mm) $^{12}$CO line maps and single spectra (from J$_{up}$ = 3 to 20) using the Heterodyne Instrument for the Far Infrared ({\tt HIFI}), the Photoconductor Array Camera and Spectrometer ({\tt PACS}), and the Spectral and Photometric Imaging REceiver ({\tt SPIRE}) onboard {\it Herschel}, and the Atacama Pathfinder EXperiment ({\tt APEX}). 
We combined the $^{12}$CO/IR flux ratios and the local thermodynamic equilibrium (LTE) analysis of the $^{12}$CO images to derive the thermal structure of the Interstellar Medium (ISM) for spatial scales raging from $\lesssim$200 pc to 2 kpc. In addition, we also present single spectra of low ($^{12}$CO, $^{13}$CO and [CI]) and high density (HCN, HNC, HCO$^+$, CS and CH) molecular gas tracers  obtained with {\tt APEX} and {\tt HIFI} applying LTE and non-LTE analyses. Furthermore, the Spectral Energy Distribution (SED) of the continuum emission from the far-IR to sub-mm wavelengths is also presented. 
}
{
From the non--LTE analysis of the low and high density tracers we derive in NGC 4945 gas volume densities (10$^3$--10$^6$ cm$^{-3}$) similar to those found in other galaxies with different type of activity. From the $^{12}$CO analysis we found clear trend in the distribution of the derived temperatures and the $^{12}$CO/IR ratios.
It is remarkable that at intermediate scales (360 pc-1 kpc, or 19\arcsec-57\arcsec) we see large temperatures in the direction of the X--ray outflow while at smaller scales ($\lesssim$200 pc-360 pc, or $\sim$9\arcsec-19\arcsec), the highest temperature, derived from the high-J lines, is not found toward the nucleus, but toward the galaxy plane. 
The thermal structure derived from the $^{12}$CO multi--transition analysis suggests that mechanical heating, like shocks or turbulence, dominates the heating of the ISM in the nucleus of NGC4945 located beyond 100 pc ($\gtrsim$5\arcsec) from the center of the galaxy. This result is further supported by the \cite{Kazandjian15} models, which are able to reproduce the emission observed at high-J ({\tt PACS}) $^{12}$CO transitions when mechanical heating mechanisms are included. Shocks and/or turbulence are likely produced by the barred potential and the outflow, observed in X--rays. 
}
{}
{}


\titlerunning{Large (parsecs) scale mechanical heating in the nearby composite AGN-SB  galaxy NGC 4945}

\authorrunning{Bellocchi et al.}
\maketitle

\begin{figure*}
\centering
\includegraphics[height=0.38\textwidth]{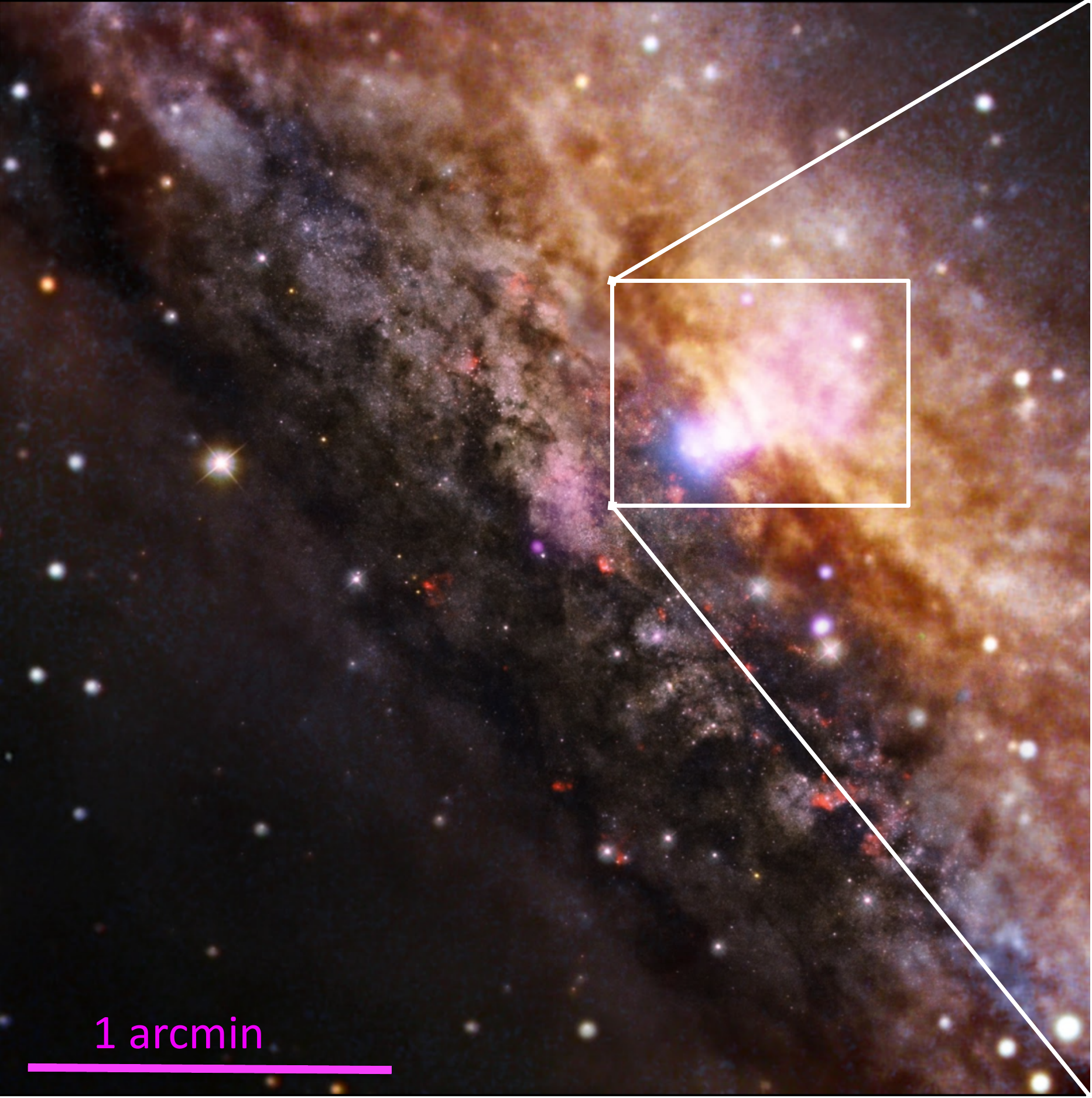}
\includegraphics[width=0.5\textwidth, height=0.38\textwidth]{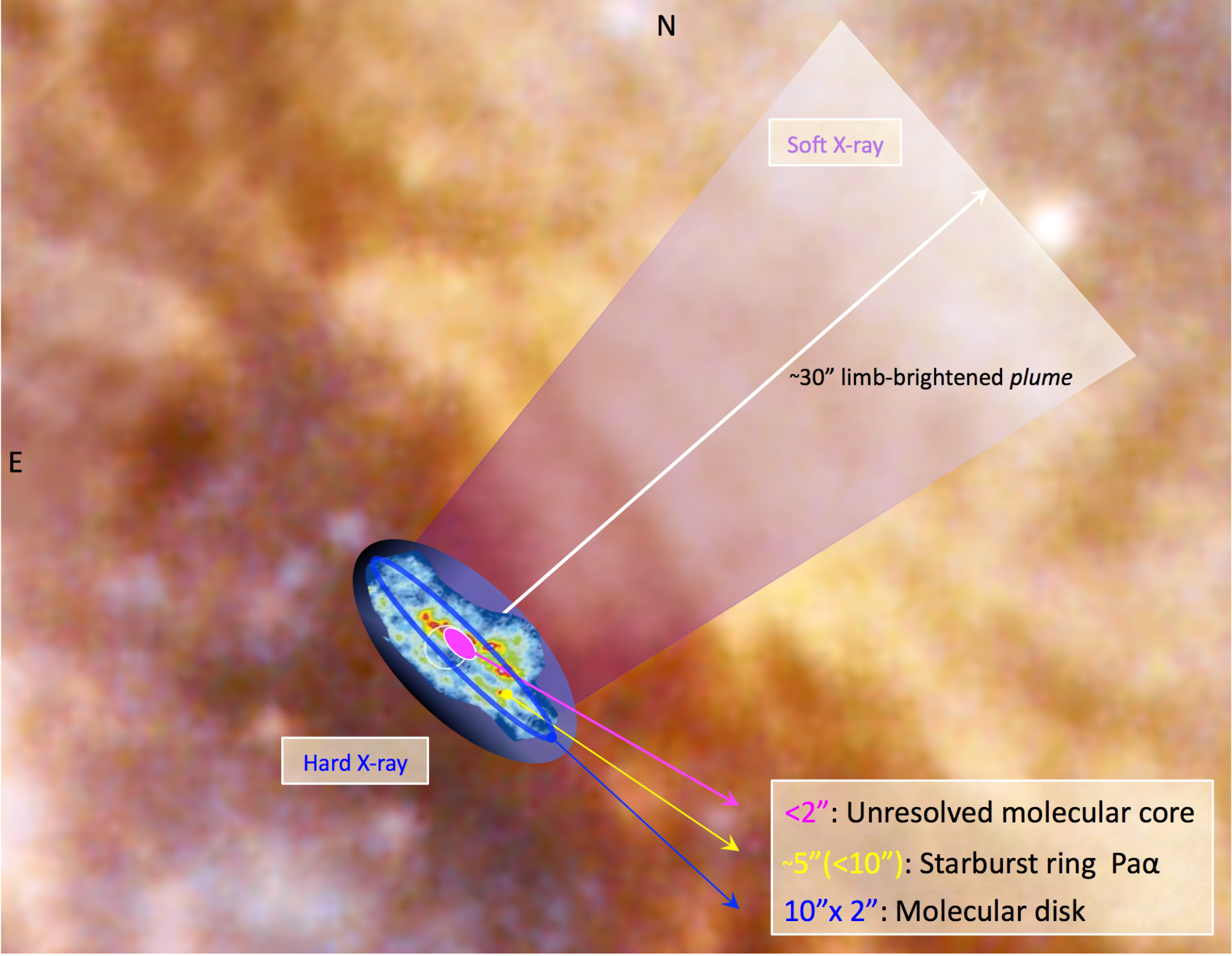}
\caption{{\it Left:} Combined image of X--rays emission from {\it Chandra} (low energy: magenta, high energy: blue), overlaid on an optical image from the European Space Observatory. Credits from NASA/CXC/Univ degli Studi Roma Tre/\cite{Marinucci12}, Optical: ESO/VLT \& NASA/STScI. {\it Right:} 
Cartoon of the central region ($<$1 kpc) of NGC 4945. The optical image shown in the left panel is used as the background. The size (diameter) of the different components observed in the (soft and hard) X--ray, near-IR and radio bands are highlighted: the limb-brightening `plume' in soft X--ray as well as the nuclear hard X--ray emission region (\citealt{Marinucci12}), the starburst ring in Pa$\alpha$ (\citealt{Marconi00}) and the nuclear molecular disk along with the unresolved molecular core (\citealt{Henkel18}).}
\label{mix}
\end{figure*}

\section{Introduction}
\label{Intr}

Galaxy interactions and mergers play important roles in the formation and evolution of galaxies, able to trigger massive starburst ({SB}) and also feed super massive black hole (SMBH). The study of the active galactic nuclei (AGN) and starburst phenomena is a key point in order to understand the relationship between the star formation and AGN activity in galaxies. 

The presence of powerful outflows are believed to play an important role in the evolution of galaxies, able to regulate both the star formation and the growth of the SMBH through `positive' or `negative' feedback in young galaxies (e.g., \citealt{Hopkins09, Cresci15}). 
Recently the evidence of massive molecular outflows in AGN and SB galaxies strongly support the study of outflowing molecular gas as a process able to quickly remove from the galaxy the gas that would otherwise be available for star formation (`negative feedback' on star formation; \citealt{Sakamoto09, Alatalo11, Chung11, Sturm11, Spoon13, Cicone14, GB14}).

The molecular gas plays not only a key role as fuel in the activity process but should also, in turn, be strongly affected by the activity. Depending on the evolutionary phase of the activity, different physical processes can be involved, changing the excitation conditions and the chemistry: strong ultraviolet ({UV}) radiation coming from young massive stars (i.e., photon dominated region or {PDR}; e.g., \citealt{Wolfire10}), highly energetic X--ray photons coming from an AGN (i.e., X--ray dominated region or {XDR}; \citealt{Meijerink06}), as well as shocks and outflows/inflows (see \citealt{Flower10}). X--rays can penetrate more deeply into the ISM than UV photons (\citealt{Maloney96, Maloney99, Meijerink05}): X--rays are able to heat more efficiently the gas, but not the dust, and they are less effective in dissociating molecules (\citealt{Meijerink13}). On the other hand, PDRs are more efficient than XDRs in heating the dust. For this reason, AGNs are suspected to create excitation and chemical conditions for the surrounding molecular gas that are spatially quite different from those in SB environments. The knowledge of the composition and properties of the molecular gas in such environments is essential to characterize the activity itself, and to differentiate between AGN and SB mechanisms.

We focus our analysis on the nearby (D$\sim$3.8 Mpc; \citealt{Karachentsev07}), almost edge-on ($i$ = 78$^\circ$) galaxy, NGC 4945, known to be a remarkable prototype of AGN-SB  composite galaxy. Its proximity (1\arcsec$\sim$19 pc) makes this object an excellent target for studies of molecular gas at the center of an active galaxy. It is also one of the closest galaxies in the local universe that hosts both an {AGN} and a starburst. 
The black hole mass estimated from the velocity dispersion of 150 km s$^{-1}$ obtained from the water maser is around $\sim$10$^6$ M$_\odot$ similar to that of our own Galaxy and a factor of 10 smaller than the black hole hosted in the Sy2 galaxy NGC 1068 (1.5 $\times$ 10$^7$ M$\odot$; \citealt{GG97}). Together with Circinus, it contains a highly obscured Seyfert 2 nucleus (\citealt{Iwasawa93, Marinucci12, Puccetti14}) with associated dense molecular clouds, bright infrared emission, compact (arcsec) radio source, bright H$_2$O `megamaser' ($\sim$15 mas; \citealt{Greenhill97}), strong Fe 6.4 KeV line and variable X--ray emission (\citealt{Schurch02}). 
These observations have revealed a Compton--thick spectrum with an absorbing column density of N$_H$$\sim$2.4--4$\times$10$^{24}$~cm$^{-2}$ (\citealt{Guainazzi00, Itoh08}). The nucleus of NGC 4945 is one of the brightest extragalactic sources at 100 keV (\citealt{Done96}), and the brightest Seyfert 2 AGN at $>$ 20 keV (\citealt{Itoh08}), whose emission is only visible through its reflected emission below 10 keV, due to the large column density that completely absorbs the primary nuclear emission. The emission at higher energy is still visible, though heavily affected by Compton scattering and photoelectric absorption. The nuclear emission between 2-10 keV is enclosed in a region of 12\arcsec$\times$6\arcsec, consistent with the starburst ring observed using molecular gas tracers (e.g., \citealt{Moorwood96, Marconi00, Curran01, Schurch02}).

From IRAS observations we know that about 75\% of the total infrared luminosity of the galaxy (L$_{IR}$~=2.4$\times$10$^{10}$ L$_\odot$) is generated within an elongated region of $<$12\arcsec$\times$9\arcsec centered on the nucleus (\citealt{Brock88}). This structure, as shown in high-resolution HST-NICMOS observations of the Pa$\alpha$ line, is consistent with a nearly edge-on starburst ring of $\sim$5\arcsec-10\arcsec\ (100-200 pc; radius $\sim$2.5\arcsec-5\arcsec, \citealt{Marconi00}). 

Recently, the very inner regions of NGC 4945 have been studied in radio by \cite{Henkel18}, who found a complex structure, composed by a nuclear disk\footnote{According to the results obtained by \cite{Marinucci12}, the nuclear emission between 2-10 keV enclosed in a region of 12\arcsec$\times$6\arcsec\ (i.e., `cold X--ray reflector') is in good agreement with the molecular disk observed by \cite{Henkel18}.} of 10\arcsec$\times$2\arcsec\ enclosing a spatially unresolved molecular core of $\lesssim$2\arcsec, consistent with the X--ray source size observed with {\it Chandra} (\citealt{Marinucci12}). Furthermore, using high density gas tracers (e.g., HCN, CS), they also observed two bending spiral-like arms connected by a thick bar-like structure, extending in the east-west direction from galactocentric radii of $\sim$100 pc out to 300 pc.

A conically shaped wind-blown cavity has been observed to the north-west at different wavelengths, extending out of the galaxy plane from the nucleus, probably produced by a starburst driven wind (\citealt{Moorwood96}). In particular, it has been detected at soft X--ray (i.e., the `plume'\footnote{This structure is observed at soft X--ray band, showing a limb-brightened morphology in the 1-2 keV band, which well correlates with the H$\alpha$ emission. The limb-brightened structure can be attributed to highly excited gas with a low volume-filling factor, produced by an interaction between the starburst-driven wind and the dense ISM surrounding the outflow (as in NGC 253 in which the plume is apparent down to 0.5 keV; see \citealt{Strickland00}). The uniform emission observed below 1 keV might be a direct proof of a mass-loaded superwind (e.g., \citealt{Strickland09}) coming out from the nuclear starburst (\citealt{Schurch02}).}), optical and IR wavelengths (\citealt{Nakai89, Moorwood96, Schurch02, Mingozzi19}). The extension of the outflow ranges from $\gtrsim$~2\arcsec\ in the X--ray band from {\it Chandra} (\citealt{Marinucci12}) reaching $\sim$30\arcsec\ in the optical band, observed with {\it MUSE/VLT} (\citealt{Venturi17}), and in the X--ray band (\citealt{Schurch02}). 

In Fig.~\ref{mix} (left panel) we show a composite view of this galaxy using optical and X--ray emissions from \cite{Marinucci12}. In the right panel we also present a sketch of the observed structures in the inner regions of NGC 4945 at different wavelengths.

In this work we study the molecular composition, as well as the excitation temperature and column density, of the interstellar medium ({ISM}) in the nucleus of NGC 4945. We apply a local thermodynamic equilibrium ({LTE}) multi--transition analysis to a dataset of several molecules observed using the Heterodyne Instrument for the Far Infrared ({\tt HIFI}) onboard {\it Herschel} satellite and the single dish Atacama Pathfinder EXperiment ({\tt APEX}; diameter D = 12 m; \citealt{Gusten06}) antenna. The LTE analysis was also applied to 2D imaging spectroscopy of $^{12}$CO data obtained with the Photoconductor Array Camera and Spectrometer ({\tt PACS}) and the Spectral and Photometric Imaging REceiver ({\tt SPIRE}).
We focus on the LTE analysis applied using the whole sub-mm and far-IR range for studying $^{12}$CO, which allows us to characterize the distribution of the heating at different spatial scale: from large (35\arcsec,$\sim$700 pc) down to small scales (9.4\arcsec,$\sim$200 pc). The aim of this work is to characterize the thermal and density structures at different spatial scales in NGC 4945. Furthermore, the determination of the dominant heating mechanism and the origin of the observed heating pattern in the inner regions of this object are also analyzed.
The photometric data allow us to derive the mass of dust and the corresponding mass of gas (once assumed a specific gas-to-dust ratio) and compare with the expectations from the heating mechanisms inferred from the $^{12}$CO analysis.

The paper is organized as follows: In Section~2 we introduce the observations and the data analysis applied for each instrument. In Section~3 we present the Spectral Energy Distribution (SED) results derived from analyzing photometric data obtained from different instruments, from the far-IR to sub-mm wavelengths. Section~4 is dedicated to the derivation of the column densities and the excitation temperatures obtained using the high spectral resolution {\tt HIFI} and {\tt APEX} data for all molecules ($^{12}$CO, $^{13}$CO, HCN, HNC, HCO$^+$, CS, [CI], CH) for spatial scales between 20\arcsec-30\arcsec ($\sim$400-700 pc). In Section~5 we focus our analysis on the thermal and column density structures of $^{12}$CO using 2D imaging spectroscopy: from {\tt SPIRE} ($\gtrsim$700 pc) down to smaller spatial scales using {\tt PACS} ($\lesssim$200 pc). Section~6 is devoted to the discussion of the results in order to understand the origin of both gas and dust heating mechanisms. Our main conclusions are summarized in Section~7. Appendix~\ref{App_SED} presents detailed information on the derivation of the flux densities used in the SED fitting analysis (\S~3). Throughout the paper we consider H$_0$ = 71 km s$^{-1}$ Mpc$^{-1}$, $\Omega_M$ = 0.27 and $\Omega_\Lambda$=0.73.

\section{Observations and data analysis}

\subsection{Observations}

\begin{table*}
\centering
\begin{small}
\caption{General properties of the {\tt HIFI}, {\tt PACS} and {\tt SPIRE} $^{12}$CO observations.}
\label{instrums_archive}
\begin{tabular}{c ccccc} 
\hline\hline\noalign{\smallskip}  
Instrument &  Type Obs & $^{12}$CO trans (J+1$\rightarrow$J) or band ($\mu$m) &  Obs ID (1342) & Level data& PI  \\
(1) & (2) &  (3)& (4) &(5)  & (6) \\
\hline\noalign{\smallskip} 	
{{\tt HIFI} }& {Spectroscopic } & {5$\rightarrow$4 to 9$\rightarrow$8}& 200939, 200989, 200944 & 2.5 & R. G\"usten \\ 
\hline\noalign{\smallskip} 	
{\tt PACS} & Spectroscopic& 15$\rightarrow$14; 20$\rightarrow$19 & 247789 & 2 & C. Hailey\\
{\tt PACS} & Photometric& 70; 100; 160 & 223660; 203022 &  2.5 & E. Sturm\\
\hline\noalign{\smallskip} 	
{\tt SPIRE} & Spectroscopic & {4$\rightarrow$3 to 8$\rightarrow$7}; {9$\rightarrow$8 to 13$\rightarrow$12} & 212343 & 2 & E. Sturm\\
{\tt SPIRE} & Photometric& 250; 350; 500 & 203079 & 2 & E. Sturm\\
\hline\hline\noalign{\smallskip} 	
\end{tabular}
\end{small}
\begin{minipage}{18cm}
\small
{{\bf Notes:} Column~(1): Instrument; Column~(2): type of observation; Column~(3): $^{12}$CO transitions or band; Column~(4): ID of the observation. The code `1342' has to be added before the number; Column~(5): level of the data used (see text for details); Column~(6): principal investigator of the observation.}
\end{minipage}
\end{table*}

\subsubsection{Heterodyne Instrument for the Far Infrared ({\tt HIFI}) and the Atacama Pathfinder EXperiment ({\tt APEX})}

The {\tt HIFI} observations are taken in the pointed dual beam switch (DBS) mode covering the frequency range between 480 GHz to 1270 GHz (band 1 to 5; see \citealt{Jackson05}) and from 1410 GHz up to 1910 GHz\footnote{The whole frequency range corresponds to a wavelength range between 157 to 625 $\mu$m.} (bands 6 and 7; see \citealt{Cherednichenko02}) at high spectral resolution (R = 10$^6$-10$^7$). 
The half-power beam width (HPBW) of the telescope was 37\arcsec\ and 12\arcsec\ at 572 GHz and 1892 GHz, respectively. The {\tt HIFI} Wide Band Spectrometer ({WBS}) was used with an instantaneous frequency coverage of 4 GHz and an effective spectral resolution of 1.1 MHz. Two orthogonal polarizations (horizontal, H, and vertical, V) were recorded and then combined together to end up with a higher signal-to-noise ratio (SNR). We used the standard {\it Herschel} pipeline Level 2.5 which provides fully calibrated spectra (\citealt{Graauw10}; see Tab.~\ref{instrums_archive}).
In particular, the {\tt HIFI} Level 2.5 pipeline combines the Level 2 products into final products. Single-point data products are stitched spectra for each of the polarizations and backends applicable to the observation. The spectra were produced using the pipeline version Standard Product Generation (SPG) v14.1.0 within HIPE. For further information see \cite{Shipman17}.

In addiction to the {\tt HIFI} data we obtained sub-mm data of lower transitions (J$_{up}$ = 3, 4) of $^{12}$CO, $^{13}$CO, HCN, HNC and HCO$^+$ using the {\tt FLASH+}\footnote{\url{https://www.eso.org/public/teles-instr/apex/flash-plus/}} receiver at 345 GHz at {\tt APEX} (see Tab.~\ref{list_molecules}).
The half-power beam width ({HPWB}) ranges between 21\arcsec\ down to 17\arcsec\ at 272 and 354 GHz, respectively. The spectral resolution provided by a Fourier Transform Spectrometer (FTS) was smoothed to a velocity resolution of 20~MHz. The data reduction was initially performed using {\tt CLASS}\footnote{{\tt CLASS} is a data reduction software, which is part of Gildas (e.g., \citealt{Maret11}).} and then imported in {\tt MADCUBA}\footnote{Madrid Data Cube Analysis has been developed at the Center for Astrobiology (CAB, CSIC--INTA) to analyze single spectra and datacubes: \url{http://cab.inta-csic.es/madcuba/MADCUBA_IMAGEJ/ImageJMadcuba.html}. More details in \S 2.2.} (\citealt{Rivilla16, Martin19}).

\subsubsection {Photoconductor Array Camera and Spectrometer ({\tt PACS})}

{\tt PACS} is a photometer and a medium resolution spectrometer\footnote{{\tt PACS} was developed and built by a consortium led by Albrecht Poglitsch of the Max Planck Institute for Extraterrestrial Physics, Garching, Germany. NASA is not one of the contributors to this instrument.}. In Imaging dual-band photometry, {\tt PACS} simultaneously images the wavelength range between 60-90 $\mu$m, 90-130 $\mu$m and 130-210 $\mu$m over a field of view (FoV) of 1.75$^\prime$ $\times$ 3.5$^\prime$. {\tt PACS}' grating imagining spectrometer covers the 55-210 $\mu$m spectral range with a spectral resolution between 75-300 km s$^{-1}$ over a FoV of 47\arcsec$\times$47\arcsec, resolved into a 5$\times$5 spaxels, each of them with an aperture of 9.4\arcsec.

{\tt PACS} data were provided from the {\it Herschel} archive\footnote{\url{http://www.cosmos.esa.int/web/herschel/science-archive}} using Level 2 and 2.5 products (see Tab.~\ref{instrums_archive}). 
The {\tt PACS} Level-2 spectroscopy products can be used for scientific analysis. Processing to this level contains actual spectra and is highly observing modes dependent. The result is an Image of Cube products (for further details see \citealt{Poglitsch10}). The Level-2.5 photometric products are maps (produced with JScanam, Unimap and the high-pass filter pipelines) generated by combining scan and cross-scan observations taken on the same sky field. The {\tt PACS} products were produced using the pipeline version SPGv14.2.2 within HIPE.

\begin{table}
\centering
\begin{tiny}
\caption{Line transitions and instrument used.}             
\label{list_molecules}      
\begin{tabular}{l |c c}      
\hline\hline                
Line (Transition) & Rest frequency $\nu$    & Instrument	\\    
 		&  (GHz) &   \\     
 (1) &  (2)  &  (3)   \\
\hline                                
$^{12}$CO J= 3 $\rightarrow$ 2    	&  345.79 	 	& {\tt APEX}  \\
$^{12}$CO J= 5 $\rightarrow$ 4	&  576.27 	  	& {\tt HIFI}  \\
$^{12}$CO J = 6 $\rightarrow$ 5	&  691.47 	  	& {\tt HIFI} \\
$^{12}$CO J = 9 $\rightarrow$ 8  	&  1036.91 	 	& {\tt HIFI}	  \\
\hline
$^{13}$CO J= 3 $\rightarrow$ 2 	& 330.59 	& {\tt APEX}	\\
$^{13}$CO J = 6 $\rightarrow$ 5	& 661.07 &  {\tt HIFI}		\\
$^{13}$CO  J = 9 $\rightarrow$ 8 	& 991.33 	& {\tt HIFI}	\\
\hline                                  
HCN  J = 4 $\rightarrow$ 3		& 354.50 		& {\tt APEX}	\\
HCN  J = 6 $\rightarrow$ 5		& 531.72 		& {\tt HIFI}	\\
HCN  J = 7 $\rightarrow$ 6		& 620.30 		& {\tt HIFI}	\\
HCN  J = 12 $\rightarrow$ 11		& 1062.98 	&{\tt HIFI}	\\
\hline
HNC J= 3 $\rightarrow$ 2			& 271.98 	& {\tt APEX} \\
HNC J = 4 $\rightarrow$ 3		& 362.63 	& {\tt APEX} \\
HNC J = 6 $\rightarrow$ 5	& 543.89 	& {\tt HIFI} \\
HNC J = 7 $\rightarrow$ 6		& 634.51 	& {\tt HIFI}  \\
\hline
HCO$^+$ J = 4 $\rightarrow$ 3	& 356.73 & {\tt APEX}	\\
HCO$^+$ J = 6 $\rightarrow$ 5	& 356.73 & {\tt HIFI}	\\
HCO$^+$ J = 7 $\rightarrow$ 6		& 356.73 & {\tt HIFI}	\\
\hline
CS J = 6 $\rightarrow$ 5		& 293.91 & {\tt HIFI}	\\
CS J = 7 $\rightarrow$ 6		& 342.88 &  {\tt HIFI}	\\
CS J = 10 $\rightarrow$ 9		& 489.75 &  {\tt HIFI} 	\\
CS J = 12 $\rightarrow$ 11 	& 538.69 & {\tt HIFI}	\\
CS J = 13 $\rightarrow$ 12	& 587.62 & {\tt HIFI}	\\
\hline
[CI] $^3$P$_1$ $\rightarrow$ $^3$P$_0$  		& 492.16 		& {\tt HIFI}   \\
\hline
CH $^2\Pi_{1/2}$ J = 3/2--1/2  		& 532.72 	 & {\tt HIFI}\\
CH $^2\Pi_{1/2}$ J = 3/2--1/2 			& 536.76 	 & {\tt HIFI}\\
CH $^2\Pi_{3/2}$ J = 5/2--3/2 			& 1656.97& {\tt HIFI}\\
\hline\hline
\end{tabular}
\end{tiny}
\vskip2mm
\begin{minipage}{9cm}
\small
{{\bf Notes:} Column~(1): line and rotational transition (J); Column~(2): frequency of the molecule in giga hertz (GHz); Column~(3): instrument used for the observation.}
\end{minipage}
\end{table}

\begin{table}[b]
\begin{tiny}
\caption{{\tt SPIRE} beams in the photometric and spectroscopic modes. }
\label{instrums}
\begin{tabular}{lccccc} 
\hline\hline\noalign{\smallskip}  
 Subinstrument && \multicolumn{1}{c}{{Photom}} & & \multicolumn{2}{c}{{SPectr}} \\
            \cmidrule(lr){2-4} \cmidrule(lr){5-6}\noalign{\smallskip}
 	& PSW	& PMW & PLW & SSW&	SLW\\
\hline\noalign{\smallskip} 	
band ($\mu$m)	& 250 	&	350  &  500  & 192-313 & 303-671\\
beam (FWHM)  & 17.6\arcsec & 23.9\arcsec & 35.2\arcsec & 17\arcsec-21\arcsec & 29\arcsec-42\arcsec\\
\hline\hline\noalign{\smallskip} 	
\end{tabular}
\end{tiny}
\end{table}

\subsubsection {Spectral and Photometric Imaging REceiver ({\tt SPIRE})}
\label{SPIRE_resolutions}

{\tt SPIRE} consists of a three band imaging photometer and an imaging Fourier Transform Spectrometer (FTS). The photometer carries out broad--band photometry ($\lambda$/$\Delta\lambda$$\approx$3) in three spectral bands centered on approximately 250, 350 and 500 $\mu$m with an angular resolution of about 18\arcsec, 24\arcsec and 35\arcsec, respectively (see Tab.~\ref{instrums}). The spectroscopy is carried out by a FTS that uses two overlapping bands to cover 194-671 $\mu$m (447-1550 GHz) simultaneously, the {\tt SSW} short wavelength band (190-313 $\mu$m; 957-1577 GHz) and {\tt SLW} long wavelength band (303-650 $\mu$m; 461-989 GHz). 
The {\tt SPIRE}--FTS is a low spatial and spectral (1.2 GHz) resolution mapping spectrometer. In particular, the beam full width at half--maximum (FWHM) of the {\tt SSW} bolometers is 18\arcsec, approximately constant with frequency. The beam FWHM of the {\tt SLW} bolometers varies between $\sim$30\arcsec\ and 42\arcsec\ with a complicated dependence on frequency (\citealt{Swinyard2010}). 

We use {\tt SPIRE} Level-2 spectroscopic and photometric products for our analysis. These data are processed to such a level that scientific analysis can be performed. The {\tt SPIRE} Level-2 photometer products (maps) are calibrated in terms of in-beam flux density (Jy/beam)\footnote{For further details see \url{http://herschel.esac.esa.int/hcss-doc-15.0/print/pdd/pdd.pdf}.}. The photometric and spectroscopic {\tt SPIRE} data Level-2 were produced using the pipeline version SPGv14.1.0 within HIPE.

\begin{figure*}
\centering
\hskip-10mm\includegraphics[width=0.75\textwidth, height=0.65\textwidth, angle=0]{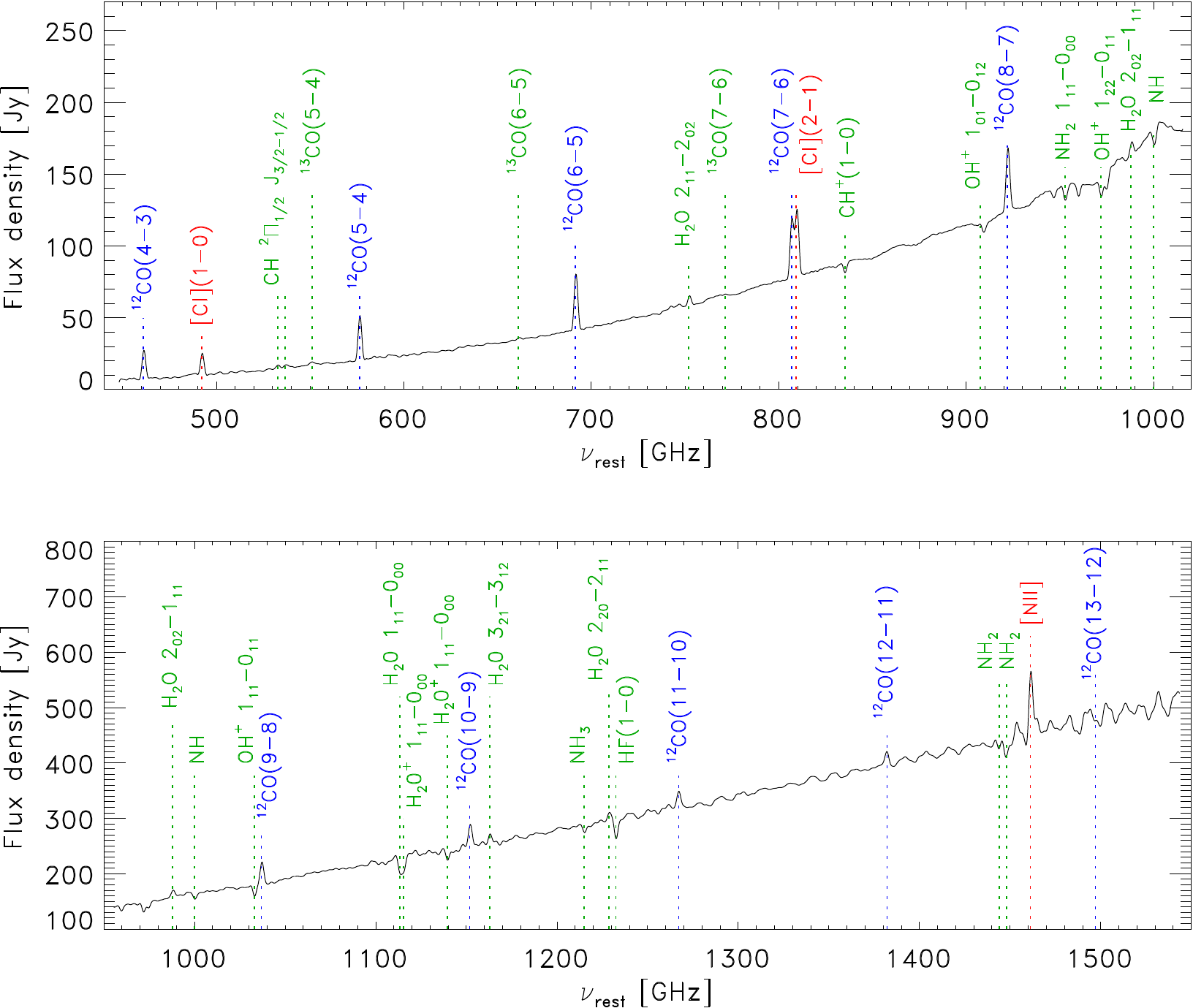}
\caption{{\tt SLW} ({\it upper panel}) and {\tt SSW SPIRE} ({\it lower panel}) spectra corresponding to the peak emission in the same FoV in the rest-frame frequency. $^{12}$CO lines are shown in blue while fine structure lines, as [CI] and [NII], are indicated in red. Other molecular species observed in emission and/or in absorption are shown in green.}
\label{SPIRE_spectra}
\end{figure*}

Our data have been achieved with an intermediate spatial sampling: in such a case, the pixel size for the {\tt SLW}  and {\tt SSW} bolometers are 35\arcsec\ and 19\arcsec, respectively. 
The $^{12}$CO ladder (from J$_{up}$ = 4 to 13) is the most prominent spectral feature in this frequency range. These mid-J $^{12}$CO emission lines probe warm molecular gas (upper-level energies ranging from 55 K to 500 K above the ground state) that can be heated by ultraviolet photons, shocks, or X--rays originated in the active galactic nucleus or in young star-forming regions. In the {{\tt SPIRE}--FTS} range besides the $^{12}$CO transitions we also detected the prominent [CI]492 $\mu$m, [CI]809 $\mu$m and [NII]205 $\mu$m transitions across the entire system along with several molecular species observed in absorption (see Fig.~\ref{SPIRE_spectra}).
A baseline (continuum) subtraction of second or third order has been applied to these spectra. Detailed information on the {\tt SPIRE} observations are summarized in Tab.~\ref{instrums_archive}.


\subsection{Data analysis}
\label{Analysis_data}

Using high spectral resolution {\tt HIFI} and {\tt APEX} data we carried out multi-line analysis of $^{12}$CO, $^{13}$CO, HCN, HCN, HCO$^+$, CS, [CI], CH which have been all detected in emission. 
Other molecules such as NH, NH$_2$, OH$^+$, HF, H$_2$O have been detected in absorption and they will be analyzed in feature work.
{\tt HIFI} and {\tt APEX} products are calibrated in antenna temperature (T$^\star_A$). This was converted to main beam temperature (T$_{MB}$) according to the relation:

\begin{equation}
\hskip1cm T_{MB} = \frac{\eta_f}{\eta_{MB} } \hskip 3mmT^\star_A,
\end{equation}

where $\eta_f$ is the forward efficiency\footnote{The forward efficiency, $\eta_f$, measures the fraction of radiation received from the forward hemisphere of the beam to the total radiation received by the antenna.} of the telescope and $\eta_{MB}$ is the main beam efficiency. For the {\tt HIFI} data $\eta_{MB}$ ranges from 0.69 to 0.76 with a $\eta_f$ = 0.96, while for the {\tt APEX} data we used a $\eta_{MB}$ = 0.73 and $\eta_f$\footnote{The {\tt APEX} beams and main beam efficiencies are taken from the website \url{http://www.apex-telescope.org/telescope/efficiency/}} = 0.97. The main beam temperature T$_{MB}$ has been corrected for beam dilution, according to the relation:

\begin{equation}
\hskip1cm T'_{MB} = \left(\frac{\theta^2_s + \theta^2_b}{\theta^2_s}\right)\hspace{2mm}T_{MB} ,\end{equation}

where $\theta_s$ and $ \theta_b$ are the source size and the beam size\footnote{The value of the {\tt HIFI} beam are taken from the website \url{http://herschel.esac.esa.int/Docs/HIFI/html/ch05s05.html\#table-efficiencies}.}, respectively. For this object a source size of 20\arcsec\ has been considered (\citealt{Wang04}). 

The {\tt HIFI} spectra were smoothed to a resolution of 20~km s$^{-1}$. When needed, further smoothing and baseline corrections have been applied to the spectra to improve the signal-to-noise ratio (SNR).

The molecular emission was modeled with {\tt SLIM}\footnote{{\tt SLIM} stands for `Spectral Line Identification and Modelling of the line profiles'. It identifies the line using the JPL, CDMS and LOVAS catalogues (\citealt{Lovas92, Pickett98, Muller01}) as well as recombination lines.} package within {\tt MADCUBA} (\citealt{Martin19}). In the model, {\tt SLIM} fits the synthetic LTE line profiles to the observed spectra. 
The fit is performed in the parameter space of molecular column density N$_{mol}$, excitation temperature T$_{ex}$, velocity v$_{LSR}$ and width of the line (FWHM) to the line profile and source size.
{\tt SLIM} allows the presence of different components (`multi Gaussian fit'), which can be differentiated using different physical parameters (e.g., column density, excitation temperature, velocity). 
In case of multiple transitions fit, two (or more) T$_{ex}$ can be also assumed (`multiple excitation temperature', see \S~3.3.2 in \citealt{Martin19}).

To properly account for the beam dilution factor, a source size was fixed as an input parameter.

\section{Continuum analysis}

\subsection{Intrinsic source size of the dust emission from PACS and SPIRE photometry}
\label{S_size}

\begin{table}
\centering
\begin{small}
\caption{Intrinsic source size using {\tt PACS} and {\tt SPIRE} photometric data.  }
\label{Conti_mix}
\begin{tabular}{ccccc} 
\hline\hline\noalign{\smallskip}  
Band--& Nominal   &	 Intrinsic  &     Flux  &  PSF  \\
Instrument			  & pixel size &  source size  & density  &	\\
($\mu$m--)&   (\arcsec) &   (\arcsec $\times$ \arcsec)   & (Jy) &(\arcsec)  	\\
(1) &  (2) & (3) &  (4) & (5) \\
\hline\hline\noalign{\smallskip}  
 70 {\tt PACS}   	&	1.6    &      7.4$\times$3.5 (0.5)    	&	258 (5) &      5.5 	\\
100 {\tt PACS} 	&	1.6    &      8.1$\times$3.7 (0.8)    &	340 (9)  	&      7.2	\\
160 {\tt PACS}   	&	3.2    &      9.3$\times$2.8 (1.6)   	&	329 (19)   &      11.5	\\
\hline
250 {\tt SPIRE} 	&	6      &  13.9$\times$5.8 (2.0)    	& 235 (6)         &  17.6	 		\\
350 {\tt SPIRE}    &	10     &  21.7$\times$3.8 (3.3)    	& $<$95$^{(a)}$  &  23.9			\\
500 {\tt SPIRE}  	&	14     &  16.3$\times$$<$35 (4.7)  	& 34 (0.3)   	   	&  35.2		\\
\hline\hline
\end{tabular}
\begin{minipage}{9cm}
\vskip2mm{{\bf Notes:} Column~(1): Photometric band and instrument; Column~(2): nominal pixel size of the instrument for the specific band; Column~(3): intrinsic source size (and uncertainty) obtained deconvolving the observed source size for the corresponding point spread function (PSF) value; Column~(4):~flux density enclosed in the observed source size; Column~(5): PSF in the different bands. $^{(a)}$ Lower limit value due to the presence of a bad pixel enclosed in the observed source size. }\\
\end{minipage}
\end{small}
\end{table}

In this section we derived the intrinsic (deconvolved) size of the different components of the dust emission in NGC 4945, as small and large grains along with polyaromatic hydrocarbons (PAH's; \citealt{Lisenfeld02, daCunha08}), using the photometric data from {\tt PACS} (70, 100, 160 $\mu$m) and {\tt SPIRE} (250, 350, 500 $\mu$m). These photometric images have been retrieved from the {\it Herschel} archive (see Tab.~\ref{instrums_archive}). 
We measured the FWHM sizes of the peak emission and we deconvolved them with the relevant PSF sizes assuming gaussian shapes for both. At these moderate resolutions the galaxy shows the presence of a compact source plus a disk component: at these wavelengths the contribution of the compact source emission dominates over the disk component within the beam. The results of the intrinsic source size are listed in Tab.~\ref{Conti_mix}.

\begin{figure*}
\centering
\includegraphics[width=0.8\textwidth]{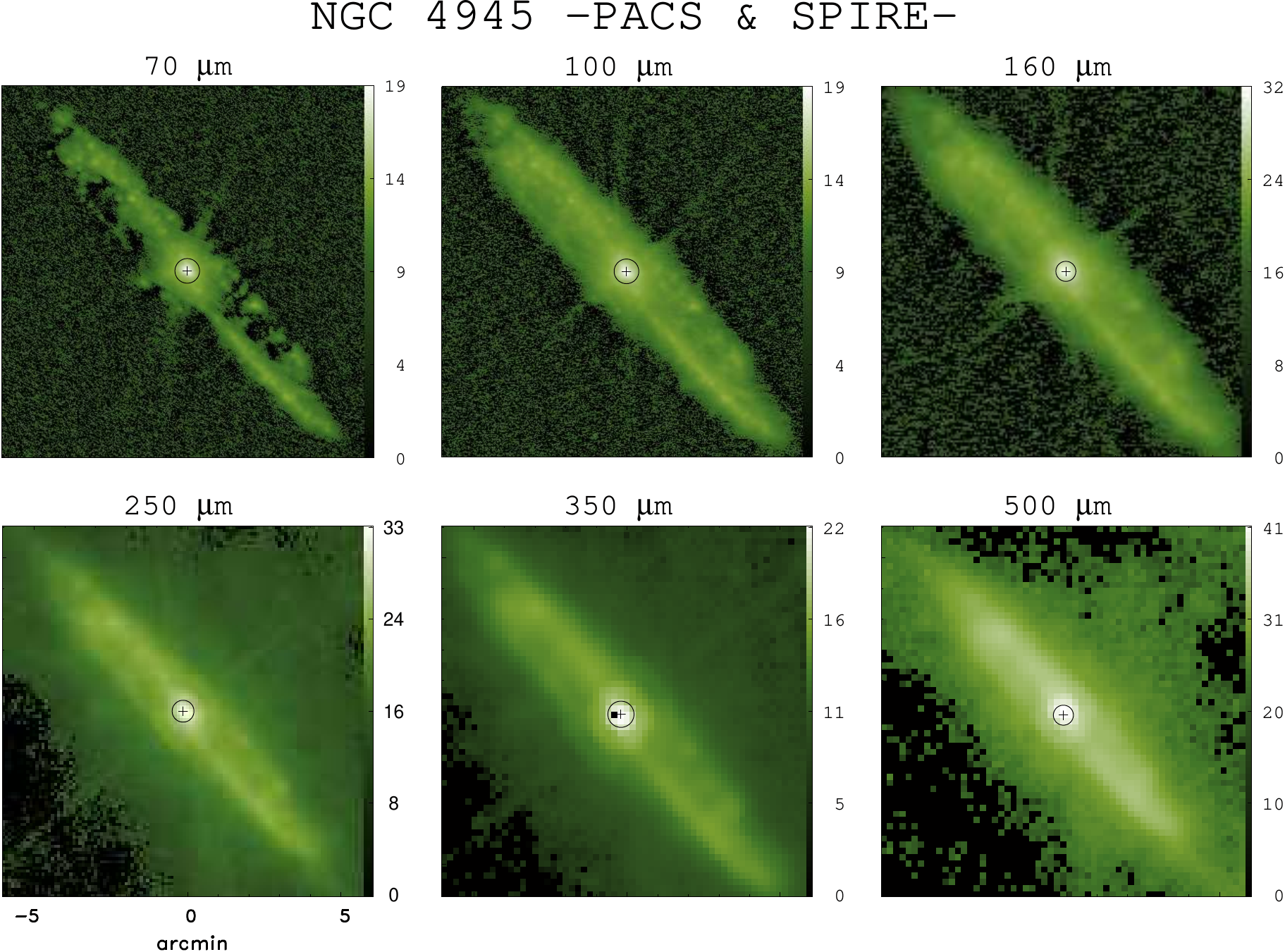} 
\caption{Photometric images of NGC 4945 at the {\tt PACS} (70, 100 and 160 $\mu$m; {\it top}) and {\tt SPIRE} (250, 350 and 500 $\mu$m; {\it bottom}) wavelengths. The flux units have been converted to {\it Jansky} for both the {\tt PACS} and {\tt SPIRE} data (see text for details). The black circle in each panel identifies a beam of 40\arcsec$\times$40\arcsec. From {\tt PACS} 70 $\mu$m to {\tt SPIRE} 500 $\mu$m wavelengths an aperture of 40\arcsec\ corresponds to 25, 25, 12.5, 6.7, 4 and 3 pixels, respectively. The black cross represents the peak emission in each band.}
\label{PACS_SPIRE}
\end{figure*}

We then computed the flux density enclosed in the observed source size. For the {\tt PACS} data the maps are in units of [Jy pixel$^{-1}$] while the {\tt SPIRE} maps (point source calibrated) are in units of [Jy beam$^{-1}$]. 
Thus, to compute the total flux density included in the (observed) source size, we treated the two dataset as follows: for the {\tt PACS} data we simply sum all fluxes of each pixel within the estimated source size while for the {\tt SPIRE} data we multiply the sum of all values within the source size by a factor of $(pixel$ $size/PSF)^2$ at the corresponding wavelength (Tab.~\ref{Conti_mix}).
From these results the emission of NGC 4945 is resolved in both directions at all but one {\tt PACS} and {\tt SPIRE} wavelengths: in particular, at 500 $\mu$m the emission is resolved in one direction and unresolved in the perpendicular direction. The photometric {\tt PACS} and {\tt SPIRE} maps are shown in Fig.~\ref{PACS_SPIRE}.

\subsection{Spectral Energy Distribution of NCG 4945 }
\label{par_SED}

\begin{figure}
\hskip0mm
\includegraphics[width=0.48\textwidth]{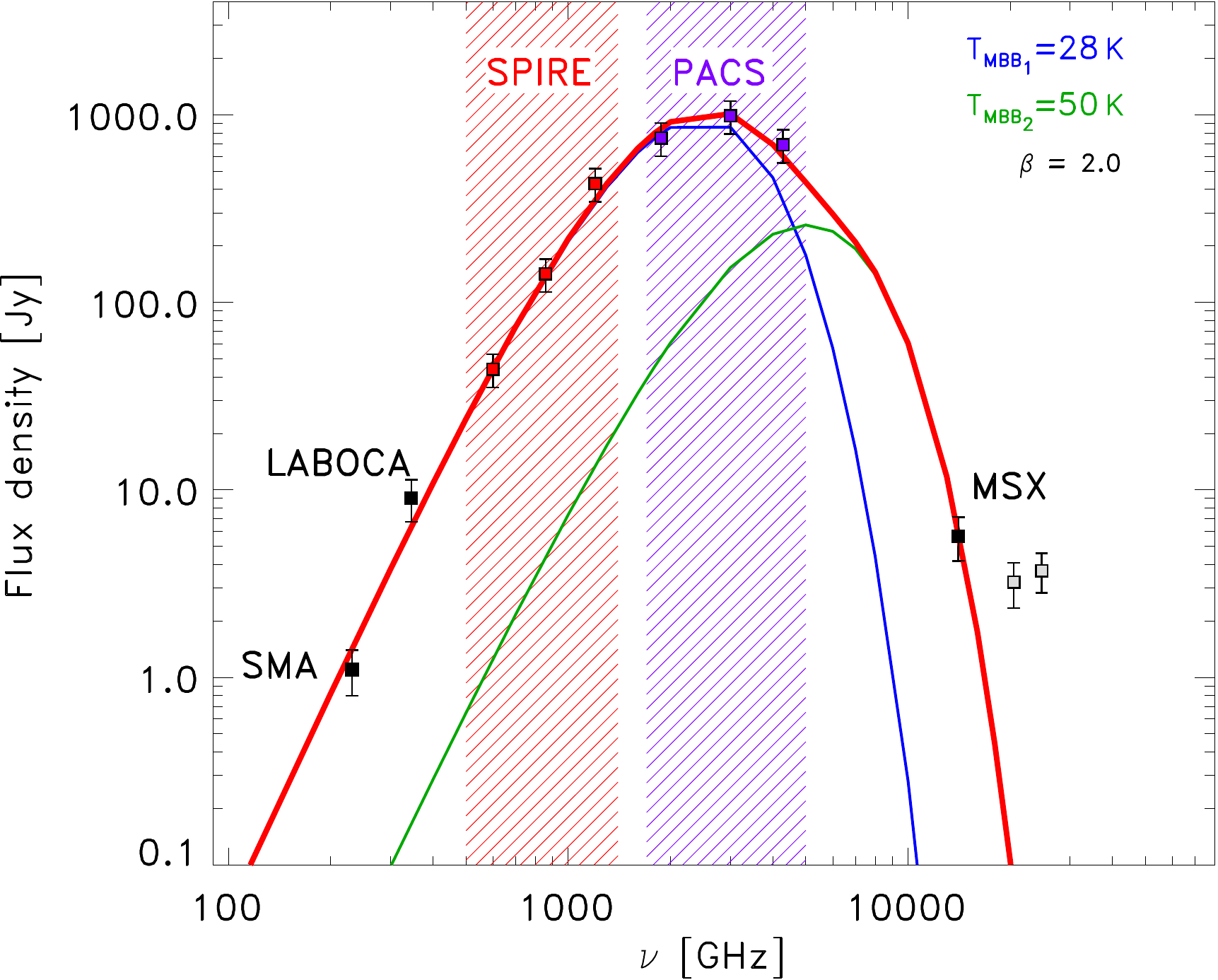} 
\caption{SED fitting results for NGC 4945. The best fit solution (red solid line) is obtained when two dust components with a temperature of 28 K (blue MBB) and 50 K (green MBB) are needed, assuming source size of 20\arcsec$\times$10\arcsec. The total dust mass obtained from the fit is $\lesssim$10$^7$ M$_\odot$. The {\tt SPIRE} and {\tt PACS} data are shown using red and violet squares, respectively. Additional data used in the fit are shown in filled black squares. In particular, at shorter frequencies the {\tt SMA} (\citealt{Chou07}) and {\tt LABOCA} (\citealt{WEIB08}) data were added (see text for details) obtaining a $\beta$ parameter of 2.0. At higher frequencies the {\tt MSX} data point is considered in the fit while two {\tt IRAC} data points are only shown for completeness (in gray).}
\label{SED_4945}
\end{figure}

We derived the spectral energy distribution (SED) combining {\tt PACS} and {\tt SPIRE} data with those obtained at sub-mm wavelengths from \cite{WEIB08} using Large {\tt APEX} Bolometer Camera ({\tt LABOCA}) and from \cite{Chou07} using the the Submillimeter Array ({\tt SMA}) within an aperture of 40\arcsec$\times$40\arcsec\ (see Fig.~\ref{SED_4945}). This is a reasonable value to consider most of the emission from the inner regions of the galaxy at all wavelengths: the aperture considered is shown in Fig.~\ref{PACS_SPIRE} for both {\tt PACS and {\tt SPIRE}} bands.
The flux density obtained in \cite{WEIB08} in an aperture of 80\arcsec $\times$ 80\arcsec\ has been scaled to our aperture, deriving a flux density of 9.05 ($\pm$ 1.3) Jy (see App.~A for details). 
We also added one far-IR data point from {\tt MSX} at $\sim$20 $\mu$m. Other data at shorter wavelengths were available from {\tt MSX, IRAC} and {\tt 2MASS} catalogues but they were not included in this analysis because their emission (in the range $\sim$3-17 $\mu$m) is strongly affected by several emission features from PAH molecules (see \citealt{Povich07, PBeaupuits18}). In Tab.~\ref{cont_SED} the derived flux densities are shown.

\begin{table}[h]
\centering
\begin{tiny}
\caption{Continuum flux density values in the far-IR and sub-mm wavelengths range.}
\label{cont_SED}
\begin{tabular}{ccccc} 
\hline\hline\noalign{\smallskip}  
Data 	&	 Wavelength &  Frequency   &  Flux  		& Flux density \\
	 &  &  &  density & MBB \\
	 & ($\mu$m) &  (GHz)	& (Jy)   &  (Jy)\\
(1) & (2) &(3) &(4) & (5)\\
\hline\noalign{\smallskip} 	
{\tt MSX}	 & 21.34 &14058	& 6 (2)	&   7	\\
{\tt PACS}	 & 70 &4286 & 694 (138) 	&  620 \\
{\tt PACS}	 & 100 & 3000 & 988 (198)	&	1020	\\
{\tt PACS}	 & 160 & 1875 & 753 (151)	&	 845  \\
{\tt SPIRE}	 & 250 &1200 & 430 (86)	& 370	\\
{\tt SPIRE}	 & 350 & 857 & 142 (28)	& 150 \\
{\tt SPIRE}	 & 500 & 600 & 44 (9)	& 45 	\\
{\tt LABOCA}	 & 870 &345 & 9.1 (1.3)	&  7	\\
{\tt SMA}	 & 1300 & 230 &  1.0 (0.3) 	& 1.4	\\
\hline\hline
\end{tabular}
\begin{minipage}{8.7cm}\vskip2mm
{\bf Notes:} Column (1): Instrument; Column (2): central wavelength in {\it $\mu$m}; Column (3): values of Column (2) in frequency, given in {\it GHz}; Column (4): flux densities (and uncertainty) in {\it Jansky} computed in an aperture of 40\arcsec$\times$40\arcsec. For the {\tt SPIRE} and {\tt PACS} data we consider uncertainties of 20\% of the flux density. Column (5): flux density values of the total modified black body (MBB) modeled emission (red solid line in Fig.~\ref{SED_4945}).\\
\end{minipage}
\end{tiny}
\end{table}

From the SED fitting we are able to constrain the source size, $\Omega_s$, the dust temperature, T$_{d}$, and the total mass of dust, M$_{dust}$ (as done in \citealt{WEIB08}). To properly fit the dust emission an attenuated black body function (i.e., modified black body) is considered. The source function, S$_\nu$, of the dust is related to the Planck's blackbody function (B$_\nu$) at the dust temperature (T$_d$), the dust opacity ($\tau_\nu$) and the source solid angle ($\Omega_s$) according to the formula: 
\vskip-4mm
\begin{equation}
\hskip3mm S_\nu = B_\nu(\nu, T_d) \times (1-e^{-\tau(\nu)}) \times \Omega_s, 
\label{GreyB}
\end{equation}

while the dust optical depth was computed as
\vskip-4mm
\begin{equation}
\hskip3mm \tau_\nu~=~\kappa_d(\nu)\times\frac{M_{dust}}{D^2 \Omega_s}.
\end{equation}

D is the distance to the source and $\kappa_d(\nu)$ is the dust absorption coefficient, in units of {\it m$^2$ kg$^{-1}$} (\citealt{Krugel94}). $\kappa_d$($\nu$) is related to the $\beta$ parameter according to the relation:~$\kappa_d$($\nu$) = 0.04 $\times$ ($\nu$/250\hskip0.5mmGHz)$^\beta$. In this work $\beta$ has been computed using {\tt SPIRE, LABOCA} and {\tt SMA} data, obtaining a value of 2.0 from the linear fit. A source size of 20\arcsec$\times$10\arcsec\ has been assumed.

In Fig.~\ref{SED_4945} the best fit SED ($\chi^2_{min}$$\sim$~4.6; red solid line) is derived when two component temperatures dust model are considered: a cold dust component at 28 K to fit the shorter frequencies and a warm component at 50 K to fit the higher frequencies. A total mass of dust of $\sim$8 $\times$10$^6$ M$_\odot$ is derived. Assuming a gas-to-dust ratio in between 100 and 150 (see \citealt{WEIB08}), we derived a total gas mass of 7.6-11.4$\times$10$^8$~M$_\odot$.

We found a good agreement with the results obtained from previous works. Indeed, \cite{WEIB08} derived a total mass of gas in the central region using an aperture of 80\arcsec$\times$80\arcsec\ of 1.6$\times$10$^9$ M$_\odot$. Comparing their results with the one we derived we can conclude that most of the total emission (70\%) is included in a region of 40\arcsec$\times$40\arcsec.

On the other hand, \cite{Chou07} estimated the mass of molecular gas from the inferred dust emission at 1.3 mm (i.e, 1 Jy; see Fig.~\ref{SED_4945}), assuming a gas-to-dust ratio of 100. They assumed a dust temperature T$_{dust}$$\sim$40 K as inferred from far-infrared measurements (\citealt{Brock88}) then deriving, according to their Eq.~1, M$_{gas}$$\approx$3.6 $\times$ 10$^8$ M$_\odot$, which corresponds to a mass of dust in the range $\sim$2.4-3.6~$\times$10$^6$~M$_\odot$. The mass of dust derived by \cite{Chou07} is a factor of 2--3 lower than that derived in our work, and it can be considered as lower limit. 

The results of our SED modeling are summarized in Tab.~\ref{sourcesize}.

\begin{table*}
\centering 
\begin{small} 
\caption{Results obtained from the SED fitting and from the literature. }
\label{sourcesize}
\hskip-5mm
\begin{tabular}{cccccccc} 
\hline\hline\noalign{\smallskip}  
  	&  T$_{dust}$		 &	 M$_{dust}$  & M$_{gas}$ & GDR	&	Notes\\
  & (K)&(10$^6$ M$_\odot$)   	& (10$^8$ M$_\odot$)	&		\\
 &(1) & (2) & (3)  &   (4)\\
\hline\noalign{\smallskip}    	
This work			& 28$\pm$1, 50$\pm$2   	 &	 7.6$\pm$0.3 &  11.4$\pm$0.5 & 	150		&\\
\hline\noalign{\smallskip}  
Chou 2007 			& 40   	 &	 {\it 2.4 - 3.6}   &   3.6$\pm$0.7 & 100	&a\\
Chou 2007 			& 30   	 &	 {\it 3.1 - 4.7}   &   4.7$\pm$0.9  & 100	&	 a\\
\hline\noalign{\smallskip}
Wei$\beta$ 2008		& 20   	&	 {\it 8 - 12}  &  15.8$\pm$1.6  &	150	&		b \\
\hline\hline
\end{tabular}
\vskip2mm
\begin{minipage}{18cm}
\small
{{\bf Notes:} Column~(1): temperature of the dust component in {\it kelvin}; Column~(2): mass of dust in units of 10$^6$ M$_\odot$. In italic font are shown the two M$_{dust}$ values derived applying a gas-to-dust ratio of 100 and 150, respectively, to the M$_{gas}$ values taken from literature. Column~(3): mass of gas in units of 10$^8$ M$_\odot$. Column (4): gas-to-dust ratio considered; Column (5): notes with the following code: (a) the gas mass has been derived using the dust emission at 1.3 mm according to the \cite{Hildebrand83} formula (assuming a gas-to-dust ratio of 100); (b) the gas mass of the central region has been derived considering the cold (20 K) and warm (40 K) contributions in an aperture of 80\arcsec$\times$80\arcsec. }
\end{minipage}
\end{small}
\end{table*}

\section{Density and temperature determination. Resolved spectra from {\tt HIFI} and {\tt APEX}} 
\label{densities_temps}

\begin{table*}
\centering
\begin{tiny}
\caption{Line parameter results derived using {\tt MADCUBA}. }            
\label{list}      
\begin{tabular}{c |c c c |c c}        
\hline\hline               
Line & $v_{LSR}$ 	& $\Delta v_{1/2}$ & T$^{peak}_{MB}$  & Area &Notes\\    
 	 & (km s$^{-1}$) & (km s$^{-1}$) &  (K)  & (K km s$^{-1}$) & \\    
 (1) &  (2)  &  (3)  &  (4)  & (5)  & (6)   \\
\hline    
 CO(3--2)   		& 451; 566; 706 & 90; 109 (20); 90  		& 3.05; 3.58; 2.28 		&  294; 456 (75); 220  &  a\\
 CO(5--4)  		&  				&   						& 0.19; 0.26; 0.13  		&  18; 33 (6); 13 	  & a\\
 CO(6--5)   		&  				&  						& 0.052; 0.08; 0.035  	&  5; 10 (3); 3   & a \\
 CO(9--8) 		&  				&   						& --						& --			& a, d \\
 
 CO(3--2) 		& 455; 578; 683 &  90; 119 (16); 90 	& 0.64; 0.32; 0.64  		&  57.4; 34.4 (5); 57.4 	  & b\\
 CO(5--4)  		&  &  								& 0.35; 0.18; 0.35 		&  33.4; 21.1 (3); 33.4	  & b\\
 CO(6--5)   		&  &   								& 0.44; 0.24; 0.44 		&  42.4; 27.7 (4); 42.3  & b\\
 CO(9--8) 		&  &  								& 0.344; 0.21; 0.34  		&  33.0; 24.4 (4); 32.6 	& b \\
\hline
$^{13}$CO(3--2) 		& 446; 566; 685 &  90; 148 (21); 90 &  0.36; 0.31; 0.30			& 35; 48 (9); 29     & \\
$^{13}$CO(6--5) 		&  &   							&  0.028; 0.015; 0.020	& 2.7; 2.4 (0.40); 1.9 	  & \\
$^{13}$CO(9--8) 		& &   						&  $<$ 0.06						& $<$ 8    & c \\
\hline
HCN(4--3) 		& 446; 574; 683 	&  90; 169 (19); 90 	& 0.12; 0.095 ; 0.13 			& 11.33; 17.04 (3); 12.39    			 & \\
HCN(6--5) 		& 	&  								& 0.008; 0.004 ; 0.008 		& 0.90; 0.84 (0.3); 0.90 		 & \\
HCN(7--6) 		&	&  								& 0.0035; 0.0012; 0.0035		& 0.34; 0.22; 0.34			 &\\
HCN(12--11) 	&  		&  & $<$0.1 	&  $<$ 1 & c\\
\hline
HNC(3--2) & 448; 572; 683 	&  90; 138 (18); 90 	& 0.064; 0.043; 0.063 		& 6.2; 6.4 (0.9); 6.0  	 &\\
HNC(4--3) & 	&  									& 0.056; 0.04; 0.059 			& 5.4; 5.9 (0.96); 5.7  	 &\\
HNC(6--5) &  					&   				& 0.002; 0.002; 0.002 		& 0.22; 0.34 (0.09); 0.22   &\\
HNC(7--6) &	&  									& $<$ 0.06 					& $<$ 6				&	c \\
\hline
HCO$^+$(4--3) 		& 450; 580; 685 &  90; 176; 90 	& 0.13; 0.11; 0.14 	& 12.2; 21.3; 12.4  	 &\\
HCO$^+$(6--5) 		& 		 &   						& 0.004; 0.004; 0.004 	& 034; 0.66; 0.34  &\\
HCO$^+$(7--6) 		&  			& 					& $<$ 0.02 	& $<$ 2   & c\\
\hline
CS(6--5) 		& 444; 564; 671 	&  90; 129 (28); 90 	& 0.023; 0.029; 0.027 			& 2.24; 3.92 (1.07); 2.61  	&\\
CS(7--6) 		&  	& 							& 0.021; 0.028; 0.027 				& 2.03; 3.82 (0.97); 2.54  	&\\
CS(10--9) 		&  	&   							& $<$ 0.02 			& $<$ 2 		& c \\
CS(12--11) 	&  	&  								&  $<$ 0.02 			& $<$ 3    	& c\\
CS(13--12) 	& 	&  								& $<$ 0.03			&  $<$ 4		& c\\
\hline
[CI] $^3$P$_1$ $\rightarrow$ $^3$P$_0$ 	& 448; 568; 688 & 90; 152 (23); 90 		& 0.24; 0.24; 0.20 	& 23; 39 (8); 20 		& \\
$[$CI$]$ $^3$P$_2$ $\rightarrow$ $^3$P$_1$ 	& &								& 0.33; 0.32; 0.38 	& 33; 52 (10); 37 & \\
\hline
CH(3/2--1/2)	& 438; 556; 698 & 90; 150 (18); 90 	& 0.041; 0.023; 0.037 	& 4.59; 4.29; 4.14 	&		\\
CH(3/2--1/2)	&  &  							& 0.041; 0.024; 0.037		& 4.59; 3.13; 4.14 	&		\\
CH(5/2--3/2)	&  &								& $<$ 0.06				&  $<$ 5	&		c\\
\hline\hline
\end{tabular}
\end{tiny}
\vskip2mm
\begin{minipage}{18cm}
\small
{{\bf Notes:} Column (1): Molecule and rotational transition (J); Column (2): centroid of the gaussian component (local standard of rest velocity, v$_{LSR}$) in {\it km s$^{-1}$}; Column (3): full width at half maximum (FWHM) of the gaussian in {\it km s$^{-1}$}. The error values have been computed only for the central component for which the FWHM has been let free to vary. For the blue and red components the FWHM has been fixed to 90 {\it km s$^{-1}$}; Column (4): main beam peak temperature of each gaussian in {\it kelvin}; Column (5): area of the gaussian component in {\it K km s$^{-1}$}; Column (6): notes with the following code: (a) cold component; (b) hot component; (c) 3$\sigma$ upper limit; (d) the cold component does not exist for this transition. }
\end{minipage}
\end{table*}

\begin{figure*}
\centering
\includegraphics[width=0.24\textwidth, height=0.35\textwidth]{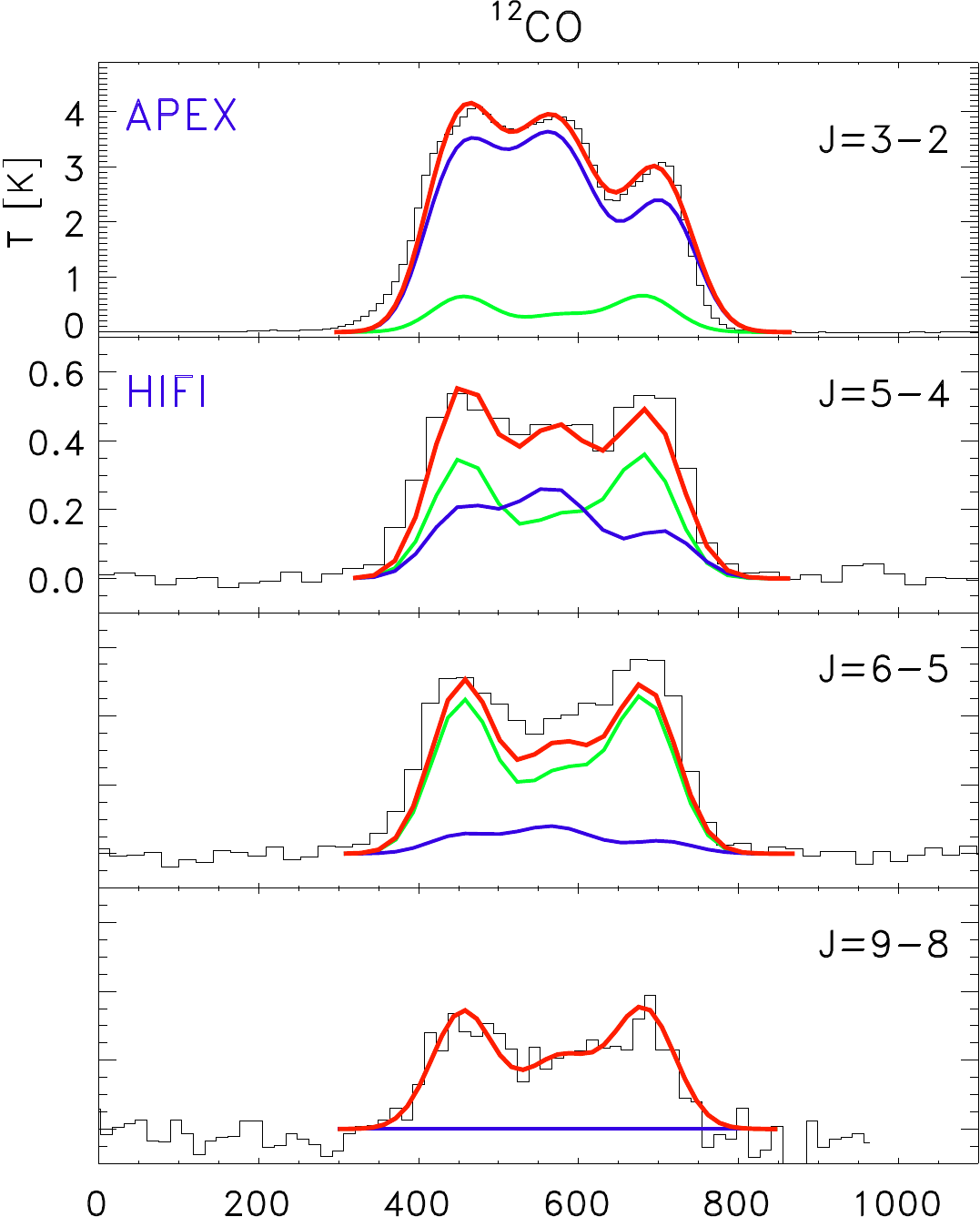}
\includegraphics[width=0.23\textwidth, height=0.35\textwidth]{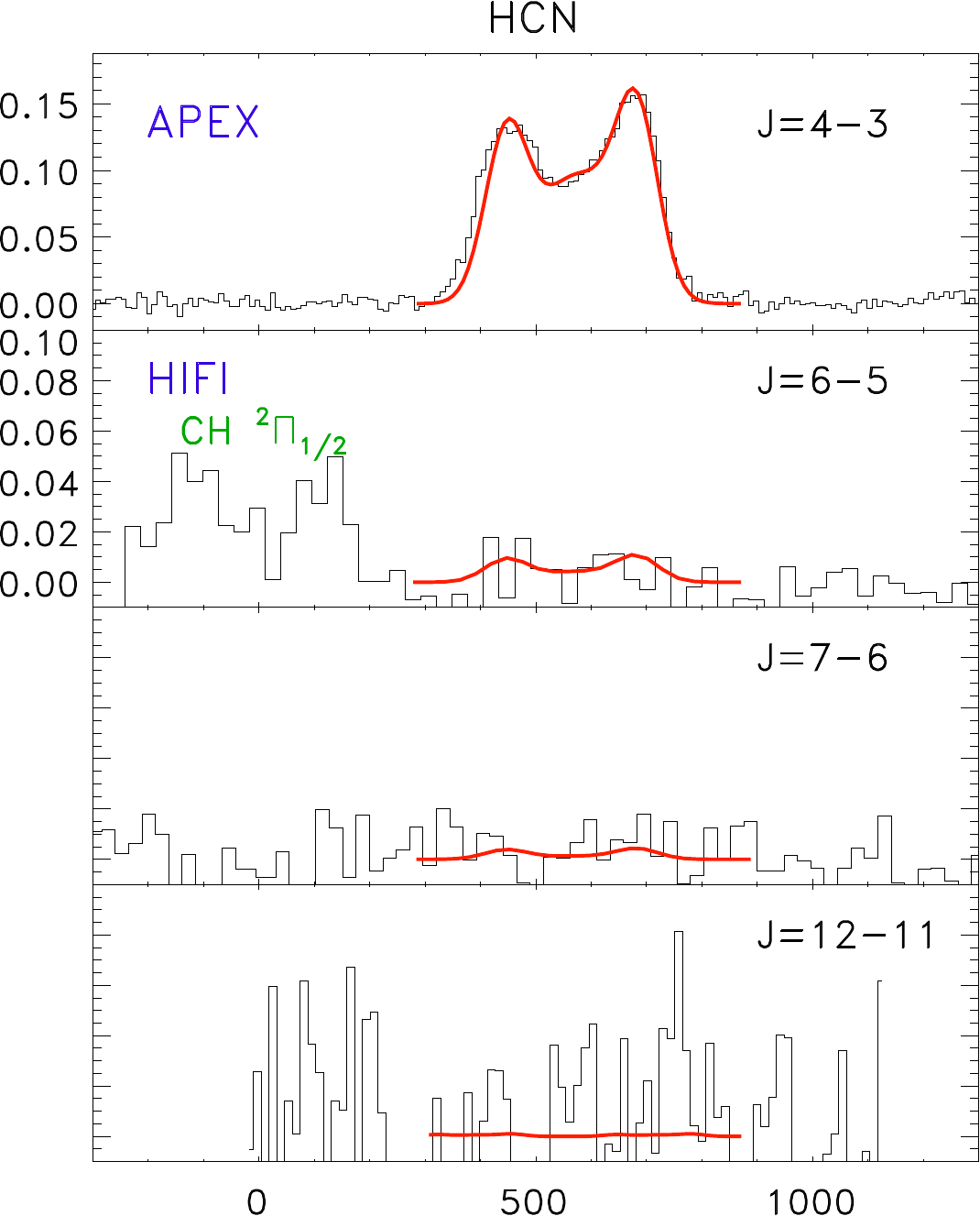}
\includegraphics[width=0.24\textwidth, height=0.35\textwidth]{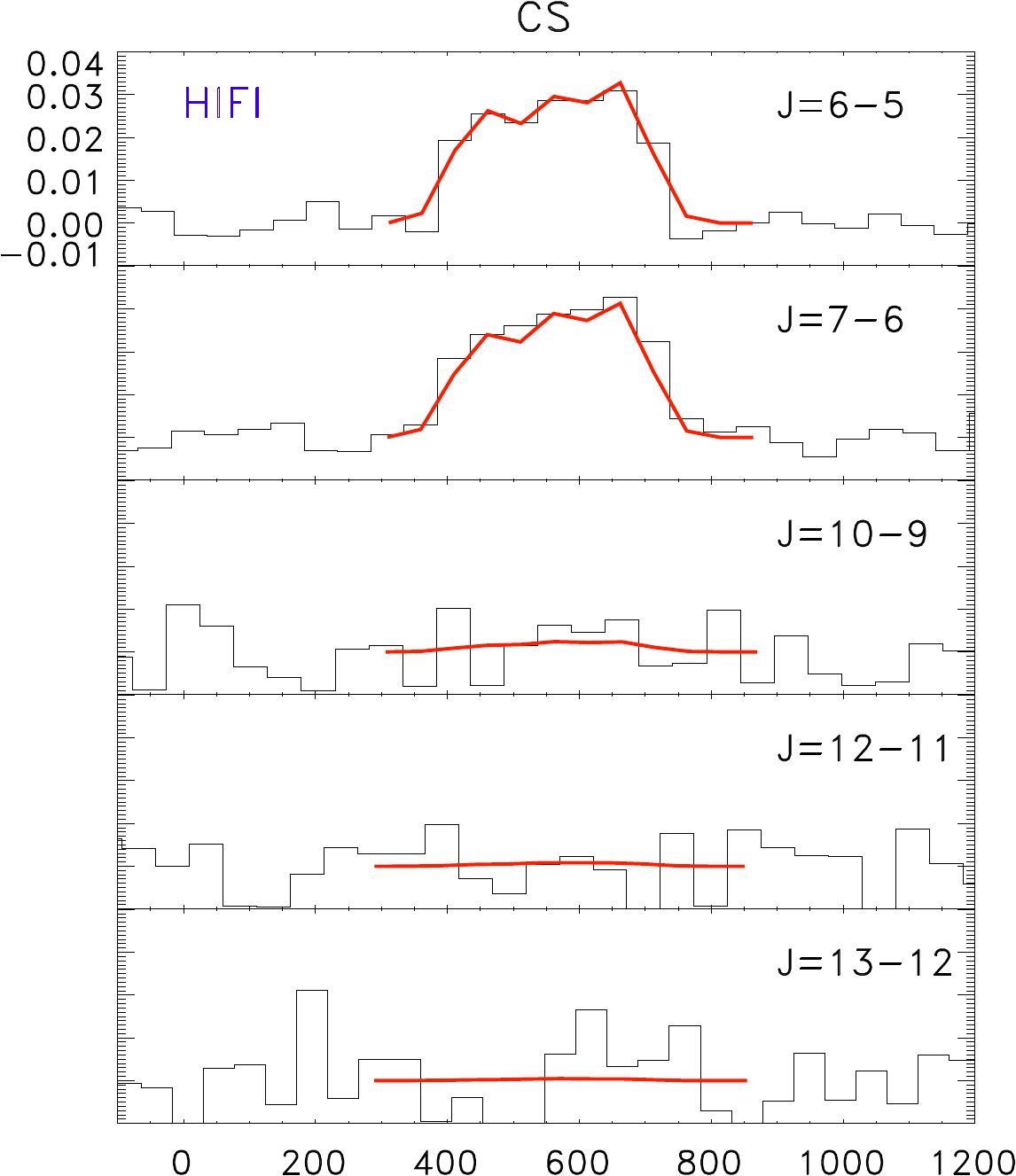}
\includegraphics[width=0.24\textwidth, height=0.35\textwidth]{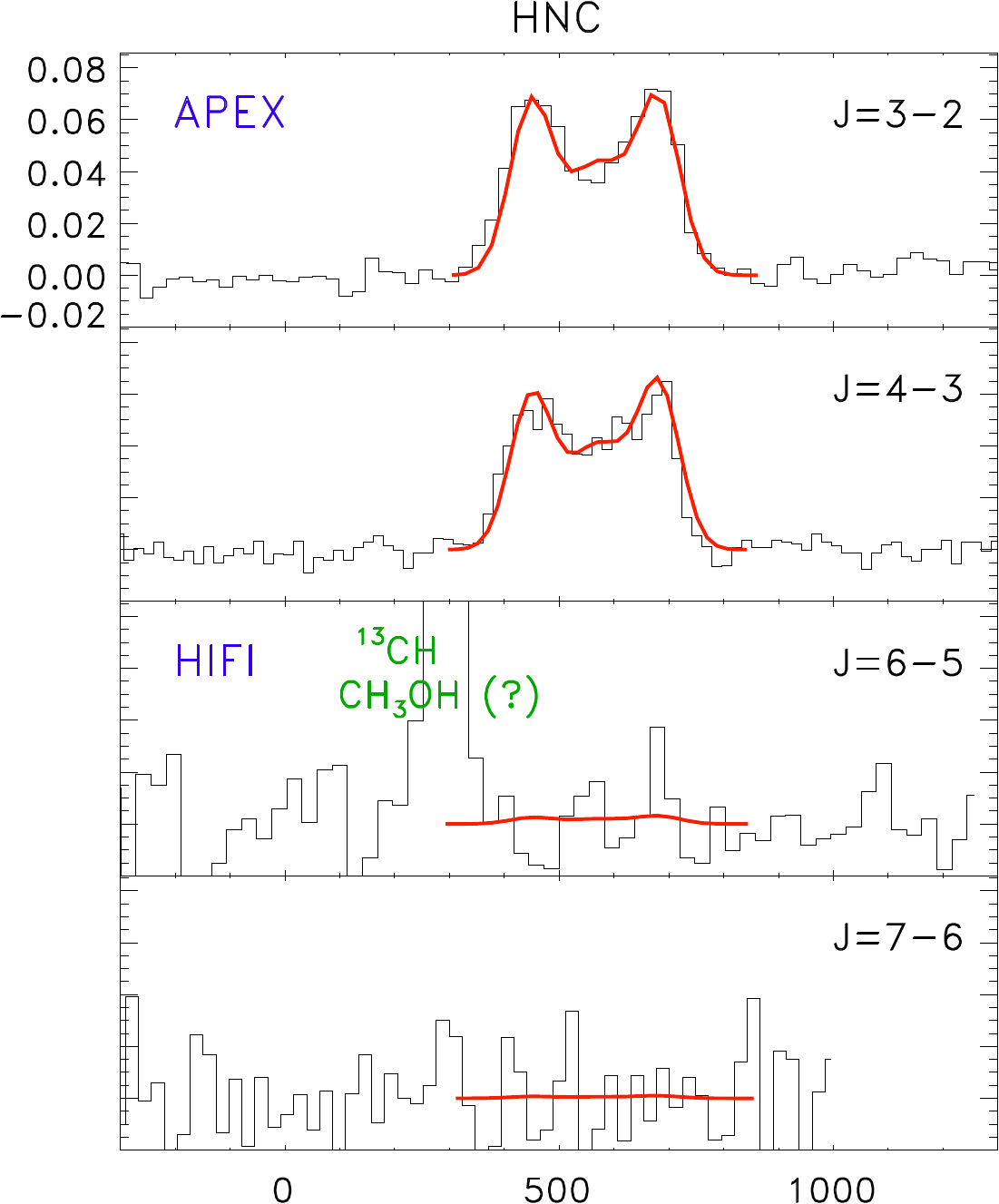}
\vskip6mm
\includegraphics[width=0.24\textwidth, height=0.32\textwidth]{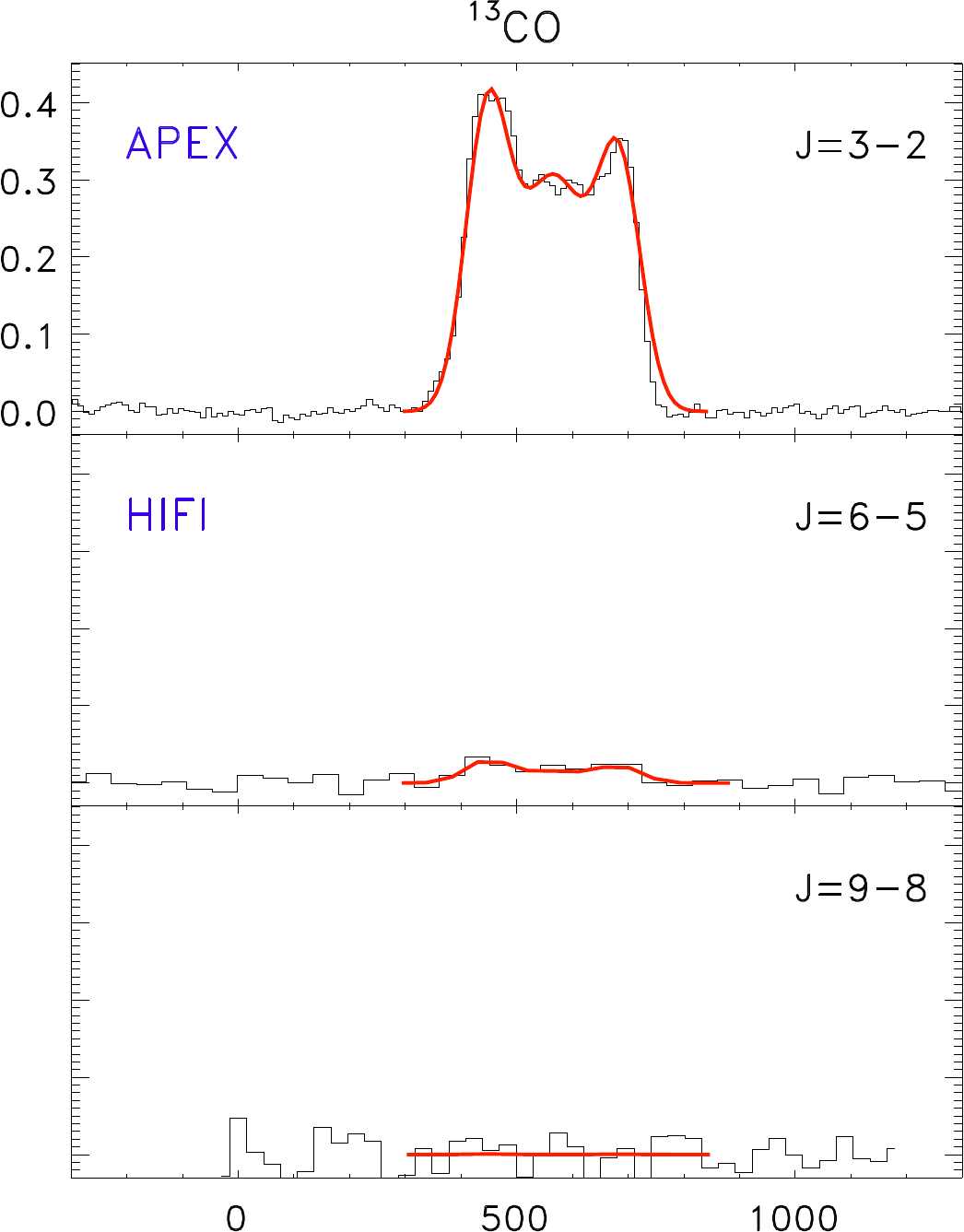}
\includegraphics[width=0.23\textwidth, height=0.32\textwidth]{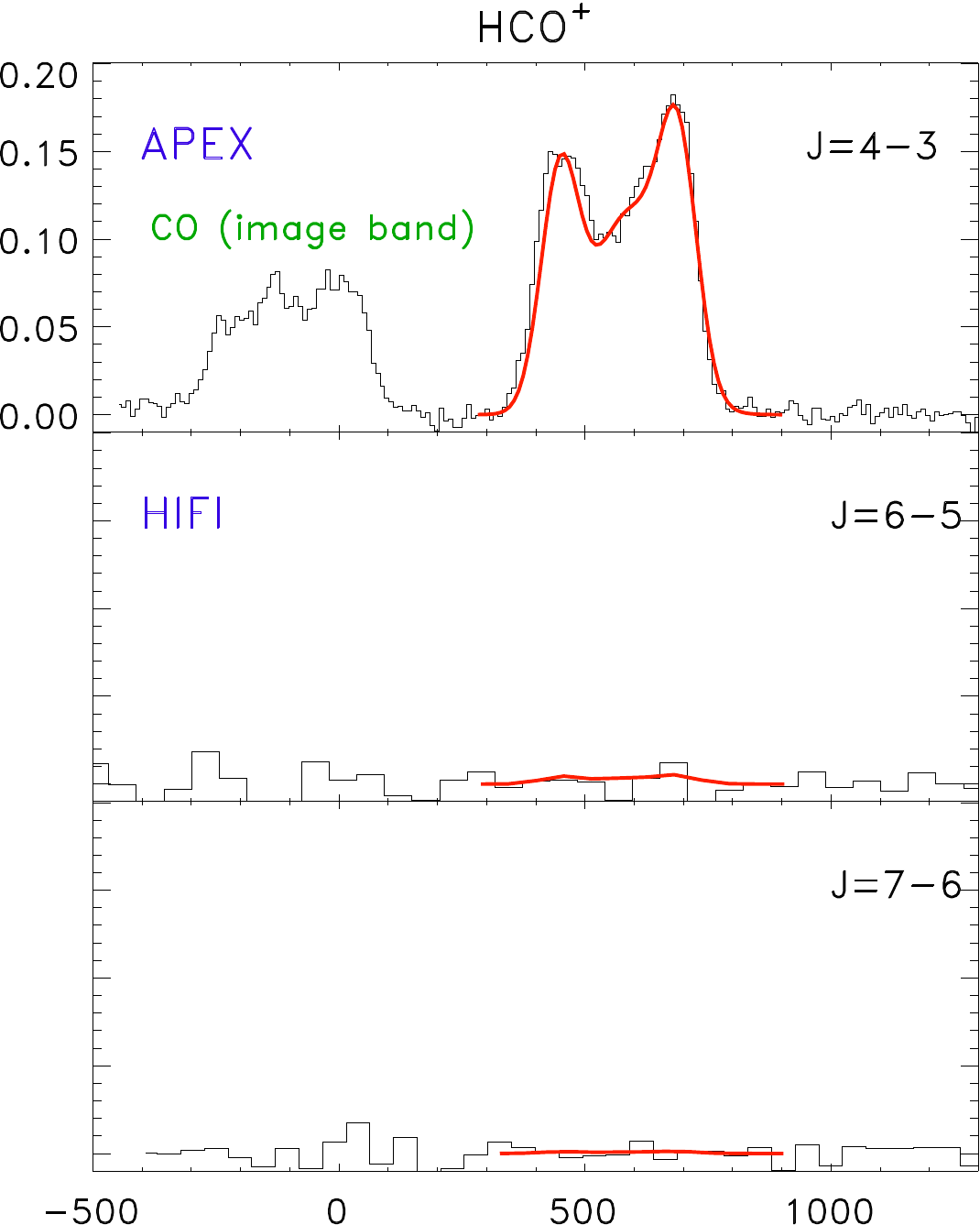}
\includegraphics[width=0.24\textwidth, height=0.32\textwidth]{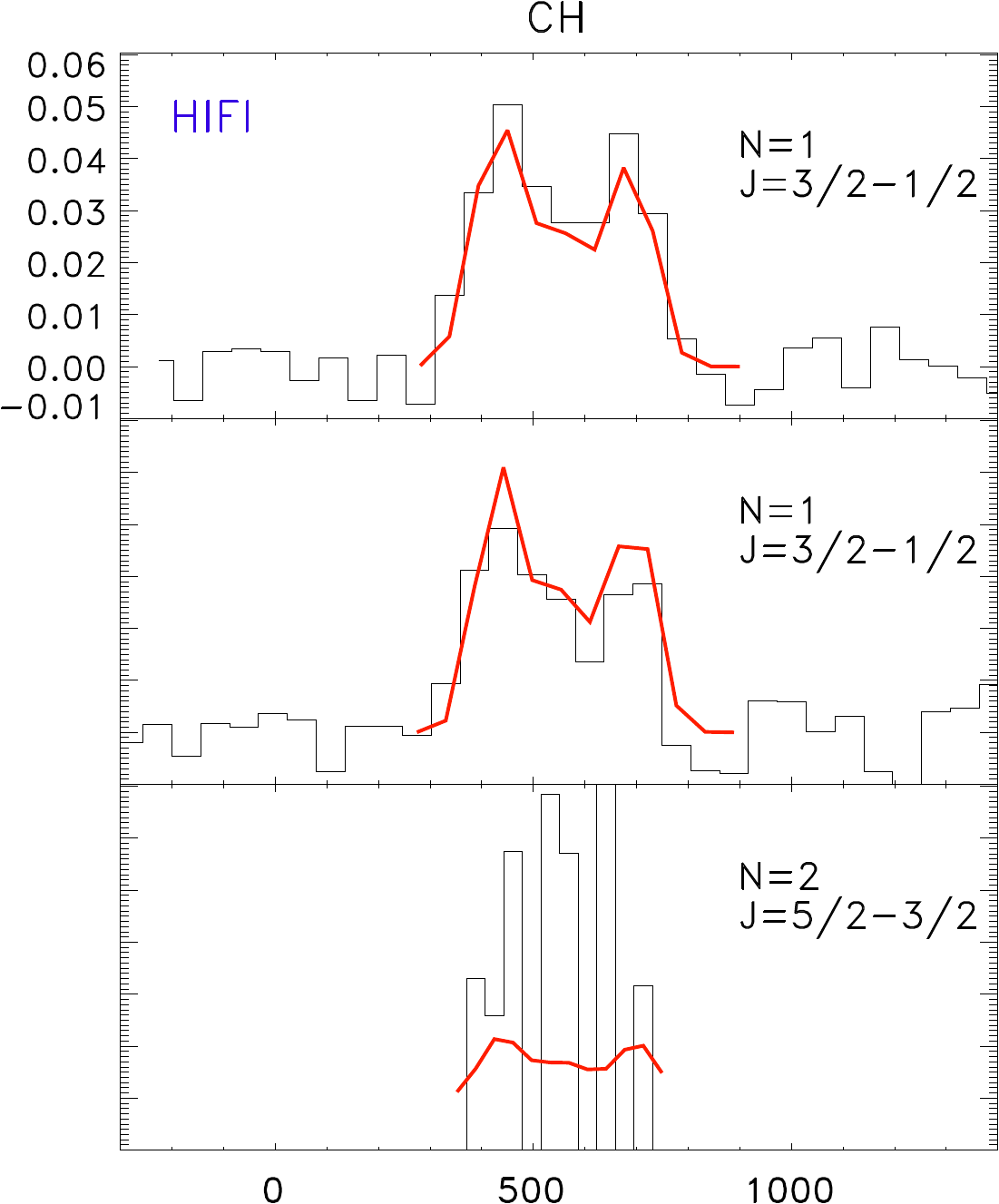}
\includegraphics[width=0.24\textwidth, height=0.325\textwidth]{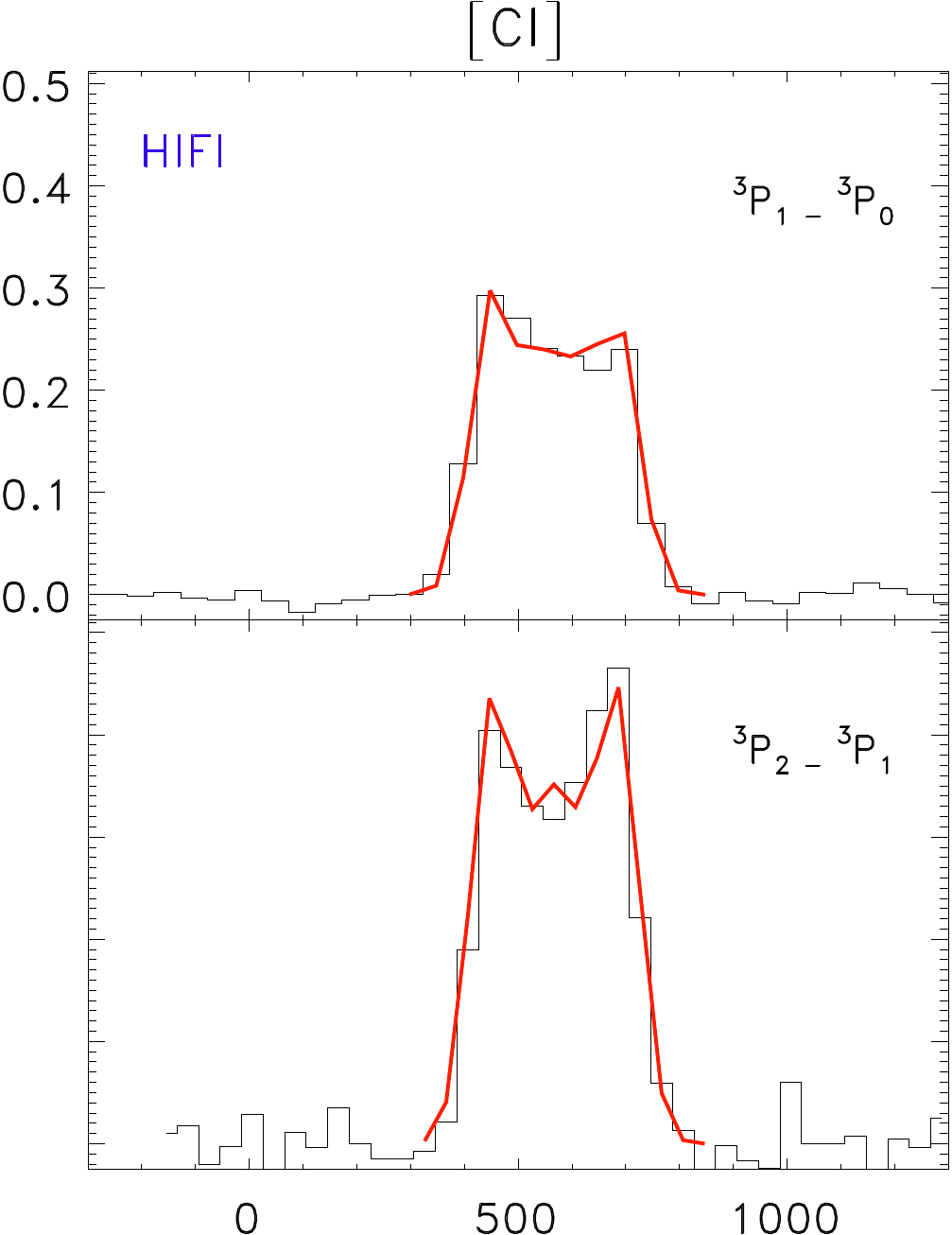}
\caption{\hskip-2mm High resolution molecular spectra from {\tt APEX} and {\tt HIFI} for all molecules analyzed in this work. In black the observed spectra, in red the LTE fit obtained using {\tt MADCUBA}. Only for the $^{12}$CO molecule we highlight a cold (blue) and a hot (green) components because two temperature components were needed to properly fit the emission. For each {\tt APEX} (J$_{up} <$ 4) and {\tt HIFI} (J$_{up} >$ 5--9) spectra the J transitions are identified. In green other molecular species like CH, $^{13}$CH and $^{12}$CO (in the image band) are detected.
For each molecule the range in temperature is the same for all J ({\tt APEX} and {\tt HIFI}) transitions with the exception of the $^{12}$CO and HCN which show different ranges to better appreciate the fainter emission of the {\tt HIFI} data. The emission is shown in main beam temperature (T$_{MB}$) in {\it kelvin}. 
}
\label{spettri_2}
\end{figure*}

\begin{figure*}
\centering
\includegraphics[width=0.45\textwidth]{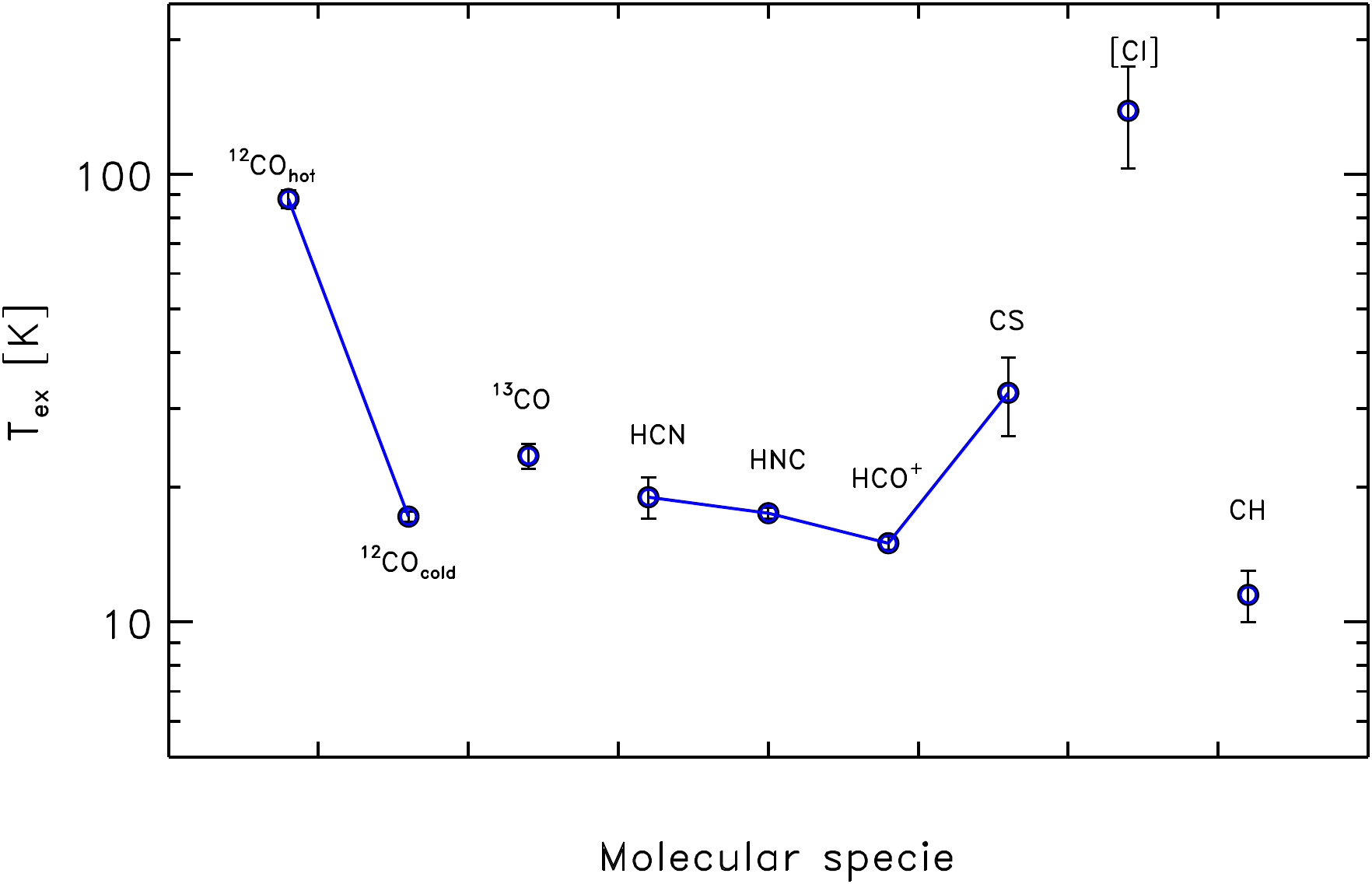}
\hskip5mm\includegraphics[width=0.45\textwidth]{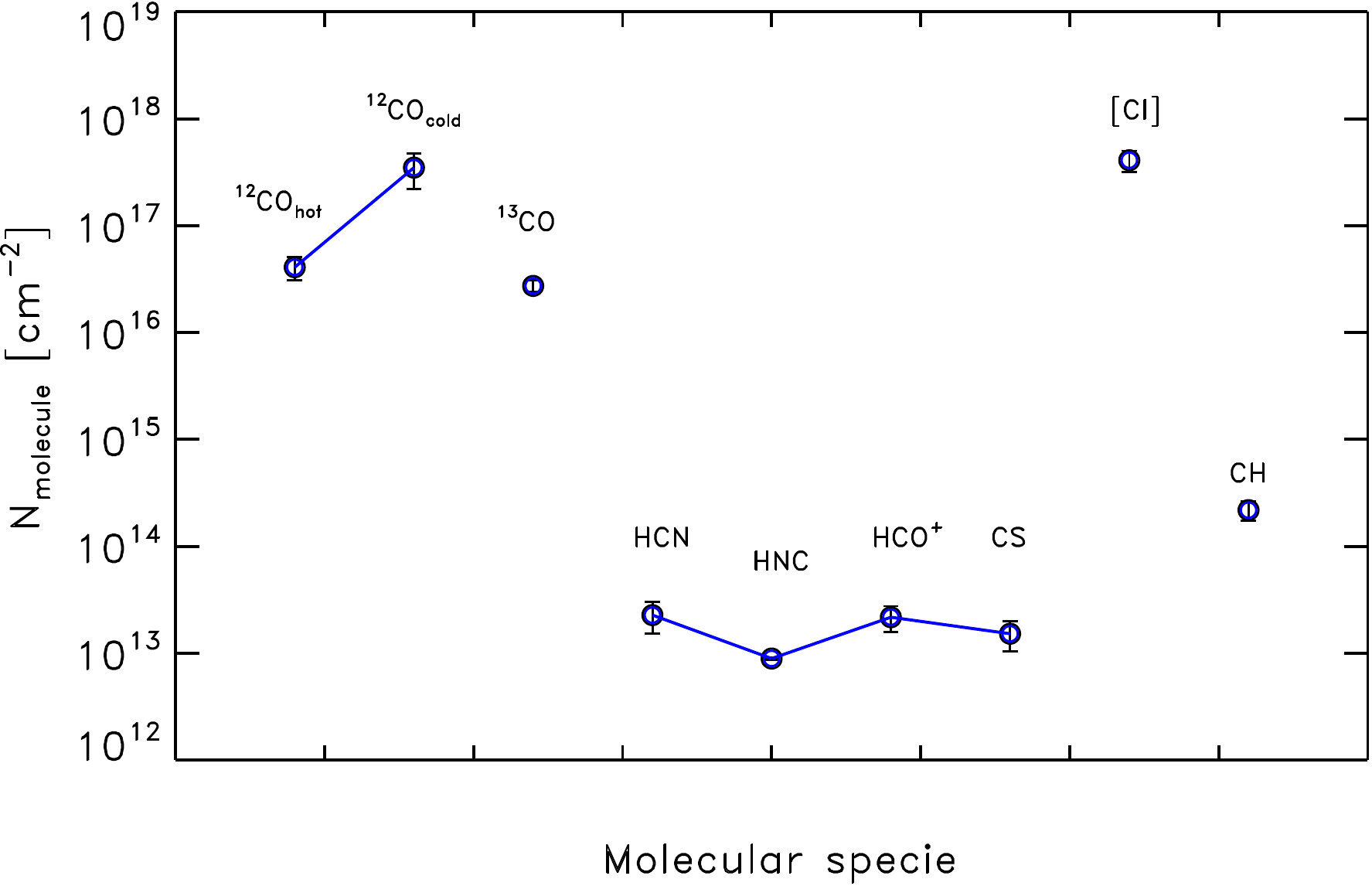}
\caption{LTE results derived with {\tt MADCUBA} for each individual molecular specie. {\it From left to right:} the excitation temperature, T$_{ex}$, in {\it kelvin}, along with the molecular column density, N$_{mol}$, in units of {\it cm$^{-2}$}, are shown. }
\label{trend_Nmol_T}
\end{figure*}

\subsection{LTE results using {\tt MADCUBA}}
\label{res_LTE_sub}

We apply the LTE analysis using {\tt MADCUBA} (\citealt{Martin19}) to $^{12}$CO, $^{13}$CO, HCN, HNC, HCO$^+$, CS, [CI], CH molecules observed using the high spectral resolution {\tt HIFI} and {\tt APEX} data. A source size of $\theta$=20\arcsec\ has been assumed (see \S ~\ref{Analysis_data}). The observed spectra and the simulated emission from the LTE model are shown in Fig.~\ref{spettri_2}. 
All molecules except $^{12}$CO have been properly fitted using one temperature component. Indeed, in the specific case of $^{12}$CO, the emission has been fitted using two temperature components (top left panel in Fig.~\ref{spettri_2}): the one cold and more dense while the other warm and less dense. The cold component ($\sim$20~K; in blue) dominates the emission characterizing the low J transitions while the warm one ($\sim$90 K; in green) dominates the emission at higher J. 
The need of two different (LTE) excitation temperatures T$_{ex}$ to fit all the line profiles is a clear indication of non--LTE excitation due to temperature and/or density gradients. The physical conditions required to explain the molecular excitation will be discussed in the next subsection.

The combination of low rotational transitions (J = 3--2 or 4--3) from {\tt APEX} with higher rotational transitions from {\tt HIFI} (J = 5--4 up to 9--8) allows to better constrain the molecular column density N$_{mol}$ and excitation temperature T$_{ex}$ parameters for each specie (see \S~\ref{Analysis_data}). 
The typical value of N$_{mol}$ derived for $^{12}$CO with {\tt MADCUBA} ranges from 4$\times$10$^{16}$ cm$^{-2}$ up to 3.2$\times$10$^{17}$~cm$^{-2}$, for the warm and cold components, respectively.
The N$_{mol}$ and T$_{ex}$ values for the different molecules are shown in Fig.~\ref{trend_Nmol_T}.

According to the LTE analysis, we then derived the following results (see Tab.~\ref{list}) for all molecules:

\begin{enumerate}

\item Three distinct kinematic components have been found for all molecules: they identify the nuclear bulk ($\sim$560~km~s$^{-1}$) and the rotating disk structures which show one blue-- ($\sim$450 km s$^{-1}$) and one red--shifted ($\sim$690 km s$^{-1}$) components. Our result is an agreement with the kinematics derived in previous works (e.g., \citealt{Ott01, Henkel18});
 
\item All the species, except $^{12}$CO and [CI], have been properly fitted using a single excitation temperature of about 20 K. $^{12}$CO needs two components with excitation temperatures of 20 K and 90 K while [CI]\footnote{For the [CI] molecule we assumed an extended source size ($\theta_s$$>$20\arcsec).} needs one component with a high excitation temperature, T$_{ex}$~$\sim$~150~K;

\item The typical value of N$_{mol}$ derived for low density gas tracers, such as $^{12}$CO, $^{13}$CO, [CI] ranges from 3$\times$10$^{16}$ cm$^{-2}$ up to 5$\times$10$^{17}$ cm$^{-2}$. For the low density tracer CH the lowest column density is achieved (N$_{mol}$ $\sim$ 10$^{14}$ cm$^{-2}$). 
The derived N$_{mol}$ for the high density gas tracers such as HCN, HNC, HCO$^+$ and CS, has a lower value, of the order of 10$^{13}$ cm$^{-2}$. If a smaller source size was considered (i.e., $\theta_s$=10\arcsec), as in \citealt{Henkel18}, the column density values would have been increased by a factor of $\lesssim$3.

\end{enumerate}

\begin{table}
\begin{tiny}
\caption{Results derived from LTE ({\tt MADCUBA}) and non--LTE ({\tt RADEX}) analyses using {\it Herschel}/{\tt HIFI} and {\tt APEX} data.}
\label{RADEX}
\hskip-1mm
\begin{tabular}{cccccc} 
\hline\hline\noalign{\smallskip}  
Molecule &  log N$_{mol}$ &  T$_{ex}$	  &  n(H$_2$)  & log N$_{mol}$  & Notes  \\
 &&&\multicolumn{2}{c}{{\tt RADEX}}\\
            \cline{4-5}\noalign{\smallskip}
   & (cm$^{-2}$)  & (K)   & (cm$^{-3}$)   &  (cm$^{-2}$)  & \\
\noalign{\smallskip}
(1) & (2) &  (3)& (4) &(5)  & (6) \\ 	
\hline\noalign{\smallskip} 	
{$^{12}$CO} 	& 16.6 - 16.8 	&	84-92 			&  5.9$\times$10$^4$  & 16.75    & a,b \\
 		& 17.4 - 17.8 	&	16-18 		&  7.3$\times$10$^3$ 		&  17.41   	&	\\
\hline\noalign{\smallskip} 	
{$^{13}$CO} & 16.39 - 16.59 	&	22-25 		&  3.8$\times$10$^3$  &  16.35 	& c\\
\hline\noalign{\smallskip} 	
{HCN} 	& 13.28 - 13.58 	&	17-21  &  1.2$\times$10$^6$  & 13.39     & d\\
\hline\noalign{\smallskip} 	
{ HNC} 	& 13.03 - 13.05	&	17-18 & 1.4$\times$10$^6$  & 13.00	 & d \\
\hline\noalign{\smallskip} 	
{HCO$^+$} 	& 13.29 - 13.53	&	15 &  5.0$\times$10$^5$  & 13.76       &  d \\
\hline\noalign{\smallskip} 	
{ CS}	 	& 13.11 - 13.39	&	26-39 &  8.0$\times$10$^5$ &  13.15      & d \\
\hline\noalign{\smallskip} 	
{ [CI]} 		& 17.57 - 17.77 	&	103-174  &  1.0$\times$10$^5$ &  17.56 	&	e, f  \\
\hline\noalign{\smallskip} 	
{ CH}		 & 14.32 - 14.50 	&	10-13 	 &    ---     & ---  &  e, g \\
\hline\hline
\end{tabular}
\end{tiny}
\vskip2mm
\begin{minipage}{9.cm}
\small
{{\bf Notes:} Column (1): molecule; Column (2): molecular column density (logarithmic value) in units of {\it cm$^{-2}$}; Column (3): excitation temperature in {\it kelvin}; Column~(4): hydrogen volume density obtained using {\tt RADEX} in units of {\it cm$^{-3}$}; Column (5): molecular column density derived with {\tt RADEX} in units of {\it cm$^{-2}$}; Column (6): notes with the following code: 
(a) two (i.e., cold and warm) components fit;
(b) all transition detected; 
(c) J=9--8 not detected;
(d) {\tt HIFI} transitions not detected;
(e) only {\tt HIFI} data;
(f) extended source ($>$20\arcsec);
(g) n(H$_2$) and N$_{mol}$ cannot be derived since this molecule is not present in the {\tt RADEX} online code.\\
The kinetic temperature and the source size considered in the {\tt RADEX} analysis are, respectively, T$_{kin}=$ 200~K and $\theta_s=$ 20\arcsec.}
\end{minipage}
\end{table}

\subsection{non--LTE results using the {\tt RADEX} code}
\label{res_nonLTE_sub}

As mentioned above, the need of two different LTE excitation temperatures T$_{ex}$ to fit all the $^{12}$CO line profiles (from J$_{up}$ = 3 up to 9) is a clear indication of a non--LTE excitation of this molecule. We then apply the non--LTE {\tt RADEX} code to derive the volume gas density of the collisional patter, n(H$_2$), in NGC 4945 for each molecular specie\footnote{We excluded the CH molecule because it is not available in the online version.}, and to confirm the molecular column densities, N$_{mol}$, and the excitation temperature, T$_{ex}$, values derived with {\tt MADCUBA} LTE analysis, restricted to the rotational J transitions of the specific molecule involved in the analysis (\S~\ref{res_LTE_sub}).

The {\tt RADEX} code is based on a non--LTE analysis taking advantage of the velocity gradient (i.e., N$_{mol}$/$\Delta$v, the ratio between the column density, in cm$^{-2}$, and line width, in km s$^{-1}$). 
This code was used to predict the line emission from all molecules, using all the lines simultaneously, and considering a kinetic temperature of 200~K. This assumption is based on the T$_{ex}$ derived for CO and [CI] (i.e., $\sim$150 K; see previous section). In fact, in the case of $^{12}$CO, we carried out the analysis for two different kinematic temperatures: T$_{kin}$~=~50 K when fitting the cold component and T$_{kin}$~=~200 K for the warm component. A lower T$_{kin}$ would not be able to properly reproduce the line profiles of the $^{12}$CO transitions at higher frequencies (e.g., J=6-5).
For most of the transitions observed in this work, except for $^{12}$CO involved levels with energies below 50 K, the choice of the T$_{kin}$ has a marginal effect in the derived H$_2$ densities and the molecular column densities.

To derive the H$_2$ densities and the molecular column densities, N$_{mol}$, from {\tt RADEX}  we have tried to fit all the observed $^{12}$CO lines with a isothermal and uniform cloud but we did not find a unique solution. To fit all lines it was required to have at least two clouds with different densities and/or temperatures. These results indicate the presence of molecular clouds with a range of densities and temperatures within the beam, as expected for the complexity of the NGC 4945 nucleus. The predicted non--LTE $^{12}$CO column densities for the two different H$_2$ density regimes are similar to those derived from the LTE analysis. The comparison of the predicted non--LTE T$_{ex}$ with the derived LTE values is more complicated since there is not a single non--LTE T$_{ex}$, but a range of T$_{ex}$ depending on the excitation requirements for each transition. The situation is even more complicated for the case of a non--uniform molecular cloud with H$_2$ density gradients exciting different $^{12}$CO lines in different regions. As illustrated by the non--LTE analysis, lower J lines will be more sensitive to low densities than the high--J lines. We have then compared the average of the predicted non--LTE T$_{ex}$ with the LTE T$_{ex}$ for the range of transitions that dominates the $^{12}$CO emission. We have found a reasonable agreement between both temperatures for the low-- and the high--J lines corresponding to the low and high density components, respectively.

As expected from the typical density of the ISM in galaxies over the scales of hundreds of pc, most of the high dipole-moment molecules (e.g., HCN, CS, HCO$^+$) usually have a critical density much larger than the average H$_2$ density of the ISM. We then derive subthermally excitation (T$_{kin}$$>$T$_{ex}$) for all density gas tracers.

According to our results, we derived a moderate volume gas density n(H$_2$) for most of the molecules, in the range 10$^3$ cm$^{-3}$ up to 10$^6$ cm$^{-3}$. Lower densities are obtained when considering the low density gas tracers (e.g., $^{12}$CO, [CI]), while higher densities are derived when studying the high density gas tracers, such as HCN, HNC, HCO$^+$ and CS (see Tab.~\ref{RADEX}).
We reproduce reasonable well the intensities of all transitions for each molecule with {\tt RADEX}, also when considering a non-uniform cloud (i.e., two H$_2$ densities), in agreement with the results derived using {\tt MADCUBA}, as in the case of the two components model applied to $^{12}$CO.

\section{Thermal and column density structures from the $^{12}$CO emission at different scales}

We study the distribution of the thermal balance and the column density distribution at different spatial scales using the 2D {\tt PACS} and {\tt SPIRE} data through the analysis of the $^{12}$CO emission over a wide range of rotational transitions. In particular, the $^{12}$CO transitions at wavelengths from 55 $\mu$m to 650 $\mu$m were covered: this molecule is the most abundant in the interstellar medium after H$_2$ and therefore considered a good tracer of the properties of the bulk of the molecular gas phase. As shown from the analysis of a limited number of $^{12}$CO transitions (\S~\ref{res_LTE_sub}), the wide range of physical properties expected  in the nucleus of NGC 4945 cannot be described by either LTE or simple non-LTE modeling. To deal with the full range of $^{12}$CO transitions and the wide range of spatial scales addressed in this work we will  apply a `transition limited' LTE analysis to a given  range of transitions sampling specific physical conditions (density and temperatures) of the molecular gas. This analysis will allow us to derive the spatial distribution of `transition limited' T$_{ex}$ and N$_{mol}$, which will describe the different phases of the molecular gas in NGC 4945.

\subsection{Mid-J $^{12}$CO at large spatial scale (700 pc -- 2 kpc)}
\label{SLED_spire}

\begin{figure*}
\centering
\includegraphics[width=0.8\textwidth]{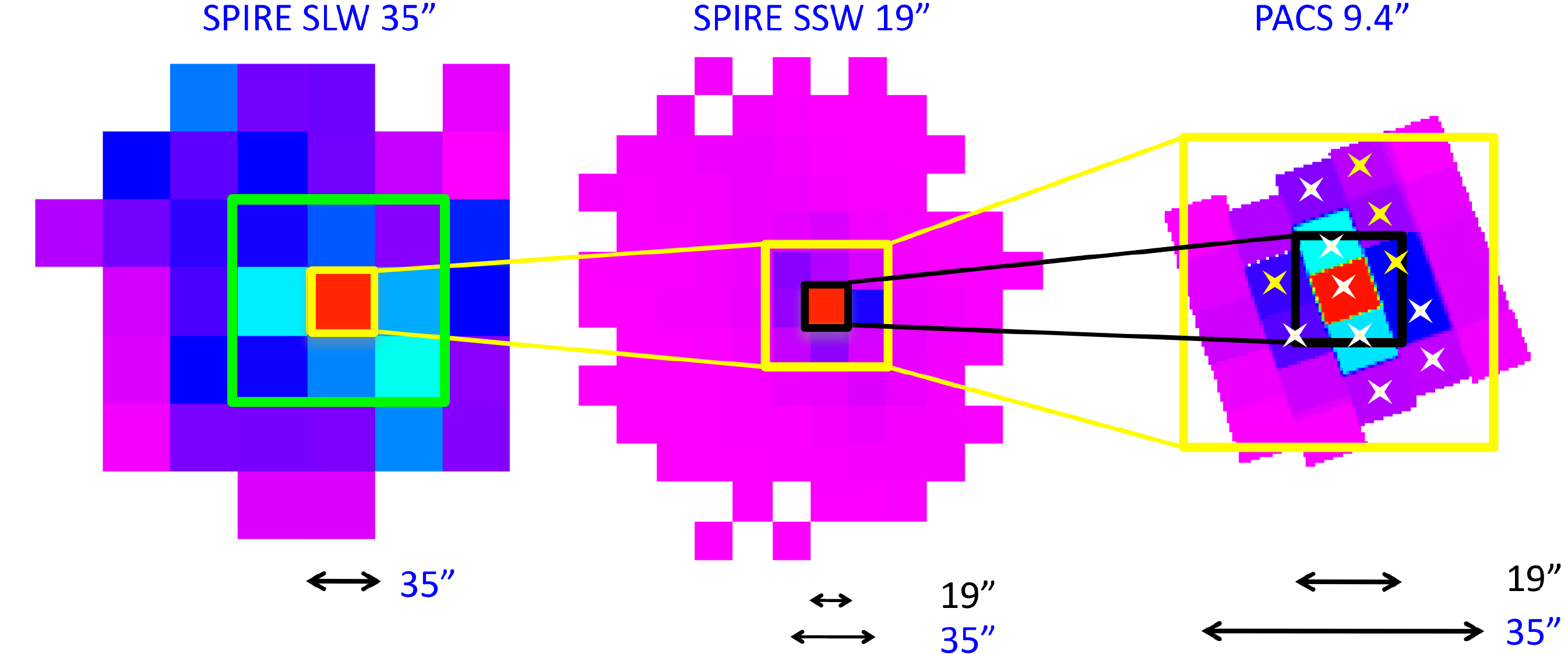}
\caption{{\it From left to right:} Schematic view of the different FoVs involved in the analysis of {\tt SLW, SSW SPIRE} and {\tt PACS} data.  {\it Left:}  In the {\tt SLW SPIRE} map the green square highlights the FoV considered in the analysis ($\sim$100\arcsec). These data are characterized by a a beam of 35\arcsec\ identified by the yellow small square. {\it Middle:} in the {\tt SSW SPIRE} map the yellow square represents the 3$\times$3 spaxels area involved in the analysis. The map is characterized by a beam of 19\arcsec\ (small black square). {\it Right:} in the {\tt PACS} map the black square identifies a FoV of $\sim$19\arcsec\ (i.e., one {\tt SSW SPIRE} spaxel) while the yellow square identifies a FoV of $\sim$35\arcsec. The star symbols represent those spaxels where the $^{12}$CO emission is observed: in particular, white stars highlight the spaxels characterized by stronger $^{12}$CO emission than that observed in the remaining spaxels marked using yellow stars. {\tt PACS} data are characterized by a beam of 9.4\arcsec.}
\label{beams_spire_pacs}
\end{figure*}

In this section we study the warm component by using the mid-- and high--J of $^{12}$CO rotational transitions from {\tt SPIRE} long wavelengths ({\tt SLW}; transitions from J$_{up}$ = 4 to 8), and from {\tt SPIRE} short wavelength ({\tt SSW}; J$_{up}$ =  9 to 13). 
For our study we mainly focus on the very central regions of the whole field of view (FoV), where the strongest $^{12}$CO emission is observed. In particular, a maximum region of 3$\times$3 spaxels at a resolution of 35\arcsec ($\sim$2$\times$2 kpc$^2$; see left panel in Fig.~\ref{beams_spire_pacs}) in the {\tt SLW} map is considered. For each {\tt SLW} spectrum we combined the contribution at higher frequencies of 3$\times$3 {\tt SSW} spectra at a resolution of 19\arcsec\ ($\sim$1$\times$1 kpc$^2$) to match the beam of the {\tt SLW} spectrum (middle panel in Fig.~\ref{beams_spire_pacs}).

\begin{figure*}
\centering
\hskip-7mm\includegraphics[width=0.65\textwidth, height=0.32\textwidth]{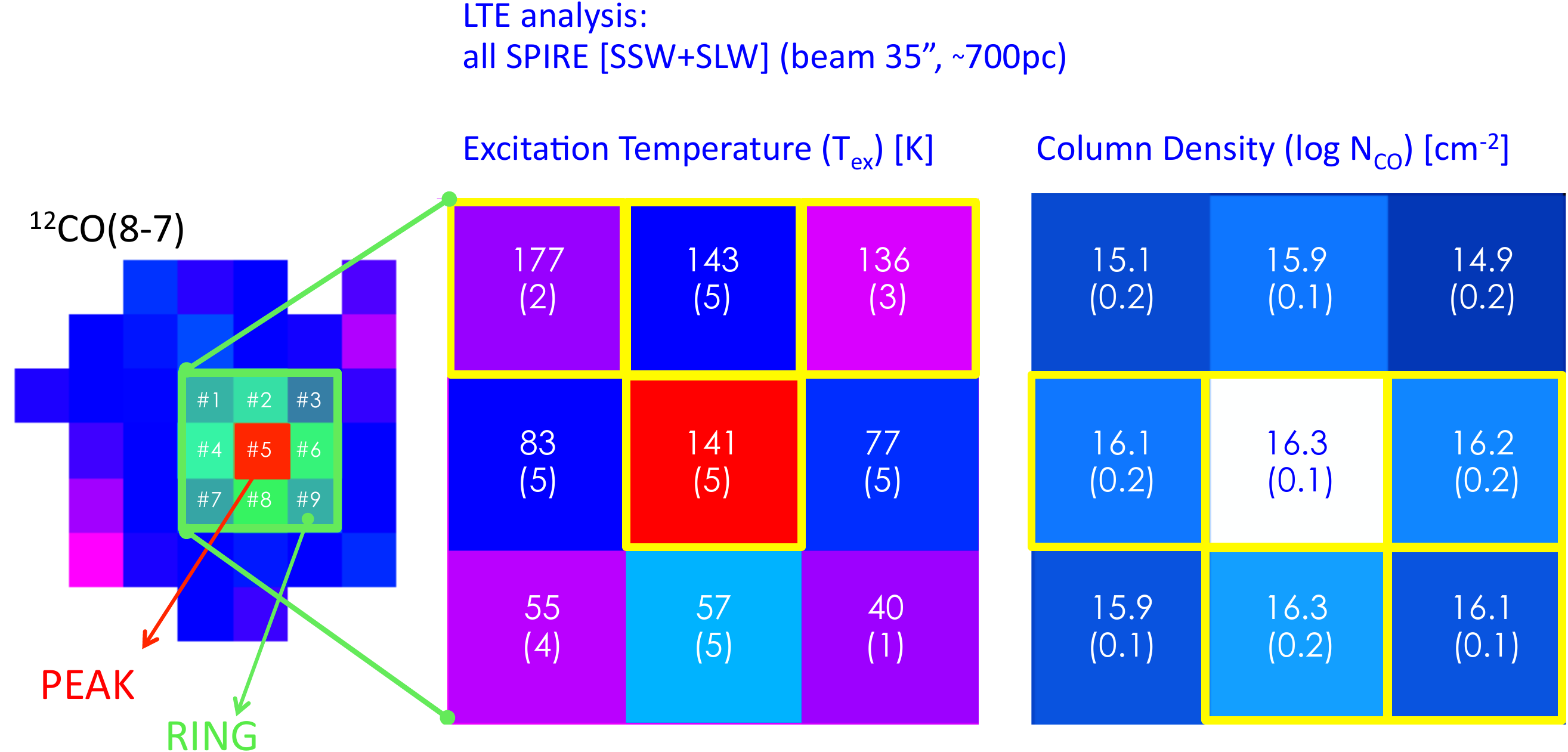}
\hskip4mm\includegraphics[width=0.3\textwidth, height=0.28\textwidth]{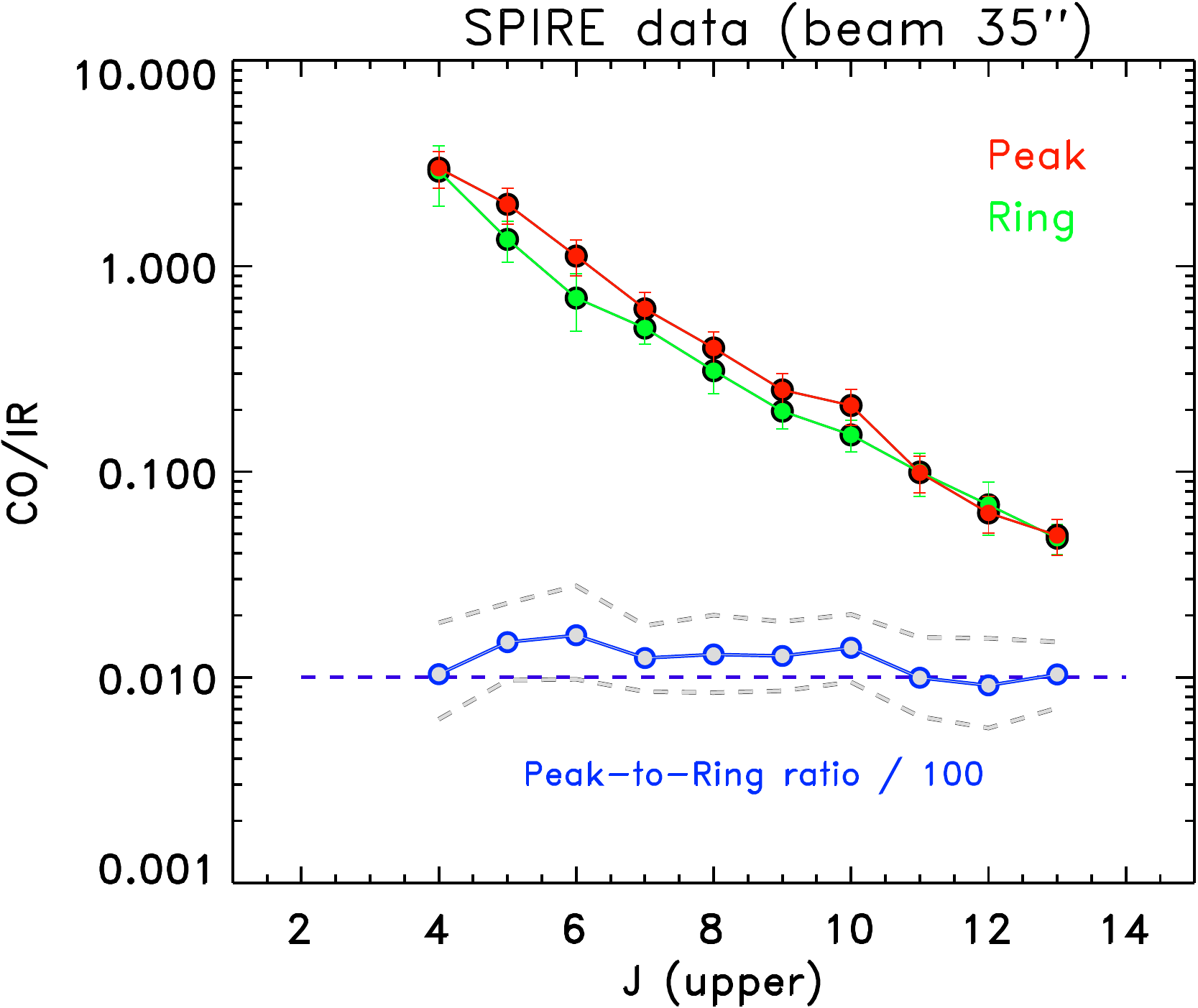}
\caption{{\it Left:} 
$^{12}$CO(8--7) emission map showing a FoV of 3$\times$3 spaxels (35\arcsec\ each; light green square) considered when combining the {\tt SSW} and {\tt SLW SPIRE} data at the same resolution (i.e., 35\arcsec, large spatial scale). The peak emission (in red) and a ring of one spaxel width around that maximum (green area) are identified. {\it Middle:} Excitation temperatures (T$_{ex}$) and the (logarithmic) column densities (N$_{CO}$) are derived for each spaxel using the rotational diagrams. The yellow boxes represent the spectra characterized by high T$_{ex}$ and N$_{CO}$ values. 
{\it Right:} $^{12}$CO/IR flux ratios computed in correspondence of the flux density peak (red) and the ring (green) for the different transitions J in the {\tt SPIRE} domain. The ratio between the maximum peak and the ring for each transition is also shown (i.e., `peak-to-ring ratio', in blue) divided by 100 (a factor of 100 has to be applied to obtain the real values). The dashed gray lines identify the lower and upper limit values of the peak-to-ring ratios computed within the errors. Higher $^{12}$CO/IR values are derived for high J (J$_{up}~\geq$~11--10) in the ring structure, implying lower ($\lesssim$ 1) peak-to-ring ratios.}
\label{CO_LIR_ring}
\end{figure*}

When applying the LTE analysis to all {\tt SPIRE} data we derived the excitation temperature and column density for each (combined) spectrum at the resolution of 35\arcsec\ ($\sim$700 pc).
From this analysis, we found higher T$_{ex}$ in the center and in the north part of the galaxy possibly affected by the presence of the outflow at large scales (see \S~\ref{Intr}). In the remaining regions lower temperatures are found. The column density peaks in the center showing higher values in the south (middle panels in Fig.~\ref{CO_LIR_ring}). At this spatial resolution the LTE analysis gives good results when using one temperature component\footnote{In the specific case of the central spaxel, an excitation temperature of T$_{ex}$=141 K and column density of log~N$_{CO}$ = 16.3 are derived. For this spectrum a good fit would be also achieved when including a secondary component, characterized by lower T$_{ex}$ and N$_{CO}$ similar to that derived for the main component. We finally considered one component temperature because the flux contribution of the secondary component was irrelevant (i.e., $\lesssim$10\% of the main component flux).}.

In order to study the spatial distribution of the heating in this galaxy we compare the emission in the peak with the emission integrated in an annular ring around the peak (one spaxel width) using the same beam for all transitions. We computed the ratio between the $^{12}$CO flux density peak and the corresponding IR continuum emission at each frequency of the $^{12}$CO as a function of J (hereafter, CO/IR ratios) as shown in Fig.~\ref{CO_LIR_ring} (right panel). For a spectroscopically unresolved line, as in this case, the flux density peak (given in W m$^{-2}$ Hz$^{-1}$, or Jy) is proportional to its line total flux (W m$^{-2}$). 
We then multiply the IR flux density by the spectral resolution to derive the total integrated IR continuum flux at the wavelength of each line. Since all $^{12}$CO transitions have similar line widths\footnote{In the {\tt SPIRE} spectra the spectral resolution (FWHM) is proportional to the wavelength (given in micron), according to the formula: FWHM [km s$^{-1}$] $\sim$1.45$\times$($\lambda$/[$\mu$m]) (see SPIRE manual).}, the CO/IR ratio corresponds to the flux ratio of each emission line: i.e., Flux(CO)/Flux(IR Continuum). Thus, the ratio is a dimensionless quantity.
Instead of using the total infrared flux, like usually considered in the literature (see \citealt{Meijerink13}), we consider the continuum underlying each $^{12}$CO transition to characterize the $^{12}$CO/IR ratio at the specific continuum value and specific frequency to take into account changes in the shape of the SED.

We found that the CO/IR values derived in the peak position are higher (of a factor of $\lesssim$ 2, within the errors) than those derived in the ring for all rotational transitions up to J$_{up}$ = 10. This trend changes for J$_{up}$~$>$~11 transitions where the emission in the ring becomes higher (or similar) to that of the peak.
This result suggests the presence of mechanisms able to increase the emission of $^{12}$CO at higher frequencies. In what follows we will study in detail this issue using higher resolution data, moving from intermediate to small scales in order to unveiling the origin of this mechanism.

\subsection{Mid- and high-J $^{12}$CO data within the inner 700 pc (large - intermediate scales) }
\label{CO_pacs_spire}

We now combine {\tt SPIRE (SSW} and {\tt SLW}) and {\tt PACS} data at a resolution of 35\arcsec\ ($\sim$700 pc). These instruments have different PSFs (19\arcsec\ and 35\arcsec\ for {\tt SPIRE SSW} and {\tt SLW} data, respectively, and  9.4\arcsec\ for {\tt PACS}). To properly analyze all the $^{12}$CO spectra over the whole frequency range we smoothed all data to the the largest PSF (35\arcsec). The reference spectrum in the {\tt SLW SPIRE} data cube is the one corresponding to the $^{12}$CO peak emission (central spaxel).

\begin{figure}
\centering
\hskip-5mm\includegraphics[width=0.47\textwidth]{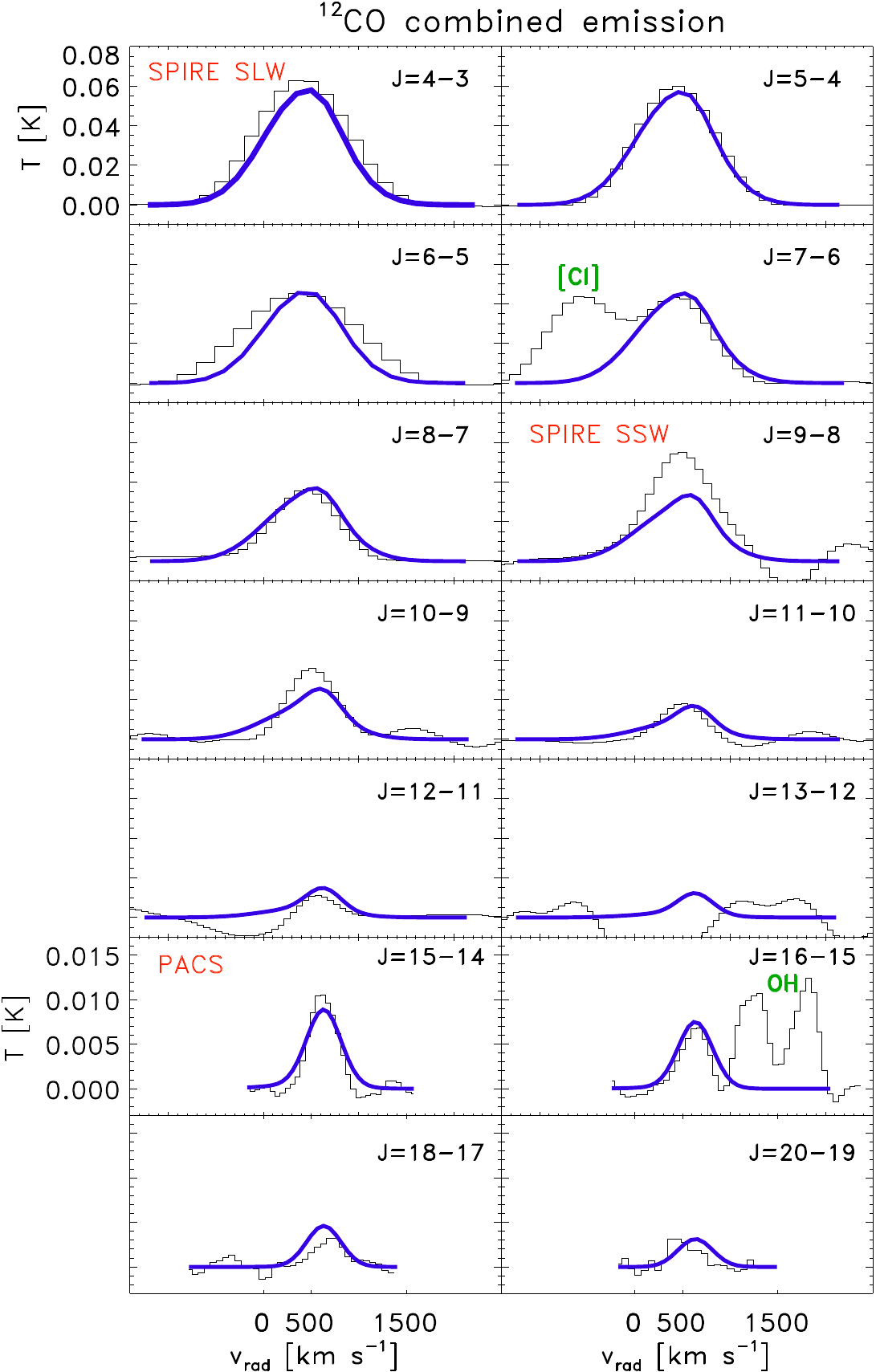}
\vskip3mm
\includegraphics[width=0.48\textwidth]{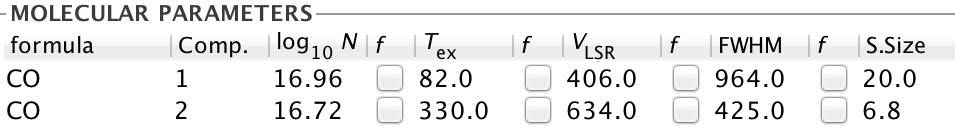}
\vskip2mm\caption{{\it Top:} Averaged {\tt SLW, SSW SPIRE} (J$_{up}$ = 4 up to 13) and {\tt PACS} $^{12}$CO spectra (J$_{up}$ = 15 up to 20) of the central region combined together at the same resolution (35\arcsec). The original spectra are shown in black while the total simulated $^{12}$CO emission obtained from the LTE approach are shown in blue. In green other molecular species like [CI] and OH are identified in emission. {\it Bottom:} Table from {\tt MADCUBA} showing the output parameter values (i.e., column density N$_{CO}$, excitation temperature T$_{ex}$, v$_{LSR}$, FWHM of the line as well as the source size) deriving different source sizes for the cold and hot components.}
\label{combi_spectra}
\end{figure}

The {\tt SPIRE} data have been combined as explained in the previous section while for the {\tt PACS} data we only combined the emission observed in eight\footnote{The $^{12}$CO emission is observed in 12 spaxels in the {\tt PACS} FoV, as shown in Fig.~\ref{beams_spire_pacs} using star symbols, but we only combined those spaxels for which the $^{12}$CO emission is stronger (white stars).}spaxels. In Fig.~\ref{combi_spectra} the {\tt SPIRE} and {\tt PACS} $^{12}$CO spectra (rotational transition from J$_{up}$ = 4 up to J$_{up}$ = 20) of the central spaxel at a resolution of 35\arcsec\ are shown. 

The LTE analysis well reproduces the observed $^{12}$CO emission (see Fig.~\ref{combi_spectra} and table below) finding the following results: 

\begin{enumerate}
\item two temperature components are needed to properly fit the spectra: one warm at $\sim$80 K and the other hot at 330 K. The hottest temperature is characterized by the lowest molecular column density (N$_{CO}$$\sim$5.2$\times$10$^{16}$ cm$^{-2}$) while the warm component is characterized by a higher column density value (N$_{CO}$$\sim$9$\times$10$^{16}$ cm$^{-2}$);
\item two different source sizes characterize the warm and hot components: for the warm component a source size of 20\arcsec\ has been assumed (see \S~\ref{Analysis_data}) while for the hot component a source size of $\sim$7\arcsec\ has been derived from the fit. 
\end{enumerate}

\subsection{Heating at intermediate scales (360 pc -- 1 kpc) from {\tt SSW SPIRE} and {\tt PACS} data}
\label{SSW_spire}

\begin{figure*}
\centering
\includegraphics[width=0.645\textwidth, height=0.45\textwidth]{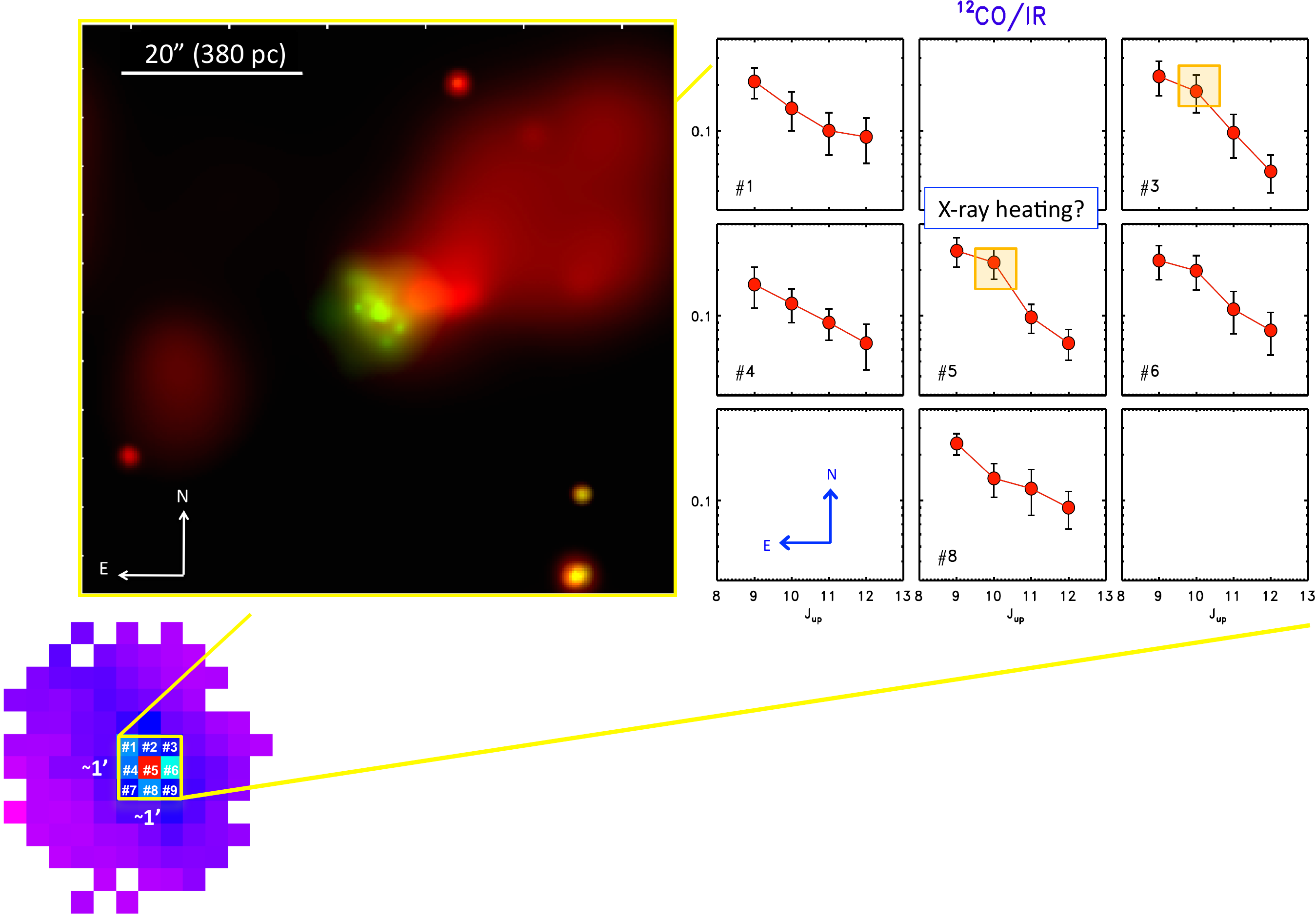}
\hskip2mm\includegraphics[width=0.305\textwidth, height=0.305\textwidth]{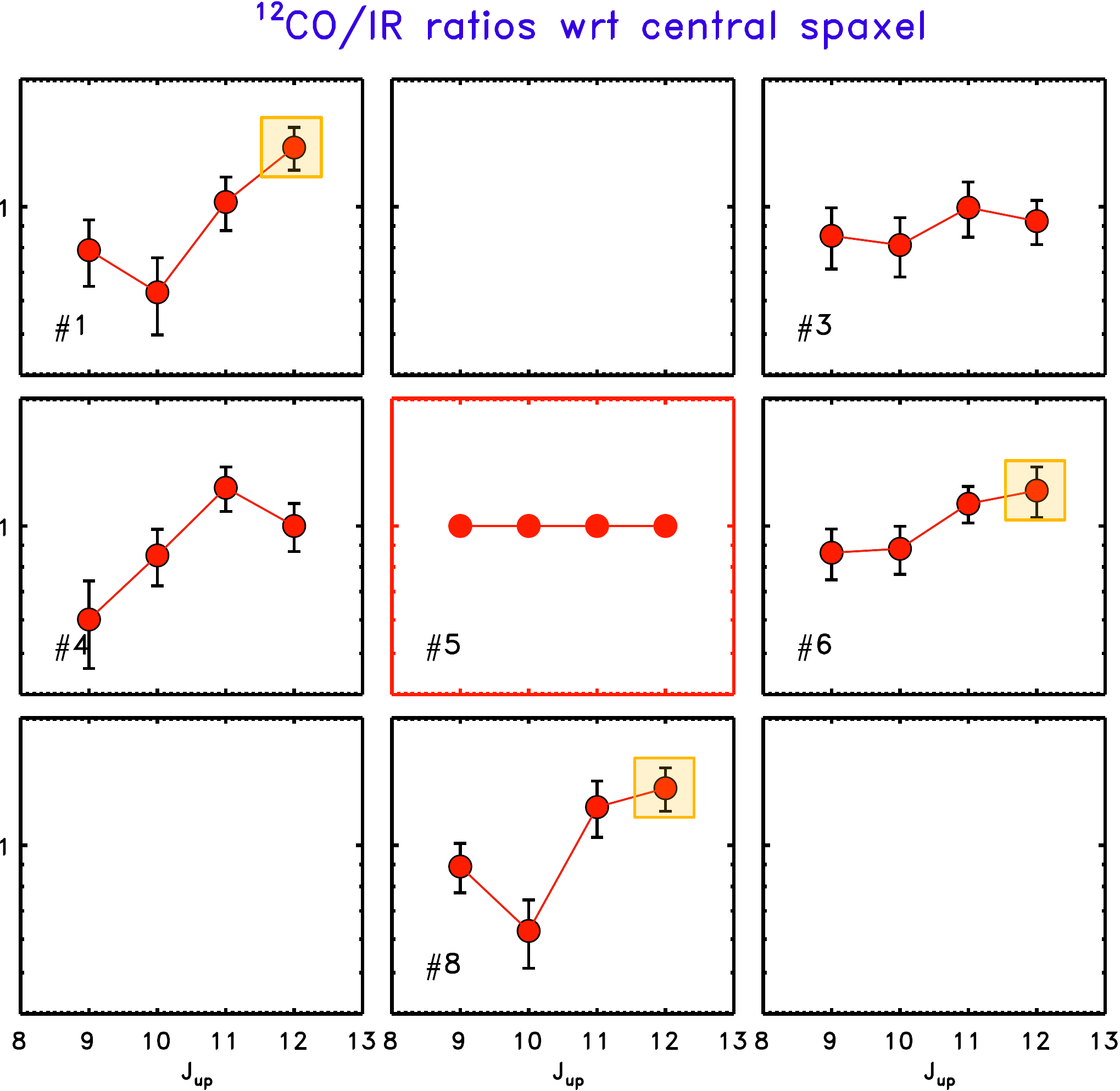}
\caption{{\it Bottom left:} {\tt SPIRE SSW} $^{12}$CO emission maps showing the nine spaxels involved in the analysis. {\it Top left:} Hard (green) and soft (red) X--ray emission from {\it Chandra} from \cite{Marinucci12} of the outflow observed in the central region ($\sim$1$^{\prime}$$\times$1$^\prime$) in NGC 4945. {\it Middle:}$^{12}$CO/IR results obtained using {\tt SSW SPIRE} data (19\arcsec\ beam) numbered following the scheme shown in the bottom left panel. {\it Right:} $^{12}$CO/IR results normalized to the emission of the central spaxel (red square). }
\label{SLED_SSW}
\end{figure*}

In this section we first focus on the analysis of the emission observed at intermediate scales described by using {\tt SSW SPIRE} data. At these scales the differences of the $^{12}$CO/IR ratios found in the ring (0.36--1 kpc, or 19\arcsec--57\arcsec) and those derived in the central spaxel ($<$360 pc) become significant. 
A FoV of 3$\times$3 spaxels ($\sim$1$^\prime$$\times$1$^\prime$) is considered, which corresponds to the observed extension of the high-J $^{12}$CO emission in this galaxy (mainly found in the disk plane).
The $^{12}$CO/IR ratio distribution in each\footnote{The spaxels have been numbered according to their position in the FoV. Those for which no data are shown implies that no $^{12}$CO emission has been detected.} spaxel is shown in Fig.~\ref{SLED_SSW}: an increase of the $^{12}$CO/IR ratio is apparent in the central spaxel (spaxel \#5) and in the north-west direction (spaxel \#3) for the rotational transitions J$_{up}$ = 9 and 10. This increase at higher J seems to follow the direction of the outflow observed in the X--ray band by {\it Chandra} (see \S~\ref{Intr}).
Assuming that the X--ray outflow is responsible of such an increase in these directions, we normalize the emission of each spaxel to the central one.
In the ring we then derived the highest $^{12}$CO/IR ratios in the disk plane of the galaxy for J$_{up}$ = 12 (i.e., north-eastern (\#1), western (\#6) and southern (\#8) spaxels; see Fig.~\ref{SLED_SSW} right panel).  
At these spatial scales the increased emission at higher frequencies (J$_{up}\ge$ 11--12) suggests that other mechanisms, like shocks, could be also at work. In principle, we excluded the (pure) PDR process to be the responsible of this increase at such high frequencies (see \S~\ref{discus_1} for further details).

\begin{figure*}
\centering
\hskip5mm\includegraphics[width=0.65\textwidth, height=0.35\textwidth]{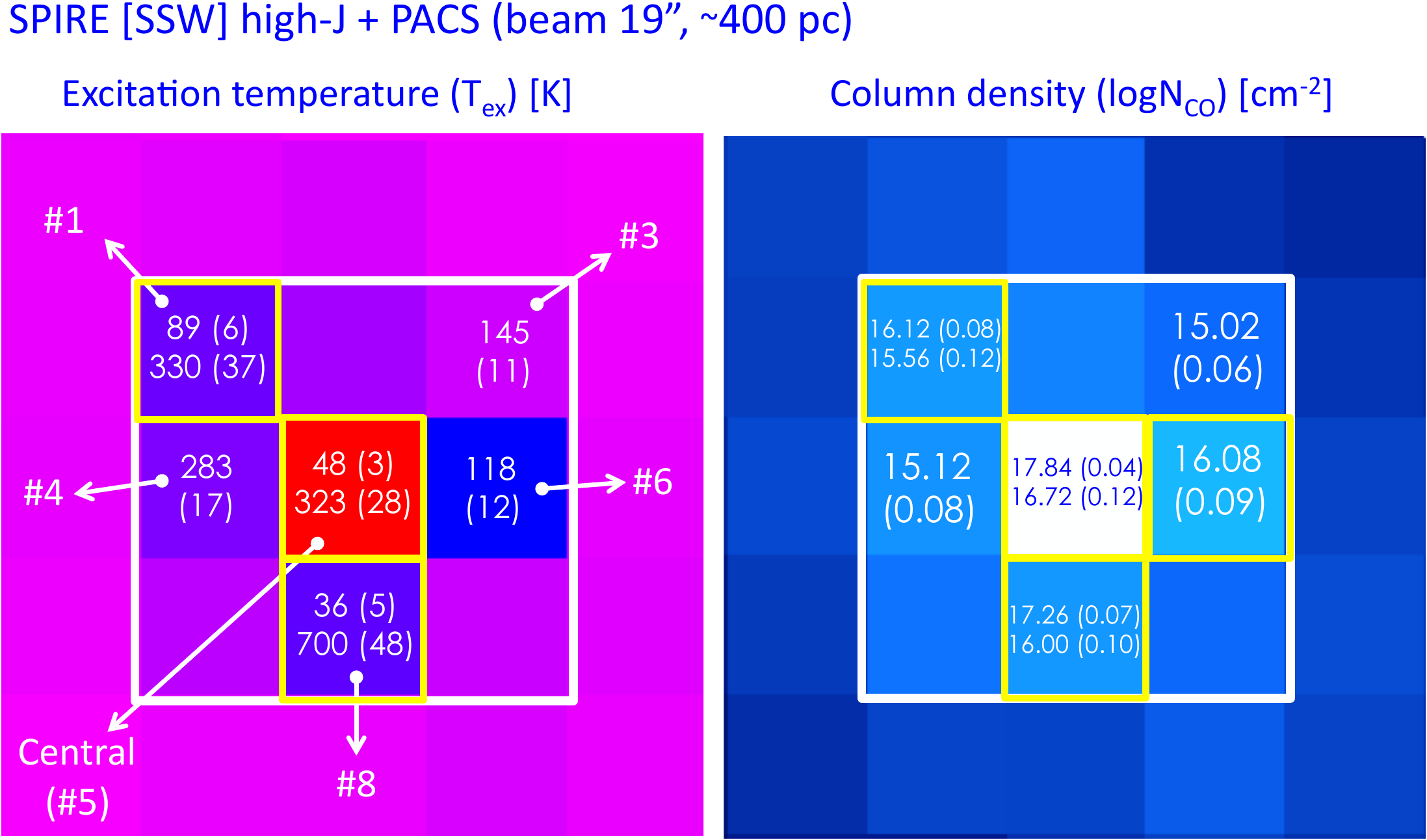}
\vskip4mm
\hskip0cm\includegraphics[width=0.26\textwidth, height=0.26\textwidth]{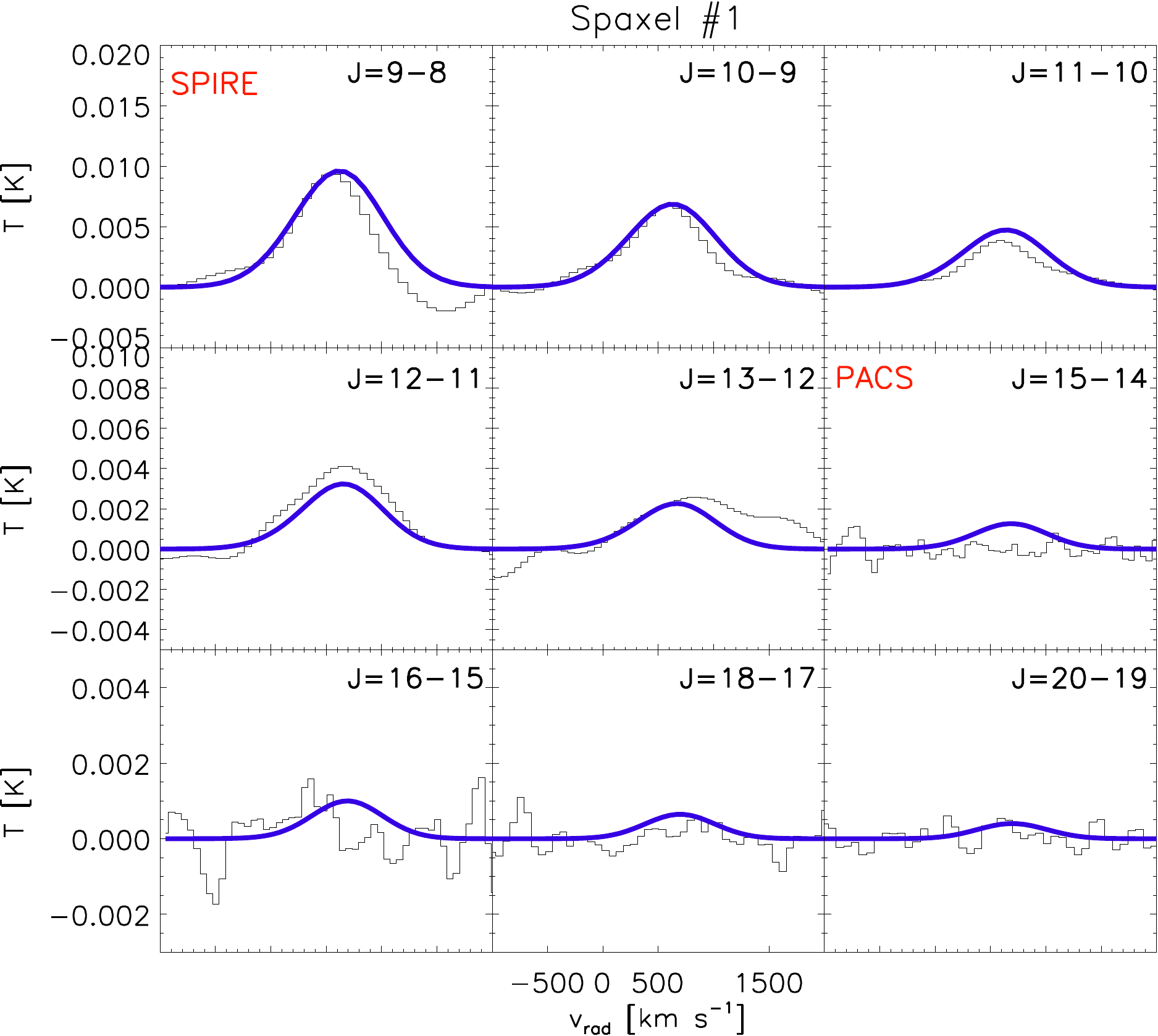}
\hskip5cm\includegraphics[width=0.25\textwidth, height=0.15\textwidth]{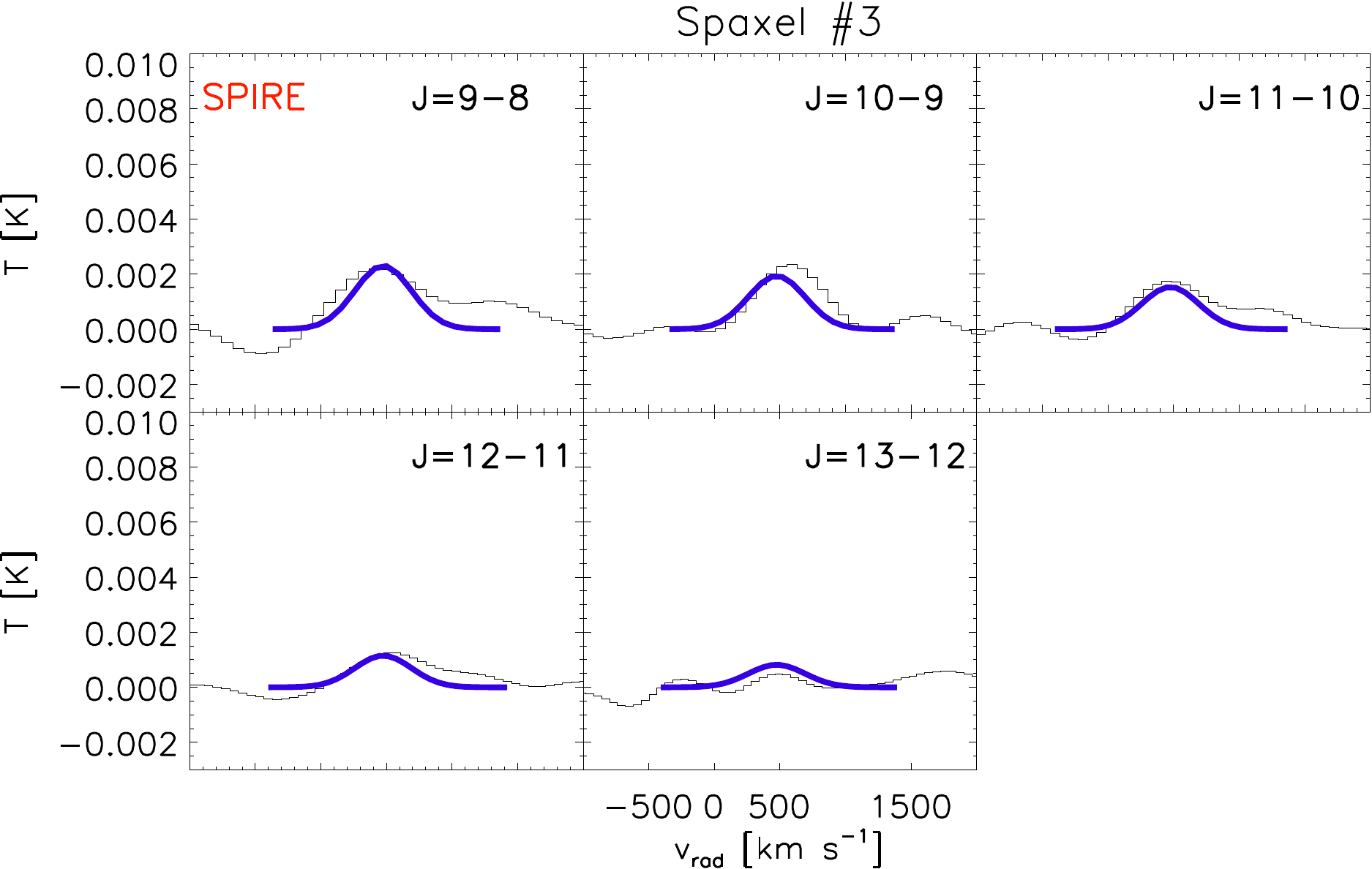}
\hspace{1cm}\includegraphics[width=0.25\textwidth, height=0.15\textwidth]{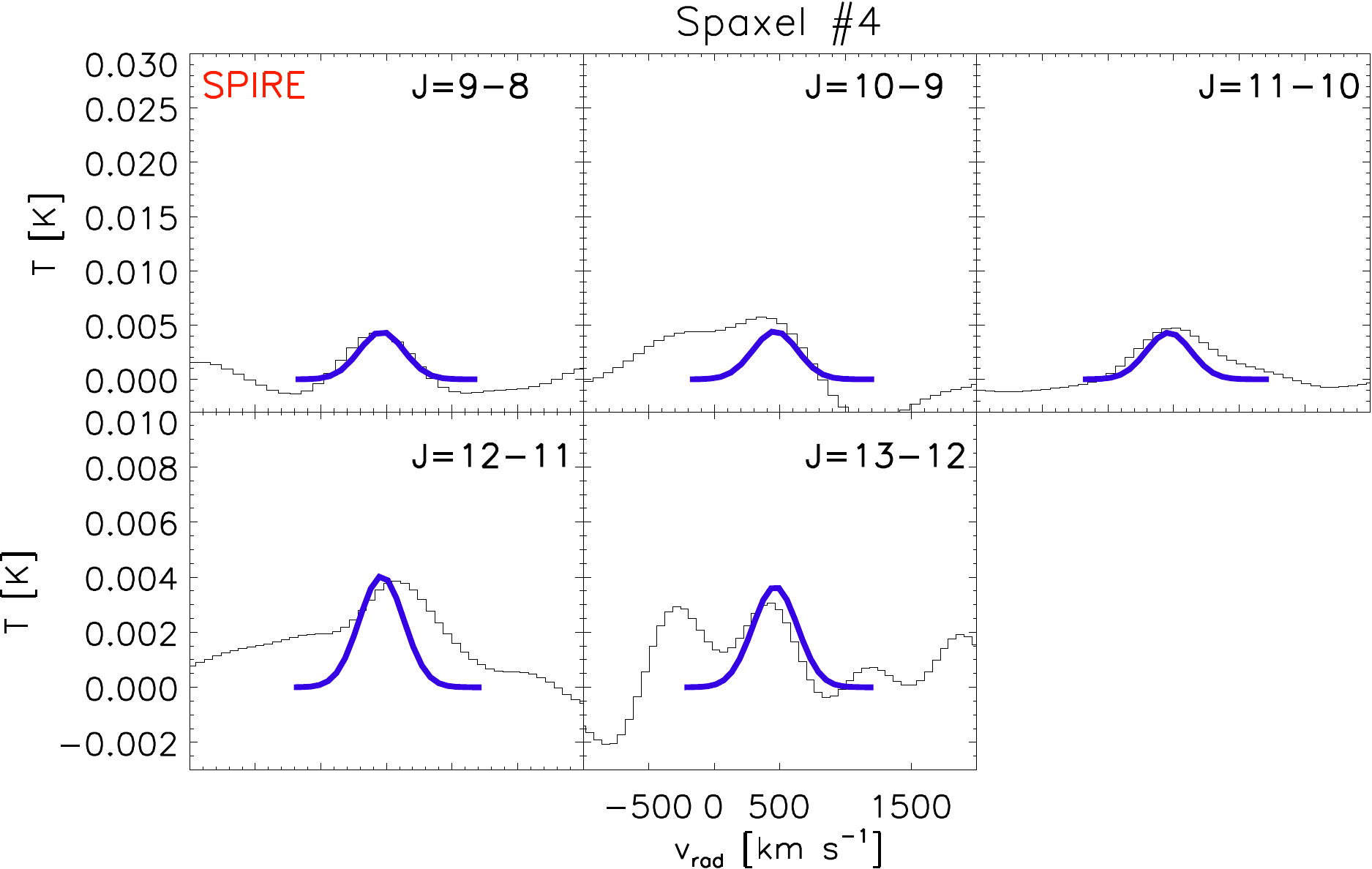}
\hskip5mm\includegraphics[width=0.26\textwidth, height=0.26\textwidth]{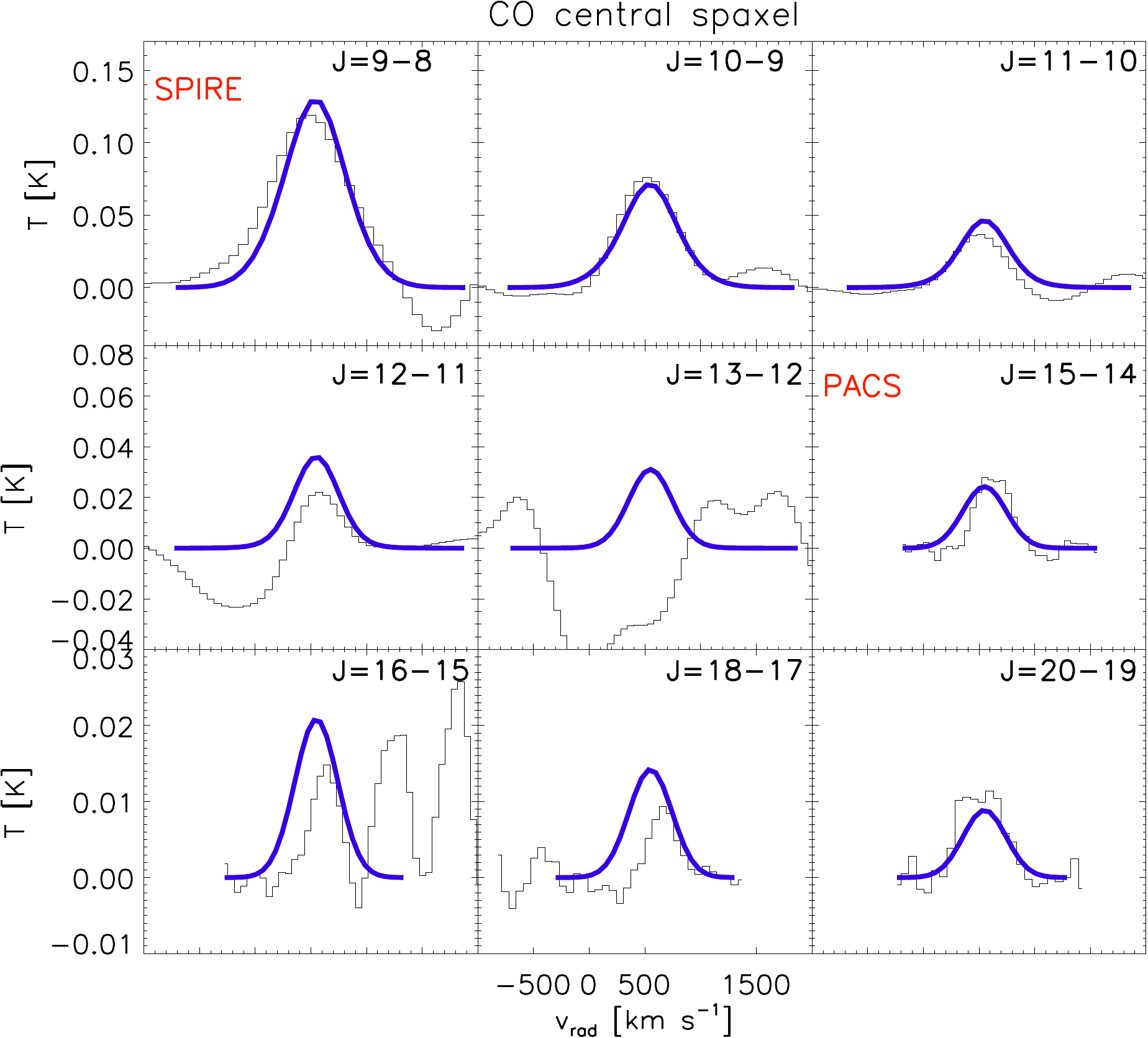}
\hskip5mm\includegraphics[width=0.25\textwidth, height=0.15\textwidth]{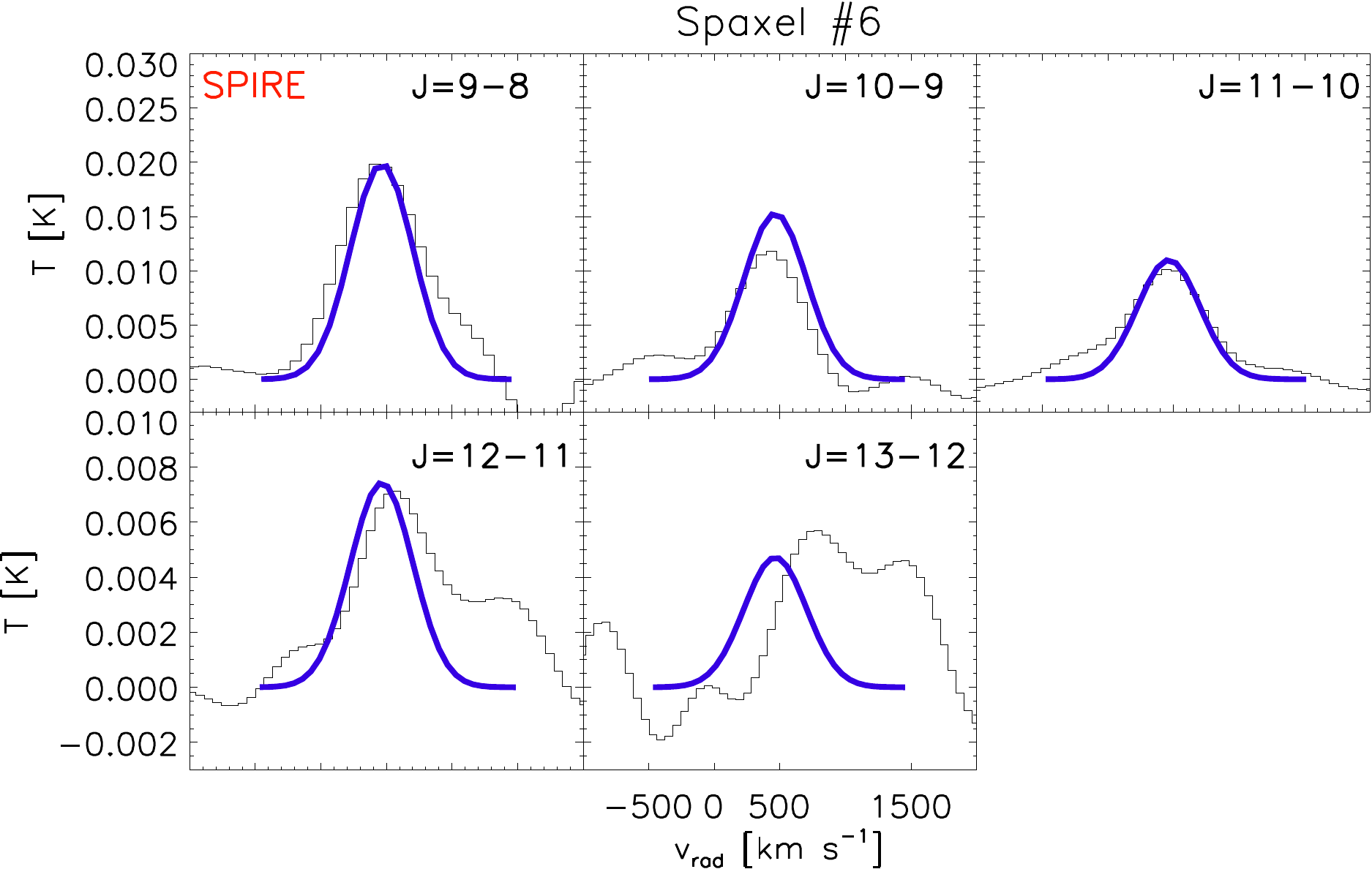}
\vskip2mm
\hskip4mm\includegraphics[width=0.26\textwidth, height=0.26\textwidth]{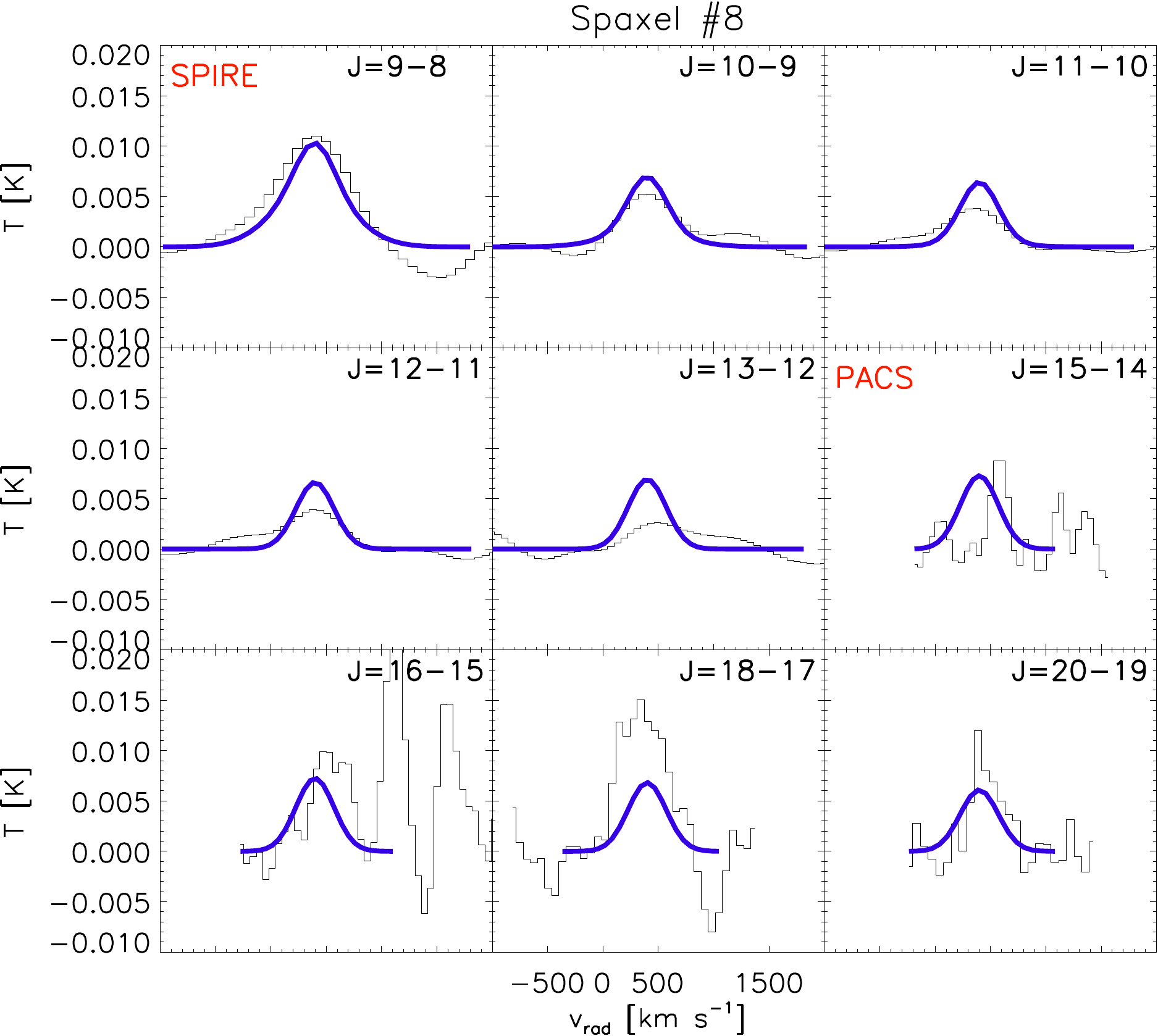}
\caption{LTE results derived combining {\tt SSW SPIRE} and {\tt PACS} spectra. {\it Top:} Distribution of the excitation temperature (T$_{ex}$, {\it top left}) and column density (N$_{CO}$, {\it top right}) derived applying {\tt MADCUBA} to the combined {\tt SSW SPIRE} and {\tt PACS} spectra. The FoV covered by 3$\times$3 spaxels  ($\sim$1\arcmin$\times$1\arcmin) is the same than that shown in Fig.~\ref{SLED_SSW}.
{\it Bottom:} Observed (black) and simulated (blue) $^{12}$CO emission spectra from combining {\tt SSW SPIRE} and {\tt PACS}. The spaxels are identified using the same number used in the top panel. For the spaxel \#3 and \#6 only {\tt SPIRE} data are available while for the spaxels \#1, \#5 (central) and \#8 {\tt SSW SPIRE} and {\tt PACS} data are combined together. The $^{12}$CO emission is found in the inner region (3$\times$3 spaxels) and mainly located in the disk, with some contribution in the perpendicular direction.  
}
\label{temp_distr_SSW}
\end{figure*}

In the next step, we combine the  {\tt SSW SPIRE} spectra with those from {\tt PACS} at higher frequencies, smoothing the {\tt PACS} data to the  {\tt SSW SPIRE} resolution (beam 19\arcsec). 
In this  case, for each {\tt SSW} spectrum we combined (averaged) $\sim$3--4 {\tt PACS} spectra. 
Unfortunately, only half of the {\tt PACS} spectra presented detections to be considered in the data.
In particular, for the spaxels \#1, \#5 and \#8 the $^{12}$CO emission from {\tt SPIRE} and {\tt PACS} were considered, while for the remaining spectra  (\#3, \#4 and \#6) we only considered the {\tt SPIRE} emission (Fig.~\ref{temp_distr_SSW}, bottom). For all of them we applied the LTE analysis which allowed us to derive the T$_{ex}$ and N$_{mol}$ parameters in each spaxel (Fig.~\ref{temp_distr_SSW}, top panel) at the resolution of 19\arcsec. 
From this analysis we found high T$_{ex}$ in the disk and in the south direction where a maximum value is found. For these spaxels (\#1, \#5 and \#8) two component temperatures are needed to properly fit the spectra. 

The column density N$_{CO}$ shows a maximum in the central spaxel for both the warm and hot components (N$_{CO}$ = 5$\times$10$^{16}$ and 6.3$\times$10$^{17}$ cm$^{-2}$) and slightly lower values in the south (N$_{CO}$ = 10$^{16}$ and 2$\times$10$^{17}$ cm$^{-2}$). In the disk plane column densities $\lesssim$ 10$^{16}$ cm$^{-2}$ are derived.

\subsection{Heating and density distribution at small scales ($\lesssim$200~pc) using {\tt PACS}}

\begin{figure}[!h]
\centering
\includegraphics[width=0.15\textwidth]{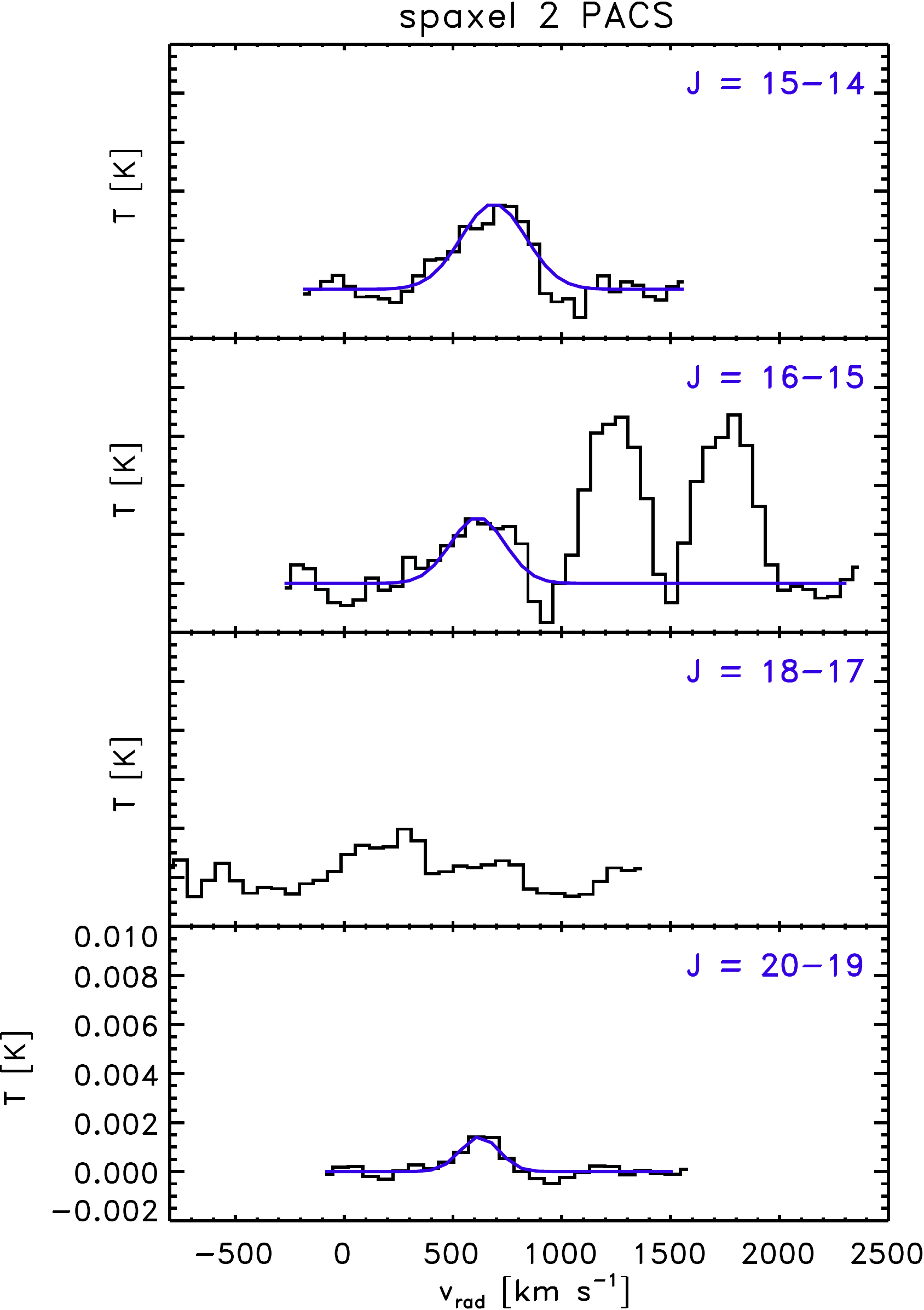}
\includegraphics[width=0.15\textwidth]{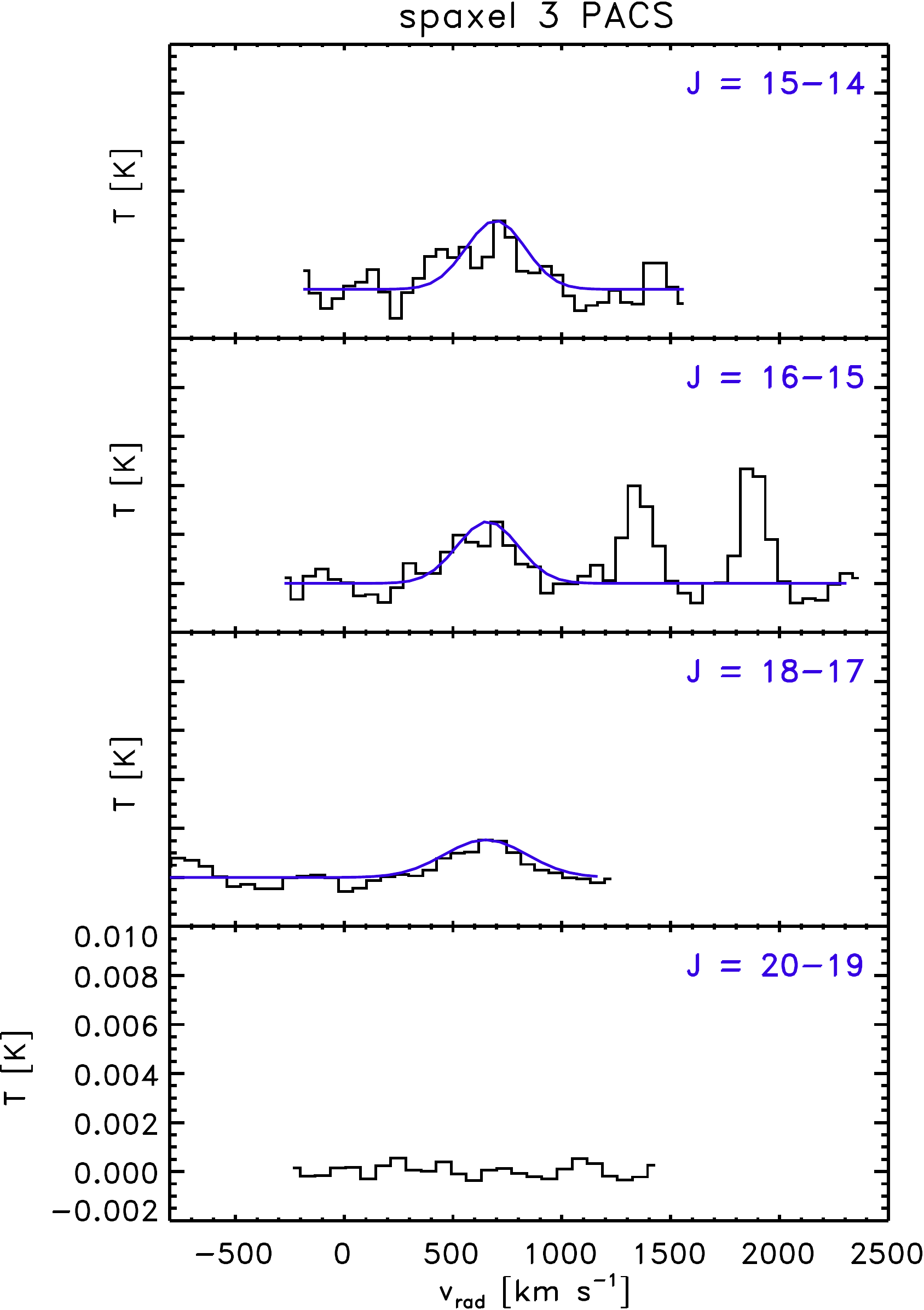}
\includegraphics[width=0.15\textwidth]{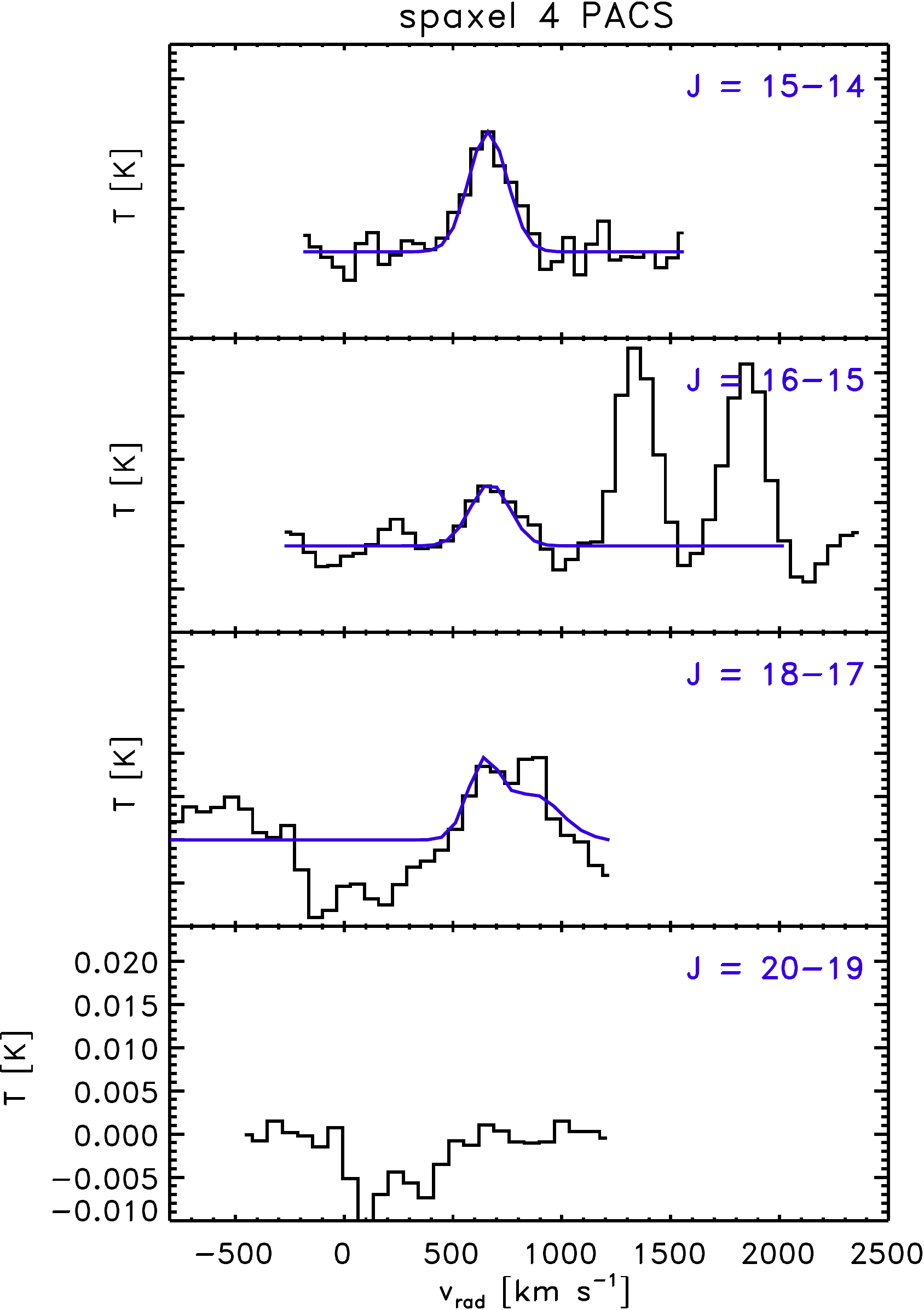}
\vskip5mm
\includegraphics[width=0.15\textwidth]{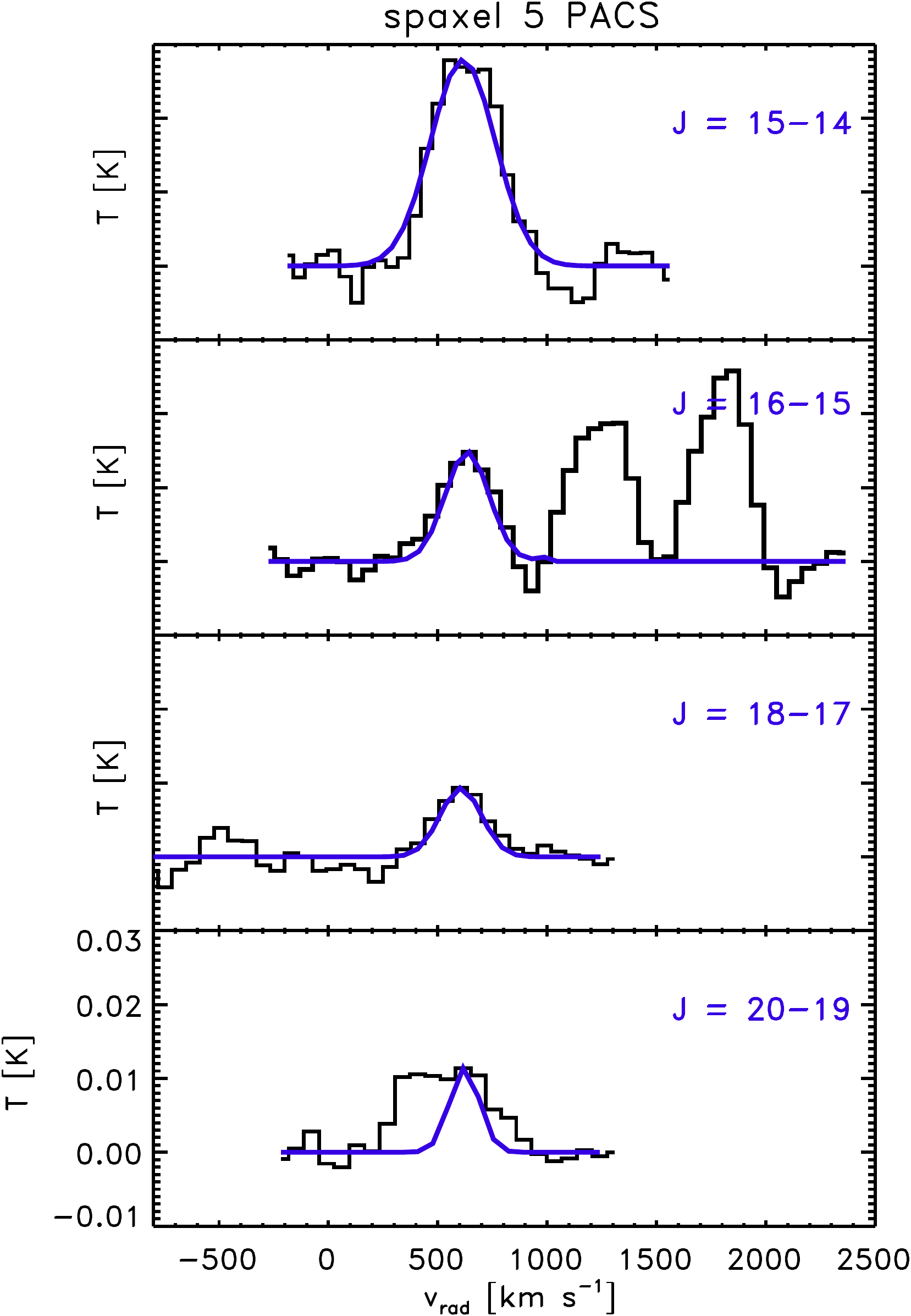}
\includegraphics[width=0.15\textwidth]{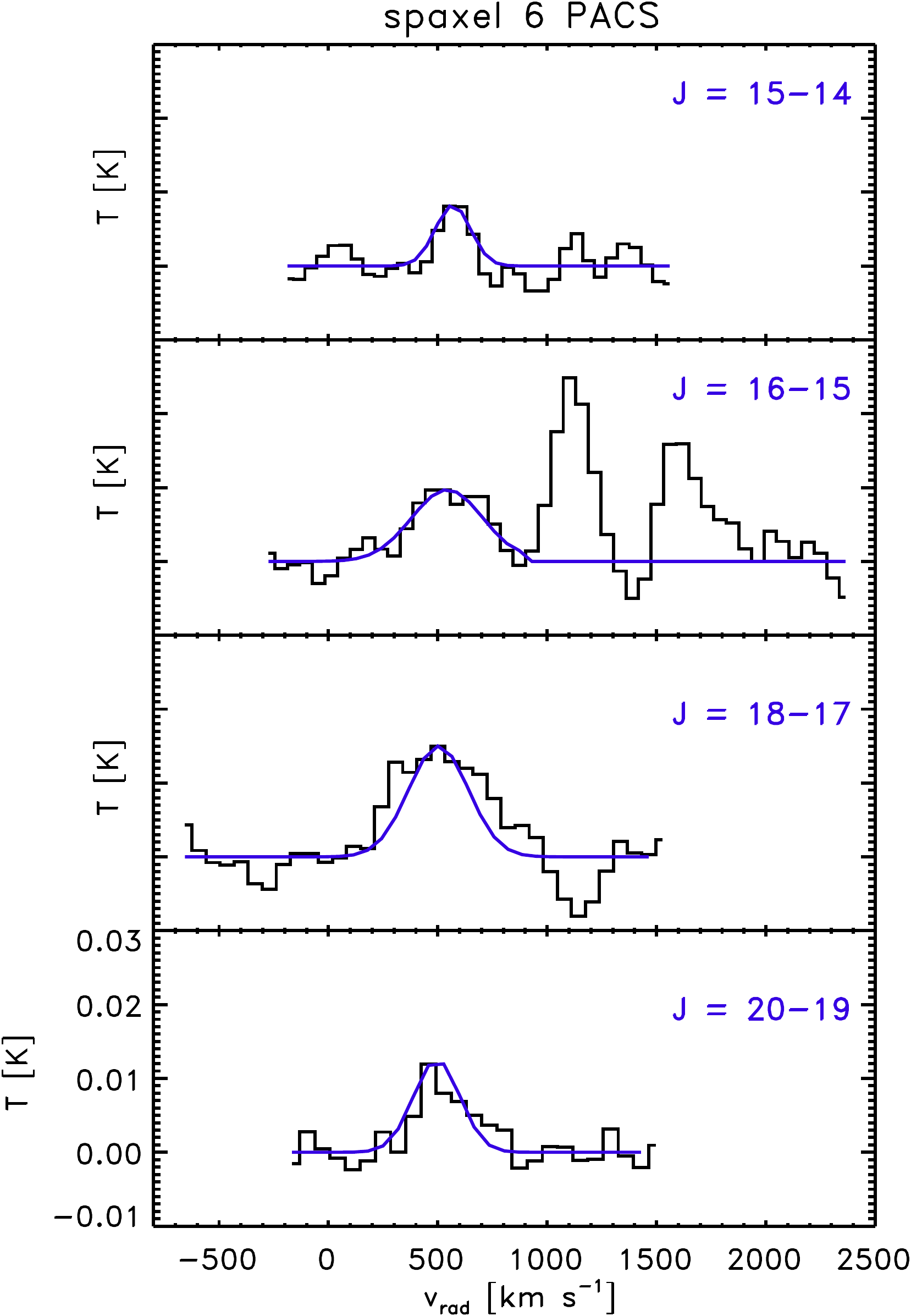}
\includegraphics[width=0.15\textwidth, height=4cm]{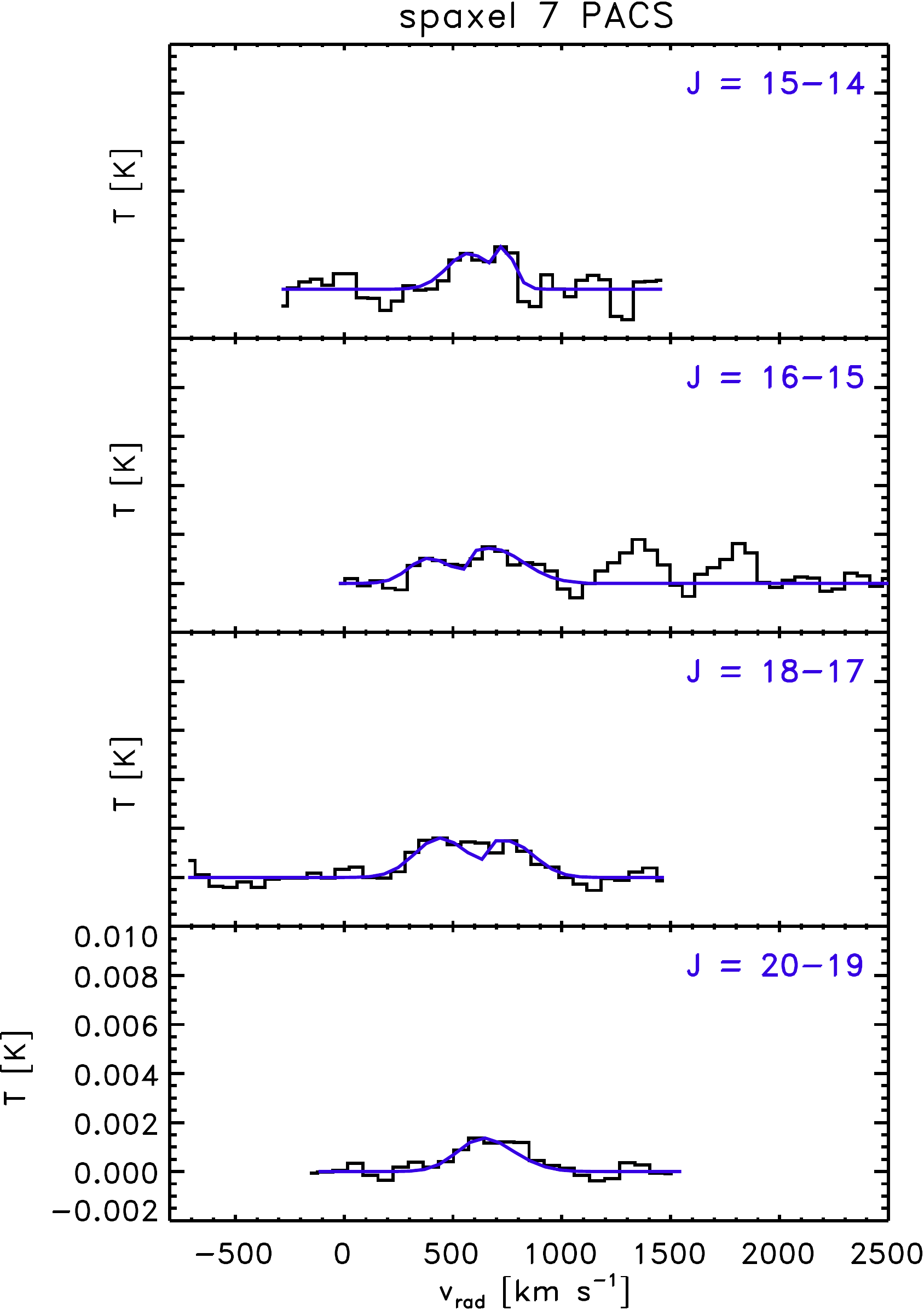}
\vskip5mm
\includegraphics[width=0.15\textwidth]{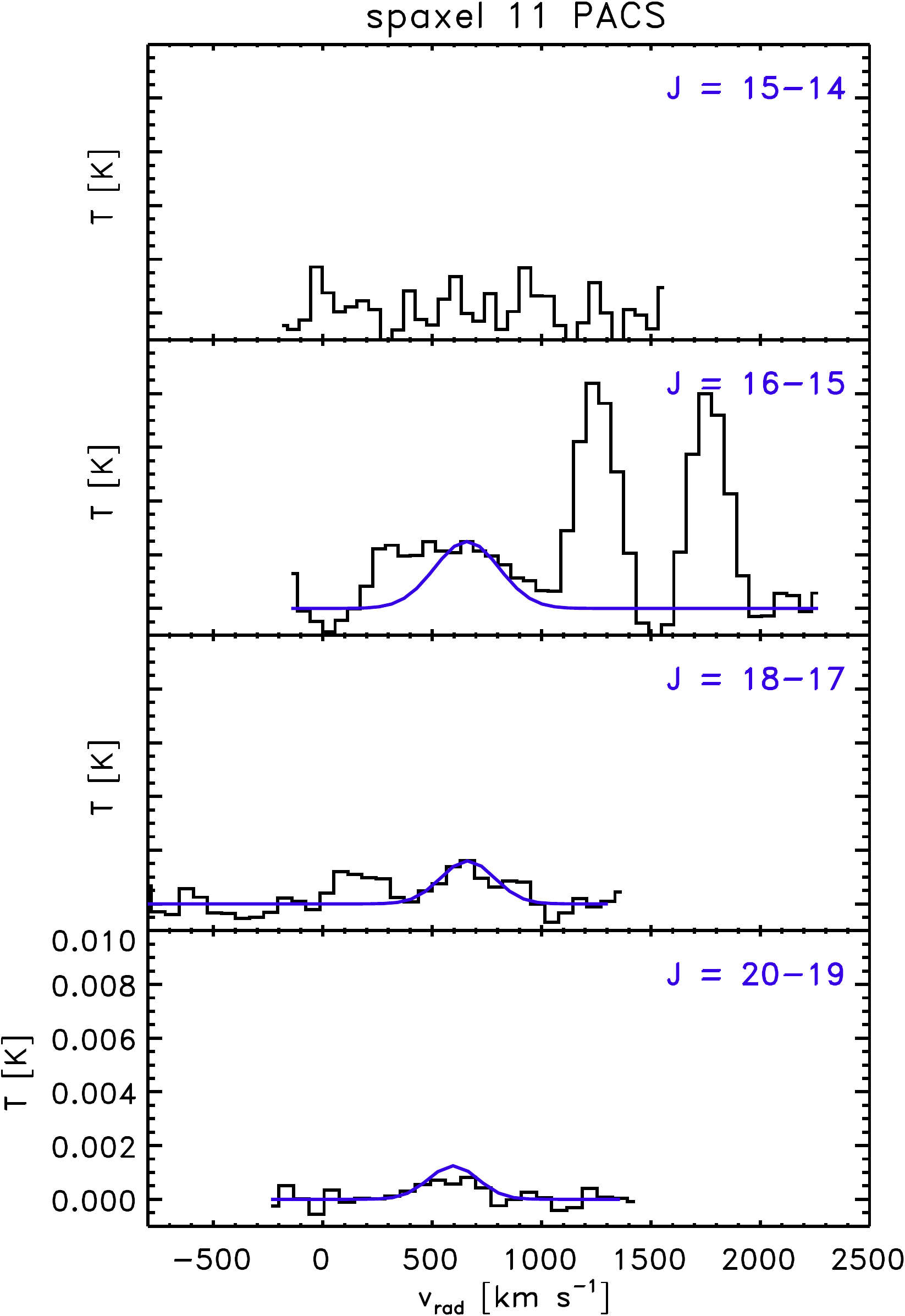}
\includegraphics[width=0.15\textwidth, height=4cm]{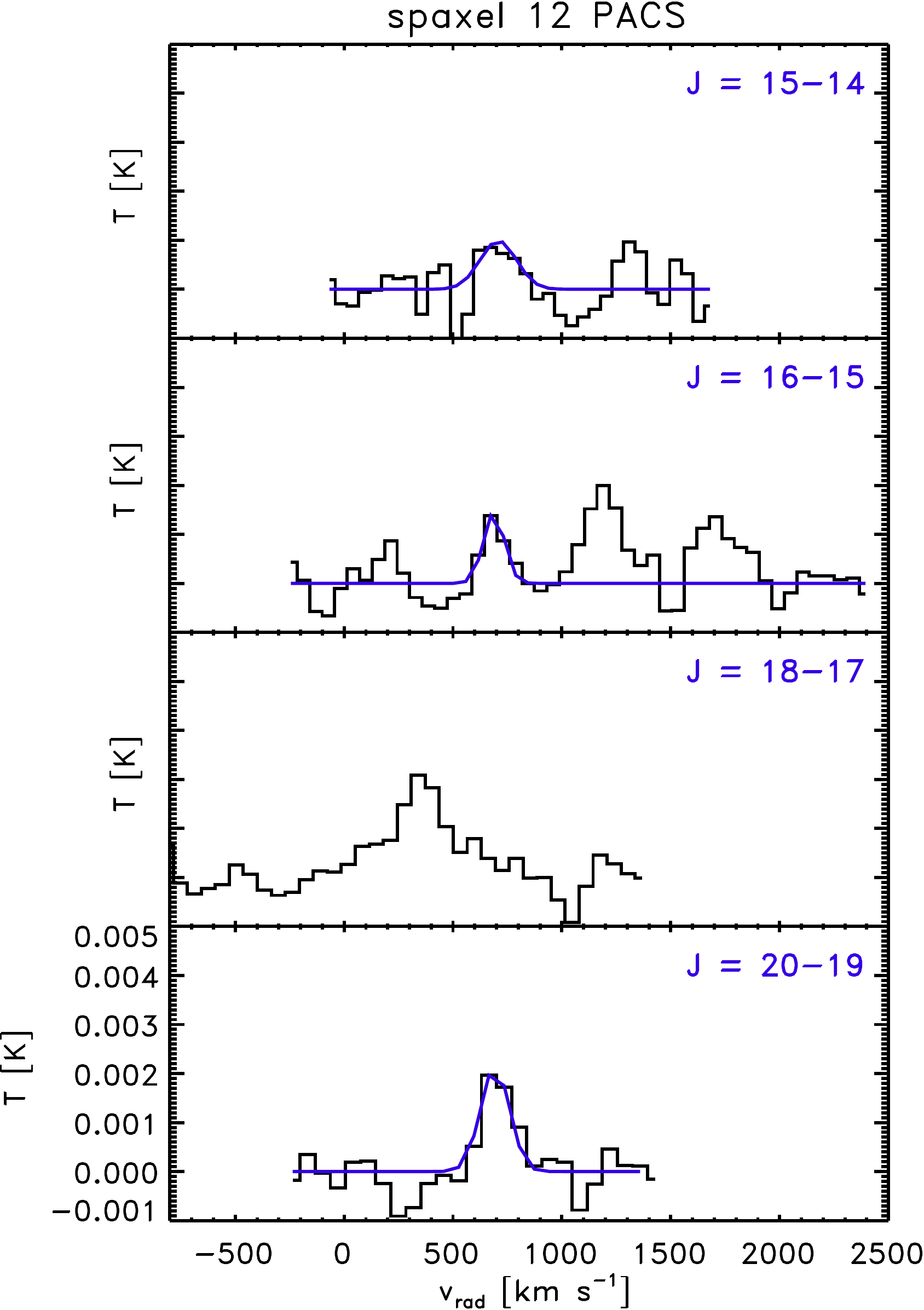}
\vskip15mm
\caption{{\it Left:} Observed $^{12}$CO {\tt PACS} spectra (black) along with the simulated Gaussian fit results (blue). For each spectrum the respective rotational transition (J+1~$\rightarrow$ J) is shown. The flux emission is shown in main beam temperature (T$_{MB}$). The OH emission lines close to the $^{12}$CO(16--15) transition are also observed (see Fig.~\ref{combi_spectra}).}
\label{PACS_rot_diag_1}
\end{figure}

\begin{figure}
\includegraphics[width=0.45\textwidth, height = 0.5\textwidth]{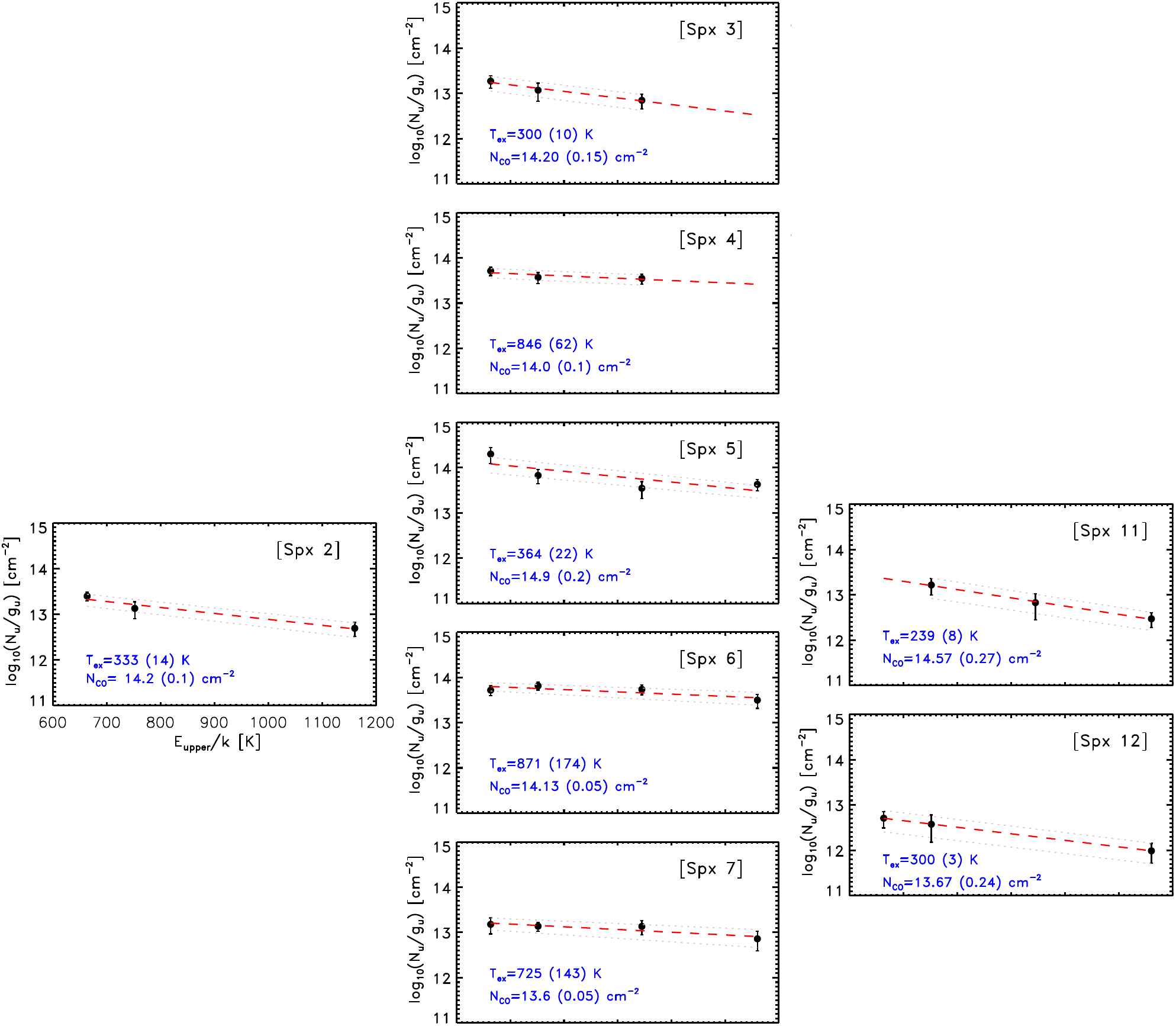}
\caption{Rotational diagrams derived for {\tt PACS} data. The excitation temperatures (T$_{ex}$) and the (logarithmic) column densities (N$_{CO}$) with their respective uncertainties are derived for each spaxel. }
\label{PACS_rot_diag_2}
\end{figure}

We now focus our attention to the $^{12}$CO emission observed at higher frequencies with {\tt PACS}. At this resolution (9.4\arcsec) we are covering spatial scales of the order of $\lesssim$200 pc. 
The observed {\tt PACS} spectra along with the simulated LTE results obtained with {\tt MADCUBA}\footnote{The fit results are obtained applying the Gaussian line fit (see \citealt{Martin19}).} are shown in Fig.~\ref{PACS_rot_diag_1}. 
From the rotational diagrams (Fig.~\ref{PACS_rot_diag_2}) we obtained the T$_{ex}$ and N$_{mol}$ for each spaxel\footnote{The uncertainties on T$_{ex}$ and N$_{CO}$ are computed considering the worst possible case (i.e., half the difference between the two extreme slopes), so they can be considered upper limit errors (3$\sigma$).}.
From this analysis we found that the highest temperatures (846 K and 871 K) are not found in the nucleus but in two spaxels located closed to the nucleus in the northern and southern spaxels. They are mainly located in the disk plane of the galaxy. On the other hand, the nucleus is characterized by T$_{ex}$$\sim$360~K. Above and below the disk plane lower T$_{ex}$ are found (from $\sim$240 K up to $\sim$330 K).

According to this result, mechanical heating seems the most probable mechanism able to explain the spatial distribution of the excitation temperature at this scale. Indeed, if the X--ray emission were dominating the nuclear region one would have expected the highest excitation temperature in the nucleus. In order to exclude the presence of a XDR in the central spaxel, we derived the intrinsic excitation temperature, correcting the observed T$_{ex}$ for the nuclear extinction. 
We thus apply the extinction law:

\begin{equation}
\hskip1mmI^{int}_{\lambda} = I^{obs}_{\lambda} \times e^{\tau_{\lambda}},
\end{equation}

where $ \tau_\lambda$ can be derived following the relation: 

\vskip-5mm
\begin{equation}
\hskip1mm \tau_\lambda = \tau_{100 \mu m}\times \left( \frac{100\hskip1mm \mu m}{\lambda} \right)^{\beta}. 
\end{equation}

The optical depth $\tau_{100 \mu m}$ is derived at 100 $\mu$m from the continuum SED fitting (i.e., $\tau_{100\mu m}\sim$1.2) with $\beta$=2.0 (\S~\ref{par_SED}), assuming that the gas is homogeneously mixed with the dust. For each {\tt PACS} spectrum of the nuclear spaxel we applied the extinction law associated to the specific wavelength. The corrected excitation temperature of the central spaxel is $\sim$470 K, far below the values obtained in the surrounding regions ($\sim$850 K). We then conclude that the dust opacity does not play an important role in our conclusions. {\it Even the AGN interaction does not seem to have a strong impact on the thermal structure of the source at large spatial scales}.

\begin{figure*}
\centering
\hskip8mm\includegraphics[width=0.86\textwidth]{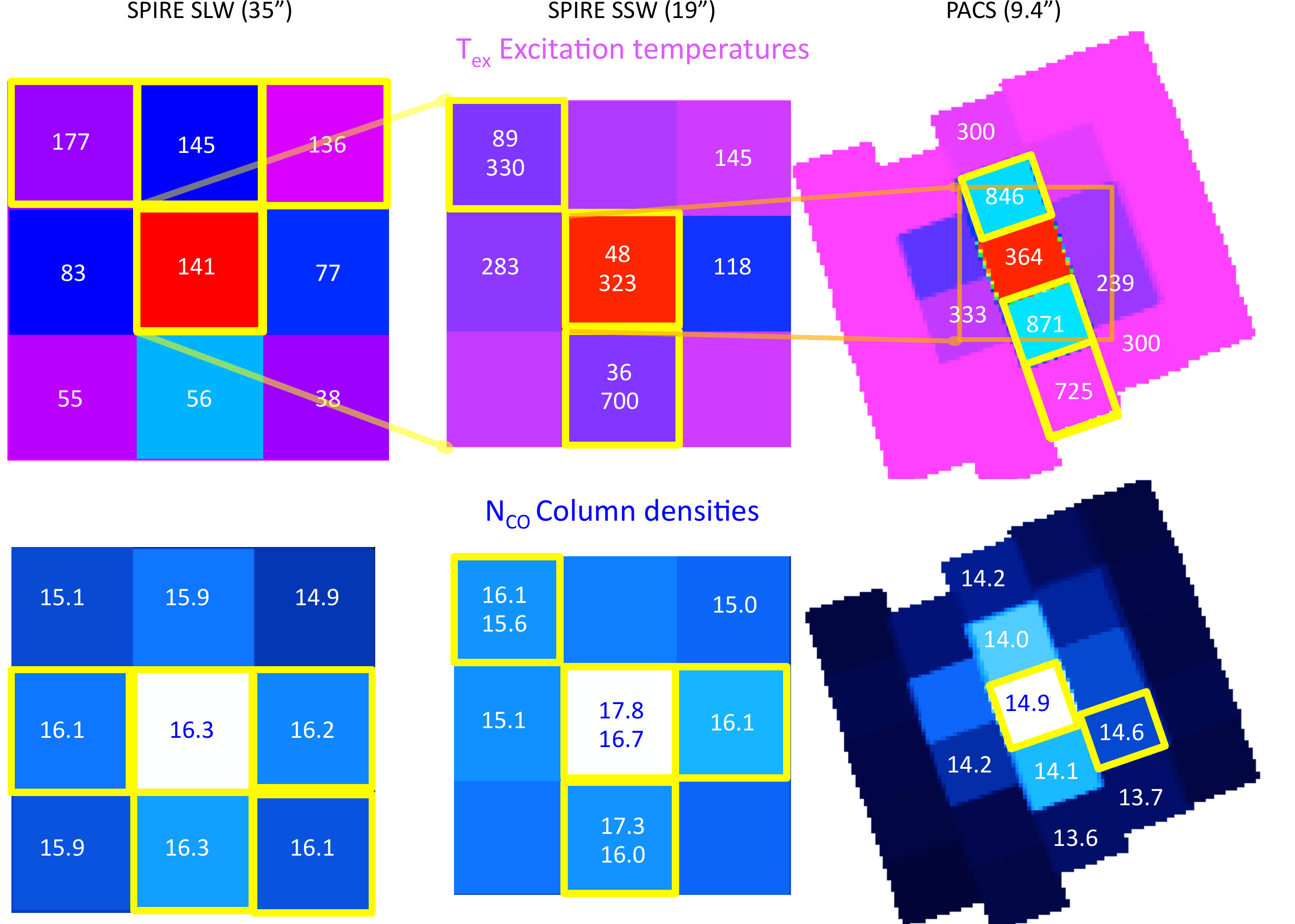}
\caption{Excitation temperature and (logarithmic) column density distributions at different spatial scales as derived using {\tt SPIRE (SLW, SSW)} and {\tt PACS} data. The yellow squares identify the spaxels with high values for each parameter.}
\label{EX_TEMP_total}
\end{figure*}

According to the results obtained from large to small scales we summarize the distribution of T$_{ex}$ in Fig.~\ref{EX_TEMP_total} (top panel). The  excitation temperature distribution of  {\tt PACS} data is in good agreement with that derived using {\tt SSW SPIRE} data.

For what concerns the molecular column density N$_{CO}$, the highest values are found in the nucleus corresponding to moderate excitation temperatures, while lower column densities are found in correspondence of maximum temperatures in the disk. In Fig.~\ref{EX_TEMP_total} (bottom panel) we reported the distribution of N$_{CO}$ at different spatial scales.

In Tab.~\ref{Summary_table_all} we summarize all the excitation temperature T$_{ex}$ and column density N$_{CO}$ values derived at different spatial scales.

\begin{table*}
\begin{tiny}
\caption{Summary of the several T$_{ex}$ and N$_{CO}$ values derived for the $^{12}$CO molecule at different resolutions.  }
\label{Summary_table_all}
\hskip-4mm
\begin{tabular}{l ccccccc} 
\hline\hline\noalign{\smallskip}  
 Spatial scale & Spatial resolution  & Instrument 	& J$_{up}$ levels 	&	Number of  	& T$_{ex}$ & N$_{CO}$ & Figure/ \\
 \hskip22mm &(arcsec, pc) &  &  &Components& (K) & (cm$^{-2}$) & Table\\ 
(1) & (2) & (3) & (4) & (5) & (6) & (7) & (8) \\
\hline\noalign{\smallskip} 	
Intermediate& 20\arcsec, $\lesssim$ 400 & {\tt APEX, HIFI}	&  3 $\rightarrow$ 9		&	2	& 17, 90   &		17.6, 16.7	&	F\ref{spettri_2}/T\ref{RADEX}					 \\
\hline\noalign{\smallskip} 	
Large& 35\arcsec, $\lesssim$ 700&  {\tt SPIRE SLW \& SSW}	& 4 $\rightarrow$ 13		&1 (2$^a$)	& 141   &   16.3	&	F\ref{CO_LIR_ring}, F\ref{EX_TEMP_total}		\\
\hline\noalign{\smallskip} 	
Large--intermediate& 35\arcsec, $\lesssim$ 700 & {\tt SPIRE SLW \& SSW, PACS}	&  4 $\rightarrow$ 20 	&	2	& 82, 330  &		$\sim$17, 16.7	&	F\ref{combi_spectra}		 \\
\hline\noalign{\smallskip} 	
Intermediate& 19\arcsec, $\lesssim$ 400 &  {\tt SPIRE SSW, PACS}	&	  9 $\rightarrow$ 20	&2	& 48, 323  &		17.8, 16.7	&	F\ref{temp_distr_SSW}, F\ref{EX_TEMP_total}			\\
\hline\noalign{\smallskip} 	
Small&  9.4\arcsec, $\lesssim$ 200 & {\tt PACS}	&	  15 $\rightarrow$ 20	&	1	& 364 (470$^b$)   &		14.9 (14.86)	&	F\ref{PACS_rot_diag_1}, F\ref{PACS_rot_diag_2}, F\ref{EX_TEMP_total}			 \\
\hline\hline\noalign{\smallskip} 	
\end{tabular}
\end{tiny}
\begin{minipage}{18cm}
\small
{{\bf Notes:} Column~(1-2): Spatial scale and spatial resolution of the data analyzed; Column~(3): instrument with which the analysis has been performed; Column~(4): (upper) rotational transition J$_{up}$ range involved in the analysis accordingly to the instruments considered, listed in Col.~(3); Column~(5): number of components (i.e., c) used in the fit;  Column~(6): excitation temperature of $^{12}$CO molecule in {\it kelvin}; Column~(7): column density of $^{12}$CO molecule in {\it cm$^{-2}$}; Column~(8): figure (F) and/or table (T) showing the results in each specific case; (a) see \S~\ref{SLED_spire} for details; (b) see \S~\ref{SSW_spire} for details.}
\end{minipage}
\end{table*}

\subsection{The  dust and gas in a multi--phase ISM}

The trend we found in our $^{12}$CO column densities, N$_{CO}$, as a function of the rotational levels J involved in the LTE analysis shows, as expected from a multi--phase molecular clumpy medium, a gradient in both the H$_2$ density, the kinetic temperature and a decreasing column density of the hot gas at increasing J. As the quantum number J of the transitions used in the analysis increases, the physical conditions required for their excitation change according to their critical densities and the energy above the ground state of the levels involved in our study.
In fact, we see large changes in temperature from 20 K for the mid J to 400 K for the high J (see Tab.~\ref{Summary_table_all}). This is consistent with the picture of the multi--phase molecular ISM described above. Therefore, the $^{12}$CO column densities, N$_{CO}$, will decrease from the {\tt HIFI} to the {\tt PACS} data analysis since the amount of dense and hot gas measure by the high-J is much smaller than the cold-warm gas measure from the low-J. 

We could use the ratios between the column densities from the different instruments to roughly estimate the fraction of the warm-hot molecular component to the cold component. In particular, the cold--warm component at $\sim$20~K is characterized by N($^{12}$CO$_{cold-warm}$) = 10$^{17.6}$ cm$^{-2}$, the warm component at 90 K shows N($^{12}$CO$_{warm}$)~$\lesssim$~10$^{17}$ cm$^{-2}$ while the hot component at $\sim$370 K is characterized by N($^{12}$CO$_{hot}$)$\sim$10$^{15}$ cm$^{-2}$. In addition, the coldest component traced by the J=1--0 and J=2--1 transitions have a $^{12}$CO column density of 9.6$\times$10$^{18}$~cm$^{-2}$ for a source size of 20\arcsec$\times$20\arcsec (\citealt{Wang04}), about one order of magnitude larger than the cold--warm component.
Thus, the ratio between the cold--warm (CW) and hot (H) components with respect to the cold (C) component is CW/C=0.05  and H/C=10$^{-4}$: these values correspond to larger column density of the cold--warm component with respect to the hot component (CW/H) of a factor of $\sim$500.

From our results we can also estimate the total molecular hydrogen column densities, N(H$_2$). 
From the SED fitting analysis we derived a total molecular of 7.6 10$^8$ M$_\odot$ for the typical GDR = 100. Then, the N$_{H_2}$ obtained for the size of the dust emission of 20\arcsec$\times$10\arcsec\ corresponds to N$_{H_2}$$\sim$ 7$\times$10$^{23}$~cm$^{-2}$. 
To properly account for the total column density N$_{CO}$ we need to consider the $^{12}$CO column densities derived for all the components discussed above and scale them (i.e, multiply by a factor of $\sim$2) to the size of the dust emission of 20\arcsec$\times$10\arcsec. 
After the correction for the different source sizes, the total $^{12}$CO column density is of $\sim$2$\times$10$^{19}$ cm$^{-2}$ which translate into a molecular hydrogen column density of N$_{H_2}$$\sim$ 2$\times$10$^{23}$ cm$^{-2}$ for the $^{12}$CO fractional abundance of 10$^{-4}$. This is within a factor of $\sim$3 lower than that derived from the SED fitting analysis which is likely within the uncertainties in the sizes, the dust absorption coefficient, the fractional abundance of $^{12}$CO and the GDR we have considered. Assuming the standard conversion:

\begin{equation}
\label{eqn:convert}
N_{H_2} = 9.4\times10^{20}\hskip2mm A_V\hskip5mm[cm^{-2}], 
\end{equation}
from \cite{Bohlin78} (see also \citealt{, Kauffmann08, Lacy17}), we derive very large visual extinction in both cases ($>$200 mag) as a results of the derived column densities.

We thus find that the hydrogen column densities derived from the SED fitting approach and $^{12}$CO analysis are in good agreement. These values are higher than those derived for some local AGN and  starburst dominated galaxies like NGC 1068 (\citealt{GB14, Viti14}) and NGC 253 (\citealt{PBeaupuits18}), but similar to those derived for Compton-thick type 2 Seyfert galaxies like Mrk 3 and NGC 3281 (\citealt{Sales14}).

Furthermore, the typical molecular fractional abundances observed in starburst galaxies derived using high density gas tracers like [HCN]/[H$_2$] (=X$_{HCN}$) is of the order of 10$^{-8}$ (see \citealt{Wang04, Martin06}). The derived column densities between $^{12}$CO and HCN from our LTE analysis are of the order of N$_{CO}$/N$_{HCN}$ $\sim$10$^{17}$/10$^{13}$$\sim$10$^4$. This is the same ratio than that obtained when considering the fractional abundances relative to H$_2$, X$_{CO}$ = 10$^{-4}$ and X$_{HCN}$ = 10$^{-8}$ (see \citealt{Martin06}).

\section{Discussion}
\label{discuss}

\subsection{Gas heating mechanisms}
\label{discus_1}

Distinguishing among the heating mechanisms, like photoelectric effect by UV photons ({PDRs}) or {XDRs} and mechanical processes (like shocks, stellar winds, outflows) is not straightforward, and in most cases a number of mechanisms coexist with different contributions depending on the spatial scale. 
Many works have addressed this issue modeling the effect of different mechanisms and comparing the predictions with $^{12}$CO observations of galaxies with different type of activity like starburst galaxies, as M82 (\citealt{Panuzzo10, Kamenetzky12}) and NGC 253 (\citealt{Rosenberg14, PBeaupuits18}), AGN-dominated galaxies, as NGC 1068 (\citealt{Spinoglio12, HaileyD12}) and Mrk 231 (\citealt{vanderWerf10, Mashian15}), and composite AGN-SB galaxies, like NGC 6240 (e.g., \citealt{Meijerink13}).
The $^{12}$CO emission is strongly affected by the specific mechanism(s) (or by the combinations of them) at work in each galaxy. Some differences can be highlighted between them. 
When PDRs dominate the emission, the $^{12}$CO emission increases up to rotational transition J$_{up}$=5 and then decreases. In presence of XDRs or shocks the contribution of the $^{12}$CO emission increases up to high (J$_{up}>$10) frequencies.

\begin{figure*}
\centering
\includegraphics[width=0.97\textwidth,height = 0.37\textwidth]{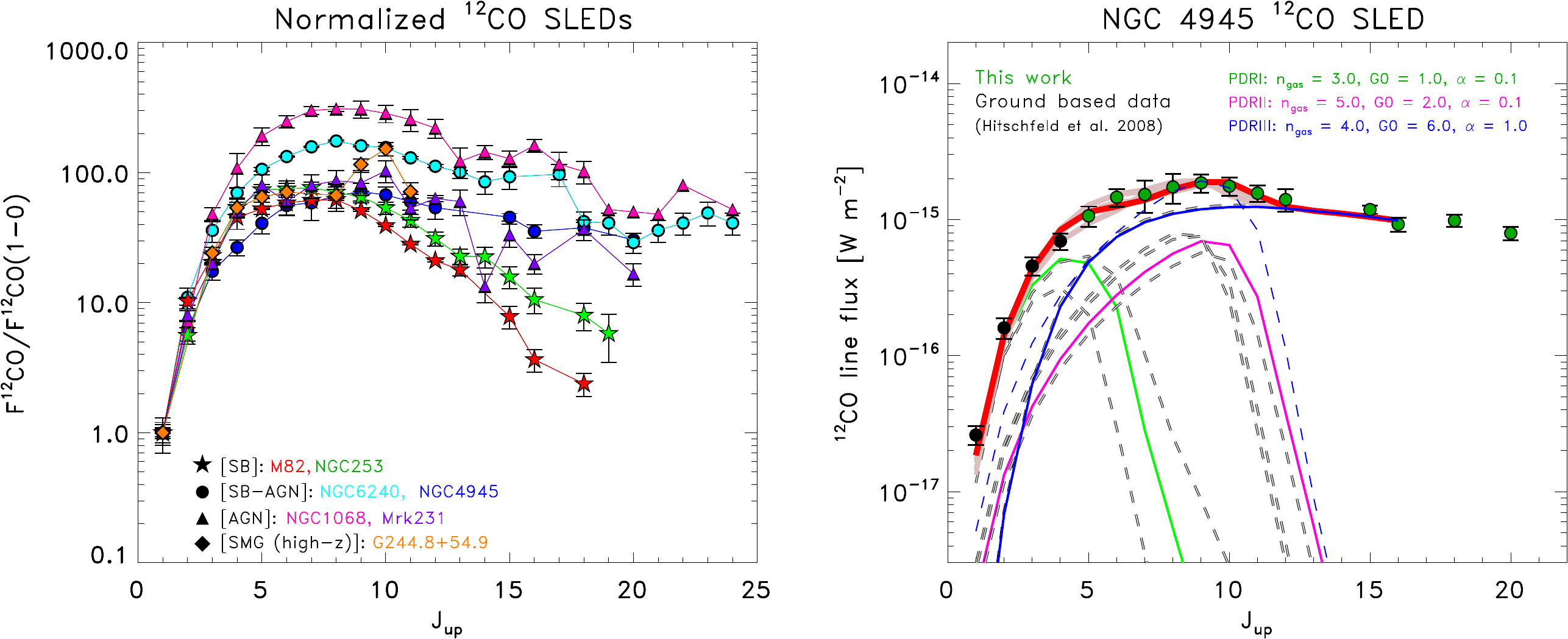}
\caption{{\it Left:} Comparison of $^{12}$CO SLEDs (normalized to the $^{12}$CO(1-0) flux of each specific source) obtained from different sources. Starburst (SB), AGN and composite (AGN-SB) sources have been considered. A $^{12}$CO SLED of a sub-millimeter galaxy (SMG) at high-z analyzed by \cite{Canameras18} has been also shown. Different colors identify the galaxies while different symbols the types of object. In each group the following galaxies have been selected: M82 and NGC 253 as starburst, NGC 6240 and NGC 4945 as AGN-SB composite galaxies and NGC 1068 and Mrk 231 as AGN-dominated galaxies. The galaxy G244.8+54.9 has been chosen as representative of SMG at high-z, for which the authors derived log~n$_{gas}$$\sim$5.1 cm$^{-3}$ and log~G$_0$$\sim$3.7 (Habing units).
{\it Right:} Results of the $^{12}$CO SLED of NGC 4945 from J = 1--0 through 20--19 applying the \cite{Kazandjian15} mechanical heating models. The black points are the observed $^{12}$CO fluxes obtained from ground based observations (\citealt{Hitschfeld08}) while in green are the data from this work. 
The fit was constrained to three PDR models, all of them with some contribution of mechanical heating (mPDR), displayed in green, magenta and blue. Dashed gray lines show the other best fit results for the three PDRs. The dashed blue line is also shown to represent a model with the same n$_{gas}$ and G$_0$ parameters as those characterizing the PDR{\tiny III} model but with no mechanical heating contribution ($\alpha$=0).
The red line represents the best fit with the minimum $\chi^2_{red}$ ($\sim$0.8) while the light red area shows the combination of other best fits with slightly higher $\chi^2$ ($<$1.4). The H$_2$ density (i.e., n$_{gas}$ in {\it log cm$^{-3}$}), G$_0$ (log Habing flux) and percentage of mechanical heating $\alpha$ values are shown in the legend for the three best fit components. }
\label{Model_alpha}
\end{figure*}

The $^{12}$CO Spectral Line Energy Distribution (hereafter, $^{12}$CO SLED) for a large variety of systems has been used in literature as a powerful tool to derive the physical parameters characterizing the molecular gas phase. 
In Fig.~\ref{Model_alpha} (left panel) the $^{12}$CO SLEDs for different kind of galaxies are shown. 
Most of the $^{12}$CO fluxes shown in Fig.~\ref{Model_alpha} (left panel) are taken from the work by \cite{Mashian15} (and references therein). This plot can be used for direct comparison of the $^{12}$CO SLEDs of the different systems up to high J transitions.

We have selected three different kinds of galaxies: (1)~AGN-dominated, (2) SB-dominated and (3) AGN-SB composite galaxies. 
Among the AGN-dominated systems, we selected the prototype (Sy2) NGC 1068 and (Sy1) Mrk~231. 
The former shows strong $^{12}$CO line emission above J$_{up}$=20 while Mrk 231 shows a $^{12}$CO SLED shape relatively flat for the higher--J transitions. In both cases, \cite{Mashian15} and \cite{vanderWerf10} claimed that the results are consistent with the presence of a central X--ray source illuminating the circumnuclear region. 
\cite{HaileyD12} also found that in NGC~1068 the gas can be excited by X--rays or shocks although they could not be able to differentiate between the two. 
As starburst galaxies, we selected M82 and NGC 253. In M82 \cite{Panuzzo10} found that the $^{12}$CO emission peaks at J$_{up}$ = 7, quickly declining towards higher J. They argued that turbulence from stellar winds and supernovae may be the dominant heating mechanism in this galaxy.
\cite{Rosenberg14}, studying NGC 253, concluded that mechanical heating plays an important role in the gas excitation although heating by UV photons is still the dominant heating source.   
The results from \cite{PBeaupuits18} are also in agreement with those presented by \cite{Rosenberg14}.
Finally, the AGN-SB composite galaxy NGC 6240 has been selected. Its $^{12}$CO SLED shows a similar shape to that of Mrk 231 but this is characterized by clear evidence of both shocks and mechanical heating (\citealt{Meijerink13}).

We then derived the $^{12}$CO SLED for our AGN-SB composite galaxy NGC 4945 (Fig.~\ref{Model_alpha}).  
Comparing the different $^{12}$CO SLEDs normalized to the $^{12}$CO(1-0) flux of each individual galaxy, NGC 4945 seems to show a behavior similar to that found in Mrk 231 (AGN-dominated object) and M82 (SB galaxy) up to J$_{up}$$\sim$8, then resembling to Mrk~231 at 8$<$J$_{up}$$<$13, finally showing a trend in between that shown by Mrk 231 and NGC 6240 (AGN-SB galaxy) at 14$<$J$_{up}$$<$20. The similarity of the $^{12}$CO SLED shape at higher J transitions with those characterizing Mrk 231 and NGC 6240 provides clues on the presence of X--ray or shocks mechanisms dominating at higher frequencies.

\subsection{The dominant heating in NGC 4945}

In order to quantify the contribution of the different heating mechanisms in NGC 4945 we applied the \cite{Kazandjian15} models to investigate the effects of mechanical heating on molecular lines.
According to their models the authors found that the emission of low-J transitions alone is not good enough to constrain the mechanical heating (hereafter, $\Gamma_{mech}$) while the emission of ratios involving high-J (and also low-J) transitions is more sensitive to $\Gamma_{mech}$. The strength of $\Gamma_{mech}$ is parametrized by using the parameter $\alpha$, which identifies the ratio between the mechanical heating, $\Gamma_{mech}$, {\it versus} the total heating rate at the surface of a pure PDR (no mechanical heating applied), $\Gamma_{surf}$.
This can assume values between 0 and 1. In particular, $\alpha$ = 0 corresponds to the situation in which no mechanical heating is present in the PDR, while $\alpha$ = 1 represents the model where the mechanical heating is equivalent to the heating at its surface (see \citealt{Rosenberg14, Kazandjian15}). 
In their models, they assume that mechanical feedback processes like young stellar object (YSO) outflows and supernovae (SNe) events are able to heat the dense molecular gas. The former injects mechanical energy into individual clouds, while the latter injects mechanical energy into the star-forming region amongst the PDR clouds, by turbulent dissipation. The mechanical energy liberated by these events is then deposited locally in shock fronts (see \citealt{Loenen08, Kazandjian15}).

For NGC 4945 we considered the observed $^{12}$CO fluxes up to J$_{up}$=16 ({\tt PACS} data). In order to properly constrain the $^{12}$CO SLED, we complement our {\tt SPIRE} and {\tt PACS} data with ground based data obtained by \cite{Hitschfeld08} al lower J transitions (J$\leq$4) using {\tt NANTEN2}. The angular resolutions used in their work for the $^{12}$CO transitions from J$_{up}$=1 to 4 vary from 45\arcsec\ to 38\arcsec\ (see Tab.~2 in their work).
We scaled all these data using the largest angular resolution of 35\arcsec\ as derived from the {\it Herschel} data.

The model which better reproduces our data has been identified by the use of the minimum reduced chi-square\footnote{The $\chi^2_{red}$ has been computed according to the formula $\chi^2_{red}$ = $\sum_{i=1}^{N_J} \left( \frac{O_i - M_i}{\sigma_i}\right)^2 / N_J$. $N_J$ are the number of degrees of freedom (i.e., the number of observed data points used in the fits), $O_i$ and $M_i$ are the observed $^{12}$CO fluxes and the flux model values for the $i$--th point, and $\sigma_i$ is the corresponding observed flux error.} value ($\chi^2_{red}\sim$0.8). 
The results are shown in Fig.~\ref{Model_alpha} (right panel). 
According to the best fit result we found that NGC~4945 is characterized by medium-high gas density (log n$_{gas}$ = 3.0-5.0 cm$^{-3}$) and FUV incident flux, G$_0$\footnote{G$_0$ is expressed in `Habing units': G$_0$ = 1.6$\times$10$^3$ erg~s$^{-1}$~cm$^{-2}$ (see \citealt{Habing69}).}, ranging from 10 up to 10$^6$. 
We properly fit the observed $^{12}$CO emission using three PDR functions (mPDR{\tiny I}, mPDR{\tiny II}, mPDR{\tiny III}), all of them needing additional mechanical heating. Two of them (mPDR{\tiny I}, mPDR{\tiny II}) have $\alpha$=0.1 while the third model (mPDR{\tiny III}) has $\alpha$ = 1.0 (or 0.75 according to the second best fit value). 
For mPDR{\tiny I} the $\alpha$ value translates into $\Gamma_{mech}$$\sim$4$\times$10$^{-24}$ erg s$^{-1}$ cm$^{-2}$ while mPDR{\tiny II} is characterized by $\Gamma_{mech}$$\sim$3$\times$10$^{-20}$ erg s$^{-1}$ cm$^{-2}$. For mPDR{\tiny III} the highest mechanical heating is achieved, $\Gamma_{mech}$$\sim$5$\times$10$^{-19}$ erg s$^{-1}$ cm$^{-2}$. According to these results, it is apparent that mechanical heating is needed to reproduce the observed data. Indeed, in contrast to the results derived for the starburst galaxies NGC 253 (\citealt{Rosenberg14, PBeaupuits18}) and Arp 299 (\citealt{Rosenberg14_2}), the shape of the $^{12}$CO ladder of NGC 4945 is flatter at higher J transitions ({\tt PACS}; left panel in Fig.~\ref{Model_alpha}). In our case the contribution of shocks (or turbulent) heating is the main source of excitation for $^{12}$CO emission at J$_{up}>$ 9--10. On the other hand, photoelectric heating seems to be the main source of heating at low and mid-J transitions (J$_{up}<$10).

These results confirm that at a resolution of 35\arcsec\ physical processes like turbulent motions or shocks are able to excite the gas mechanically in the central region of NGC 4945. In the next section we propose a plausible interpretation of the mechanical heating able to explain the emission of the high-J $^{12}$CO lines detected with {\tt PACS}.

From the work by \cite{Canameras18} a sample of sub-mm galaxies at high-$z$ has been analyzed deriving the G$_0$ and n$_{gas}$ parameters. For their sample they derived a typical gas density and a FUV radiation field in the range log~n$_{gas}$$\sim$4-5.1~cm$^{-3}$ and log~G$_0$$\sim$2.2-4.5 (in `Habing units'), comparing their values with those derived for normal star-forming galaxies for which derive log~n$_{gas}$=2-4 cm$^{-3}$ and log~G$_0$$\sim$2.5-5 (in `Habing' units) (\citealt{Malhotra01}) and local ULIRGs, with log~n$_{gas}$=4-5 cm$^{-3}$ and log~G$_0$$\sim$3-5 (in `Habing' units) (\citealt{Davies03}). The values shown in Fig.~9 in their work can be used to compare our results with those obtained for these samples. 
For instance, in the case of the starburst galaxy NGC 253 (\citealt{Rosenberg14}), a (mean) density of log~n$_{gas}$$\sim$5.1~cm$^{-3}$ and FUV radiation field log~G$_0$$\sim$5.0 can be derived.
For NGC 4945, we ended up with (mean) values of log n$_{gas}$$\sim$4.6 cm$^{-3}$ and log~G$_0$$\sim$5.5. Our results place NGC 4945 in the region covered by local starburst galaxies (as NGC 253) and ULIRGs with similar density but characterized by higher FUV radiation.

From \cite{Hollenbach91} we are also able to derive a first-order estimate of the incident FUV flux radiation, G$_0$, assuming FUV heating from the dust properties derived in \S~\ref{par_SED} following the equation:
\vskip-5mm
\begin{equation}
\hskip3mmG_0 = 3.7 \times 10^{-3} \hskip2mm \tau_{100\mu m} \hskip2mm T_{dust}^5.
\end{equation}

We used the opacity $\tau$ computed at 100 $\mu$m ($\tau_{100\mu m}$$\sim$1.2) estimated from the SED fitting assuming that the dust temperature is similar to the equilibrium dust temperature at the surface of the emitting region. 
According to the SED fitting results (\S~\ref{par_SED}), we ended up with two dust temperature components from which we derived log G$_0$$\sim$5.8, similar (slightly higher) to that found for the cold component in NGC 253 (i.e., log G$_0$$\sim$5.5; see \citealt{PBeaupuits18}) and consistent with that found from the models ($\sim$5.5).

On the other hand, the n$_{H2}$ densities derived from the LTE and LVG models applied to the $^{12}$CO molecule using {\tt HIFI} and {\tt APEX} data are in between n$_{H_2}$$\lesssim$10$^4$-10$^5$ cm$^{-3}$. These values are in good agreement with the densities derived with the PDR model (n$_{H2}$$\sim$10$^3$-10$^5$ cm$^{-3}$).

\subsubsection{Mechanical heating: the bar potential}

According to the results derived in the 2D thermal structure analysis performed at different spatial scales (from $\lesssim$200 pc to 2 kpc) focusing on the emission of the $^{12}$CO molecule from the J = 4--3 to 20--19 we can summarize what follows for the four regions:

\begin{description}
\item[$\bullet$] {\tt 700 pc--2 kpc:} The $^{12}$CO/IR ratios of the low-J lines (J$_{up}$$<$10) is larger by a factor of $\lesssim$2 in the inner 700~pc than in the surrounding region (2 kpc). However, for mid-J lines (J$_{up}$=11--13), the $^{12}$CO/IR values are similar in both regions;
 
\vskip2mm 
\item[$\bullet$] {\tt Inner 700 pc:} The LTE analysis of the inner 700 pc, which includes the whole range of $^{12}$CO lines (from J$_{up}$=4 to 20), shows that the $^{12}$CO emission can be explained by a two components model with temperatures of 80 K and 330~K and source size of about 20\arcsec\ (400~pc) and 7\arcsec\ ($\sim$150 pc), respectively;

\vskip2mm 
\item[$\bullet$] {\tt 360 pc--1 kpc:} The $^{12}$CO/IR ratios peak at the mid--J (J$_{up}$=9, 10) lines in the central and north-west spaxels: this is in very good agreement with the distribution of the X--ray outflow (i.e., perpendicular -- north-west direction -- to the disk of the galaxy). Normalizing for this contribution, we derived increased $^{12}$CO/IR ratios for the highest mid--J lines (J$_{up}$=11 and 12) in the ring structure, in particular along the galaxy plane and toward the west. This result lets us argue that shocks probably dominate at these frequencies;

\vskip2mm 
\item[$\bullet$] {\tt Inner 200 pc--360 pc:} The LTE analysis of the high-J $^{12}$CO lines confirmed the trend found at larger scales. The highest temperatures ($\sim$560 K) are found around the nucleus (T$\sim$360 K, after correcting for the nuclear extinction), in the north--south direction along the galaxy disk. 

\end{description}

We thus found a clear trend in the distribution of the excitation temperatures and the $^{12}$CO/IR ratios. At large scale ($>$700 pc), the highest temperatures are found toward the nucleus and the north, with moderate temperatures in the south. The high temperatures in the north might be related to the large scale X--ray outflow. It is remarkable that, basically at all scales $<$700 pc, the highest temperatures are not found towards the nucleus but toward the disk. Like found at large scale, at intermediate scales (360~pc--1~kpc) we also see high temperatures in the direction of the X--ray outflow.
At the smallest scales ($\sim$200 pc), we clearly see that even if the largest column density is found in the nucleus, the highest temperatures are found in the disk.

\begin{figure*}
\centering
\includegraphics[width=0.78\textwidth]{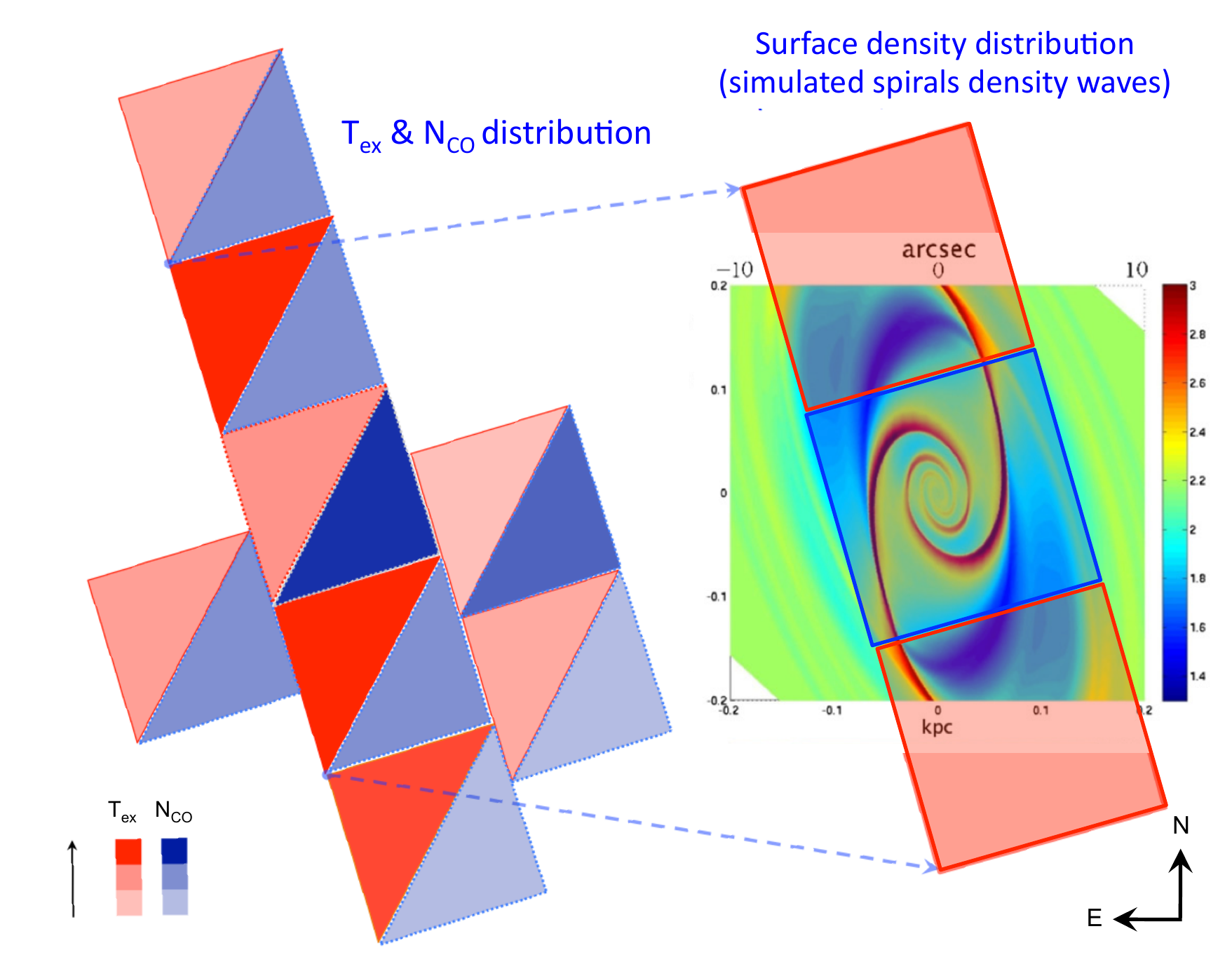}
\caption{Comparison between the results obtained at small scales using {\tt PACS} ({\it left}) with those derived in \cite{Lin11} ({\it right}). {\it Left:} Excitation temperature (T$_{ex}$; red colors) and column density (log N$_{CO}$; blue colors) values derived using {\tt PACS} data for the inner spaxels. The highest temperatures are found in correspondence of the two spaxels closed to the nucleus, in the northern and southern directions, while the peak in the column density is found in the central spaxel.
{\it Right (back):} Simulated spiral density waves overplotted to the velocity structure (\citealt{Lin11}) in a FoV of 20\arcsec$\times$20\arcsec. High density values are in red while lower values are in blue. {\it Right (front):} Corresponding {\tt PACS} spaxels overplotted onto the same FoV to show the respective excitation temperature distribution derived from {\tt PACS} data. 
Higher values of the simulated density map reveal a pair of tightly wound spirals in the center which correspond to the highest N$_{CO}$ and lower T$_{ex}$ in {\tt PACS} data. On the other hand, lower values in the surface density correspond to higher T$_{ex}$ and lower N$_{CO}$ according to {\tt PACS}. North is at the top and East to the left.}
\label{discus_fig}
\end{figure*}

This result is an agreement with that derived by \cite{Lin11}. In their work the authors analyzed in detail the $^{12}$CO(2--1) emission of the central region (20\arcsec$\times$20\arcsec) of this galaxy using Submillimeter Array (SMA) data, mainly focusing on its circumnuclear molecular gas emission.
They showed that the S-shaped structure of the isovelocity contours can be well reproduced by the bending generated by a shock along the spiral density waves, which are excited by a fast rotating bar. 
As a result, their simulated density map reveals a pair of tightly wound spirals in the center which pass through most of the ring-like (claimed to be a circumnuclear starburst ring by other authors) high-intensity region as well as intersect several Pa$\alpha$ emission line knots located outside the ring-like region (Fig.~\ref{discus_fig}). 
According to their scenario, the inner region of NGC 4945 is characterized by high column density surrounded by lower values in the nearby spaxels. As shown in Fig.~\ref{discus_fig}, at high density regions correspond low temperatures, and vice versa, in very good agreement with our {\tt PACS} results.

Our data does not have the high angular resolution required to reveal the distribution of the thermal structure of this region. However, \cite{Henkel18}, using ALMA data, found a good agreement with the simulated results presented by \cite{Lin11}. As mentioned in \S \ref{Intr}, \cite{Henkel18} found the presence of a dense and dusty nuclear disk (10\arcsec$\times$2\arcsec) which encloses an unresolved molecular core with a radius $\lesssim$2\arcsec\ (of the same size as the X--ray source observed with {\it Chandra}). 
At scales larger than those shown by \cite{Lin11}, \cite{Henkel18} also observed the presence of two bending spiral-like arms (one in the west turning toward the north-east and one in the east turning toward the south-west) at a radius of $\sim$300 pc ($\sim$15\arcsec) from the center. The arms are connected through a bar-like structure with a total length $\lesssim$ 20\arcsec\ along the east-west directions. 
They also suggested the presence of an inflow of gas from 300 pc down to 100 pc through the bar (see Fig.~26 in their work).

At the spatial scales sampled by our data, we found good agreement between the presence of a bar-like structure and tightly wound spiral arms (see \citealt{Ott01, Chou07, Lin11, Henkel18}) and our results. 
Accordingly, the mechanical heating produced by shocks, possibly driven by the outflow and by the bar potential, dominates in NGC~4945 at scales $\lesssim$20\arcsec\ ($\lesssim$360 pc).

\subsection{Dust heating }

Understanding the nature of the source that heats the dust in the nuclear regions of NGC 4945 is a topic under discussion. Many works suggested that the circumnuclear starburst, rather than the AGN activity, is the primary heating agent of the dust.

From the work by \cite{Brock88}, we know that $\lesssim$80\% of the total infrared emission of this object is enclosed in a region no larger than $<$12\arcsec$\times$9\arcsec. 
This size is comparable to the continuum source we derived from the multi-wavelength SED fitting analysis (20\arcsec$\times$10\arcsec) and characterized by two dust temperatures of 28 and 50 K. Within this area we found a total mass of dust of $\lesssim$10$^7$ M$_\odot$, in agreement with previous works (\citealt{WEIB08, Chou07}). The dust temperature we obtained is lower than the excitation temperature of the mid-J $^{12}$CO, characterized by energies from 55 K to 500 K above the ground state, as expected from mechanical heating.

Similar continuum source size at millimeter wavelengths have been derived from the works by \cite{Chou07} and \cite{Bendo16}. 
\cite{Chou07} derived a deconvolved continuum source size of 9.8\arcsec$\times$5.0\arcsec\ at 1.3 mm and slightly smaller ($\sim$7.6\arcsec$\times$2.0\arcsec) at 3.3 mm. These emissions are not aligned either with the starburst ring or with larger galactic disk, observed in Pa$\alpha$ by \cite{Marconi00}.

The size of the continuum source derived at centimeter wavelengths is much smaller (i.e., 7.4\arcsec$\times$3.4\arcsec\ at 21 cm) than that measured at 1.3 mm. Furthermore, the centimeter continuum source size and inclination are comparable with those of the starburst ring (i.e., $\sim$5\arcsec\ or 100 pc-scale and position angle $\sim$45 degree). They proposed that at cm wavelengths the star formation activity (as supernovae events) in the starburst ring can produce the nonthermal (synchrotron) emission which dominates at these wavelengths. \cite{Chou07} conclude that the dust emission dominates at 1.3 mm (and at 3.3 mm, with a radius of 4.9\arcsec), mainly heated by the star formation activity. The different origin of the dust emission at mm and cm wavelengths help us to understand the different continuum source sizes obtained.

On the other hand, \cite{Bendo16} analyzed the continuum emission at 85.69 GHz ad the H42$\alpha$ emission line deriving comparable size ($\sim$7\arcsec$\times$1.7\arcsec) with those measured by previous works at similar frequencies and emission lines (\citealt{Ott01, Cunningham05, Roy10}). 
They support the hypothesis that if the gas around the nucleus would be photoionized by the AGN they would expect to see a central peak excess in their exponential disk model (with scale length of 2.1\arcsec). 
Indeed, the spatial extent of the emission derived in their work, the absence of the peak excess and of any broad line emission suggest that both (continuum and line) emissions originate from the gas photoionized from the circumnuclear starburst, not from the AGN.

Our results suggest that star formation (in the circumnuclear starburst) is the main driving activity able to heat the dust. Unfortunately our data does not have the high angular resolution required to reveal the distribution of the thermal structure of the inner region. Continuum observations at higher angular resolutions as those feasible by using the Stratospheric Observatory For Infrared Astronomy ({\it SOFIA}; \citealt{Young12}) airborne observatory are required to obtain photometric measurements of the dust continuum in the far-IR in order to disentangle and to characterize the contributions produced by the starburst and the central AGN, respectively.

\section{Summary and conclusions}

Using data from the {\tt HIFI, SPIRE} and {\tt PACS} instruments onboard the {\it Herschel} satellite and {\tt APEX} we have been able to study the properties of the molecular ISM in the nearby galaxy NGC 4945, a remarkable prototype of AGN-SB composite galaxy. The spectroscopy data presented in this paper includes a combination of low and high density molecular gas tracers such as the $^{12}$CO, $^{13}$CO, HCN, HNC, CS, [CI], HCO$^+$ and CH. The main focus of the paper is to study the spatial distribution of $^{12}$CO emission over a large frequency range covering transitions from J= 3--2 to 20--19 combining all data available from the instruments mentioned above. We also preset the Spectral Energy Distribution (SED) of the dust continuum emission from 20 $\mu$m to 1.3 mm. 

Under the assumption of the local thermal equilibrium (LTE analysis, using {\tt MADCUBA}) excitation, we derived the excitation temperature T$_{ex}$ and the molecular column density N$_{mol}$ for $^{12}$CO, $^{13}$CO, HCN, HNC, CS, [CI], HCO$^+$ and CH combining {\tt HIFI} and, when available, {\tt APEX} data.
The highest column densities are found for $^{12}$CO, $^{13}$CO and [CI], ranging between log N$_{mol}$ $\sim$ 16.5-17.5 cm$^{-2}$. For the remaining molecules we derived a column density in the range log~N$_{mol}$=13-14 cm$^{-2}$ and T$_{ex}$ between 20-30 K. According to the non-LTE approach (using {\tt RADEX}) we derived moderate volume gas densities, with n(H$_2$) in the range 10$^3$-10$^6$ cm$^{-3}$. Lower values are derived for $^{12}$CO, $^{13}$CO and [CI] while higher values are obtained for the high density gas tracers such as HCN, HNC, HCO$^+$ and CS. From the low and high density tracers we derived in NGC 4945 gas volume densities (10$^3$--10$^6$ cm$^{-3}$) similar to those found in other galaxies with different type of activity.

We used the {\tt SPIRE} and {\tt PACS} spectroscopic data applying the LTE (multi-line transitions) analysis to derive the thermal and column density structures of the molecular gas $^{12}$CO for scales raging from $<$200 pc up to 2 kpc. We also analyzed the $^{12}$CO Spectral Line Energy Distribution (SLED) applying the \cite{Kazandjian15} models.

The main results can be summarized as follow:

\begin{description}
\item[$\bullet$] At large scale (700 pc--2 kpc), as obtained from {\tt SLW SPIRE}, we derived that high temperatures are mainly found in the northern part of the galaxy as well as in the disk plane. This suggests the presence of the outflow might affects the temperature observed in the north-west direction. The column densities are higher in the center and towards the south direction;

\vskip2mm
\item[$\bullet$] At intermediate scales (360 pc--1 kpc), as those obtained from the {\tt SSW SPIRE} data, the heating is distributed along the disk plane and in the south direction too. The column density follows a similar trend than that shown at large scales, characterized by high values in the center and towards the south;

\vskip2mm
\item[$\bullet$] When moving to smaller scales ($\sim$200 pc), using {\tt PACS} data, we found a peak in the column density in the central spaxel, surrounded by lower column density but higher excitation temperatures. This result is an agreement with that derived by \cite{Lin11} . They proposed the presence of tightly wound spirals in order to explain the distribution of the surface density and the related velocity field in a 20\arcsec$\times$20\arcsec\ region;

\vskip2mm
\item[$\bullet$] The thermal structure derived from the $^{12}$CO multi-transition analysis suggests that shocks seem to dominate the heating of the ISM in the nucleus of NGC~4945, located beyond 100 pc from the center of the galaxy. This result is confirmed by the mechanical heating models, which suggest the existence of PDRs but mainly dominated by mechanical heating (i.e., feedback from SNe) in the inner regions of NGC 4945.
We proposed that shocks are likely produced by the barred potential and the outflow, observed in the optical and X--ray bands. The high temperatures in the north and north-west directions, found at larger scales, might be related to the outflow;

\vskip2mm
\item[$\bullet$] The spectral energy distribution (SED) in NGC 4945 has been obtained combining photometric data from far-IR ({\tt MSX, PACS} and {\tt SPIRE}) and sub-mm ({\tt LABOCA}) and mm ({\tt SMA}) data. From the SED fitting analysis we derived that the source is resolved at these wavelengths. We found good results when two modified black body functions are considered, characterized by a cold (T$\sim$28~K) and a warmer temperature component of 50 K and a source size of $\Omega_s$=20\arcsec$\times$10\arcsec. 
Assuming a gas-to-dust ratio between 100 and 150, we derived in NGC 4945 a total gas mass in the range $\sim$8--11$\times$10$^8$ M$_\odot$ in a region of 40\arcsec$\times$40\arcsec. This result is consistent with the gas mass estimations derived in previous works (\citealt{WEIB08});

\vskip2mm
\item[$\bullet$] The hydrogen column density derived from the SED fitting is in good agreement with that derived by \cite{Wang04} using $^{12}$CO(1--0) and $^{12}$CO(2--1) transitions. NGC~4945 is characterized by N(H$_2$)$\sim$5$\times$10$^{23}$ cm$^{-2}$: this value is larger than those obtained for local starburst galaxies (e.g., NGC 253) and similar to those obtained for Compton-thick Seyfert 2 galaxies.

\end{description}

The results obtained in this work seem to confirm that the presence of the AGN in NGC 4945 has little impact on the thermal properties of its nuclear starburst. IR observations at higher spatial resolution are required to characterize both the dust and molecular line emissions. Spectroscopic and photometric observations like those could be achieved by the instruments onboard {\it SOFIA} (e.g., HAWC$+$, CASIMIR and/or GREAT) are needed in order to characterize the physical conditions of temperature and density as well as the structure of the emission itself in the very inner regions of NGC 4945.

\begin{acknowledgements}

We thank the anonymous referee for useful comments and suggestions which helped us to improve the quality and presentation of the manuscript. 
E.B. acknowledges the support from Comunidad de Madrid through the Atracci\'on de Talento grant 2017-T1/TIC-5213. J.M.P. acknowledges partial support by the MINECO and FEDER funding under grant ESP2017-86582-C4-1-R and PID2019-105552RB-C41. M.A.R.T. acknowledges support by the APEX project M-081.F-0034-2008. S.G.B. acknowledges support from the research projects PGC2018-094671-B-I00 (MCIU/AEI/FEDER, UE) and PID2019-106027GA-C44 from the Spanish Ministerio de Ciencia e Innovaci\'on.
\newline

This work is based on observations acquired with the {\it Herschel} Satellite, obtained from the ESA {\it Herschel} Science Archive, and with the {\it APEX} antenna, obtained from the project M-081.F-0034-2008.

This research has made use of: 1) the ESA {\it Herschel} Science Archive; 2) the NASA/IPAC Extragalactic Database (NED), which is operated by the Jet Propulsion Laboratory, California Institute of Technology, under contract with the National Aeronautics and Space Administration.

\end{acknowledgements}

\bibliographystyle{aa} 
\bibliography{biblio}

\appendix

\section{Deriving the flux density values for the SED fitting analysis}
\label{App_SED}

In this appendix we present how the flux densities, used in the SED fitting in \S~\ref{par_SED}, are computed for the different wavelengths. 
The additional data are characterized by different units, we then summarize here the several conversion factors applied to transform the flux densities in {\it Jansky} as well as the formula applied for the conversion. 
In Tab.~\ref{CF_values} we show the computed flux density values (with their associated errors), the conversion factors and the pixel sizes for each band. 

\begin{itemize}
\setlength\itemsep{0.5mm}

\vskip2mm
\item  For the {\tt MSX} catalogue we used the following formula: 

\vskip-4mm
\begin{equation}
\hskip5mmF(40\arcsec) = F_{OBS}\cdot CF \cdot \left(\frac{pixel\hskip1mmsize}{3600}\frac{2\pi}{360}\right)^2. 
\end{equation}

The conversion factor (CF) transform the [W m$^{-2}$ sr$^{-2}$] into [Jy  sr$^{-1}$] and they have been taken from the website~\url{http://irsa.ipac.caltech.edu/applications/MSX/MSX/imageDescriptions.htm}. 

\vskip2mm
\item Taking into account the {\tt LABOCA} data from \cite{WEIB08} we considered the flux density of 15.8 ($\pm$1.6) Jy computed in a beam of 80\arcsec$\times$ 80\arcsec. From their work we also know that in a beam of 19\arcsec$\times$15\arcsec\ (average beam 17\arcsec) the flux density is 7.2 ($\pm$0.8) Jy. If we assume uniform emission outside the beam of 17\arcsec\ ($<$80\arcsec) we are able to compute the flux density in a annular beam with size in between 17\arcsec\ and 40\arcsec:
\vskip-4mm
\begin{equation}
 \frac{(15.8 - 7.2)[Jy]}{(80^2- 17^2)[arcsec]} \cdot (40^2 - 17^2) [arcsec] = 1.85\hskip1mm [Jy].
\label{Conversion}
\end{equation}

This can be considered a reasonable assumption according to the {\tt LABOCA} data from which we derived that half of the total flux density emission is enclosed in a beam $\lesssim$20\arcsec, while the remaining emission is diluted in the annular beam between $\sim$20\arcsec\ and 80\arcsec.

Finally, to derive the flux density in a beam of 40\arcsec, we added the flux density value obtained in the annular beam to the flux density obtained in a beam of 17\arcsec. This gives a total flux density (in a beam of 40\arcsec) of 7.2~Jy~+~1.85 Jy = 9.05 (1.3) Jy. 

\end{itemize}

\vspace{1mm}

\begin{table}[h]
\vskip1cm
\centering
\begin{tiny} 
\caption{Flux densities derived from the archive and from the literature for the additional data.}
\label{CF_values}
\begin{tabular}{cccccc} 
\hline\hline\noalign{\smallskip}  
Data 	&  band		&	Flux  	&	Conversion  & Pixel    & Ref.\\
 	&  		&	density  	&	 factor &  size   & \\
 			& ($\mu$m)   & 	(A.U.)     & 					&    (\arcsec)   & 	\\
(1) & (2) & (3) & (4) & (5) & (6) \\
\hline\hline\noalign{\smallskip}  
{\tt MSX}		& 21.34		&      0.27 (0.07)$\times$10$^{-3}$	& 2.476$\times$10$^{13}$ & 6  &  a\\
\hline\noalign{\smallskip}  
{\tt LABOCA}		& 870		& 9.05 (1.3) &  Eq.~\ref{Conversion} &  --		& b \\
\hline\hline\noalign{\smallskip}  
\end{tabular}
\vskip2mm
\begin{minipage}{9cm}
{{\bf Notes:} Column (1): data considered; Column (2): central wavelength of the band; Column (3): flux density (and uncertainty) in the aperture in arbitrary units (see text); Column (4): conversion factor to derive the flux density in {\it Jansky}; Column (5): pixel size in {\it arcsec}; Column (6): reference considered to get the data: (a) {\tt MSX} Archive, (b) \cite{WEIB08}.}
\end{minipage}
\end{tiny}
\end{table}

\end{document}